\documentclass[11pt]{article}
\usepackage[left=1in,top=1in,right=1in,bottom=1in,head=0in,nofoot]{geometry}

\usepackage{setspace,url,bm,amsmath} 
\usepackage{scalerel}

\usepackage{xcolor}
 \usepackage{multirow}
 \usepackage{arydshln}
 \usepackage{mathtools}
\usepackage{titlesec} 
\titlelabel{\thetitle.\quad} 
\titleformat*{\section}{\bf\Large\center}

\usepackage{authblk}
\usepackage{float}
\usepackage{graphicx} 
\usepackage{bbm}
\usepackage{latexsym}
\usepackage{caption}
\usepackage[margin=20pt]{subcaption}
\usepackage{enumerate}
\graphicspath{ {./Graphs/} }

\usepackage{amsthm}
\usepackage{amssymb}
\usepackage{amsmath}
\usepackage{color}

\usepackage{comment}
\theoremstyle{definition}
\newtheorem{assumption}{Assumption}
\newtheorem*{theorem*}{Theorem}
\newtheorem{theorem}{Theorem}
\newtheorem*{rmk*}{remark}
\newtheorem{proposition}{Proposition}
\newtheorem{lemma}{Lemma}
\newtheorem{example}{Example}

\newtheorem{condition}{Condition}

\newtheorem{corollary}{Corollary}
\newtheorem*{corollary*}{Corollary}

\usepackage{natbib} 
\bibpunct{(}{)}{;}{a}{}{,} 

\usepackage{etoolbox} 
\apptocmd{\sloppy}{\hbadness 10000\relax}{}{} 

\usepackage{multibib}
\newcites{sec}{References}

\usepackage{color}
\usepackage{listings}
\usepackage{hyperref}
\usepackage{booktabs}

\usepackage{lscape}

\RequirePackage[normalem]{ulem}

 \def \cG {{\mathcal{G}}}
 
\def \P{\mathbb{P}}

\def \E {\mathbb{E}}

\newcommand{\bcol}[1]{{\color{black}  #1}}
\newcommand{\indep}{\perp \!\!\! \perp}

\usepackage{bm}

\allowdisplaybreaks

\begin{document}

\singlespacing

\title{\bf The Promises of Multiple Experiments: Identifying  Joint Distribution of Potential  Outcomes}

\author[1]{Peng Wu}
\author[2]{Xiaojie Mao\thanks{Corresponding author: maoxj@sem.tsinghua.edu.cn}}
\affil[1]{\small School of Mathematics and Statistics, Beijing Technology and Business University, 100048, China}
\affil[2]{\small School of Economics and Management, Tsinghua University, Beijing 100084, China}


\date{}

\maketitle

\begin{abstract}
Typical causal effects are defined based on the marginal distribution of potential outcomes. However, many real-world applications require causal estimands involving the joint distribution of potential outcomes to enable more nuanced treatment evaluation and selection. In this article, we propose a novel framework for identifying and estimating the joint distribution of potential outcomes using multiple experimental datasets. 
We introduce the assumption of transportability of state transition probabilities for potential outcomes across datasets and establish the identification of the joint distribution under this assumption, along with a regular full-column rank condition.  
The key identification assumptions are testable in an overidentified setting and are analogous to those in the context of instrumental variables, with the dataset indicator serving as ``instrument".  Moreover, we propose an easy-to-use least-squares-based estimator for the joint distribution of potential outcomes in each dataset, proving its consistency and asymptotic normality. We further extend the proposed framework to identify and estimate principal causal effects. We empirically demonstrate the proposed framework by conducting extensive simulations and applying it to evaluate the surrogate endpoint in a real-world application.
\end{abstract}


\medskip 
\noindent 
{\bf Keywords}: 
Causal Inference, Data Fusion, Principal Stratification, Surrogacy Evaluation

\newpage

\onehalfspacing


\section{Introduction}  \label{sec1}
Estimating the causal effect of a treatment variable on an outcome is a fundamental scientific problem, central to many fields such as social and biomedical sciences~\citep{Imbens-Rubin2015, Pearl-Mackenzie2018, Hernan-Robins2020, Rosenbaum2020}. 
Typical causal effects like average treatment effects are defined in terms of the marginal distribution of potential outcomes. 
However, in many real-world applications, causal estimands involving the joint distribution of potential outcomes are  required to enable more nuanced treatment evaluation and selection.  
 Examples include the probability of causation~\citep{pearl1999, Dawid2022, lu2023evaluating}, the effect of persuasion~\citep{Jun-Lee2023, Jun-Lee2024}, treatment benefit rates and treatment harm rates~\citep{shen2013treatment, Yin-etal2018, 2022NathanFacct, 2022nathan, li2023trustworthy,  Wu-etal-2024-Harm}, and so on. 

Identifying the joint distribution of potential outcomes from a single dataset presents a significant challenge. This difficulty arises from the fact that only one potential outcome is observed for each individual, which is known as the fundamental problem in causal inference \citep{Holland1986}. Consequently, the joint distribution of potential outcomes is generally unidentifiable, even in randomized controlled trials  \citep{Pearl-etal2016-primer}. Instead of identifying the joint distribution, many studies focus on deriving bounds for causal estimands involving the joint distribution \citep[e.g., ][]{tian2000probabilities, Zhang-etal2013, Yin2018-second, li2022probabilities, 2022nathan}.  

In this article, we propose a novel framework for identifying and estimating the joint distribution of potential outcomes using multiple experimental datasets.  
Specifically, we make the following three contributions. 
\bcol{
First, our primary contribution is to study identification of the \emph{joint
distribution of potential outcomes}, which in turn defines many important causal quantities. For binary outcomes, we introduce the assumption of
transportability of state transition probabilities, namely that the conditional law of the treated potential outcome
$Y^1$ given the untreated potential outcome $Y^0$ is invariant across trials, together with a full-column rank condition
requiring sufficient cross-trial variation in the untreated potential outcome
distribution. Under these conditions, we establish identification of the joint distribution of $(Y^1, Y^0)$ for each trial. We further extend this result to general
categorical outcomes. In an overidentified setting, the same identifying restrictions
also generate observable implications that can be assessed empirically. The resulting
identification strategy is analogous to instrumental-variable reasoning, with the
dataset indicator serving as a source of identifying variation.

Second, we propose a simple least-squares-based framework for estimation and
inference. The key idea is that the observed treated-arm outcome distribution in
each trial can be written as a linear function of the untreated potential outcome
distribution and the common transition parameters. Building on this representation,
we derive least-squares estimators for the invariant transition probabilities and the induced joint
distributions, establish their consistency and asymptotic normality, and develop an
overidentification test for the observable restrictions implied by the transportability
assumption. The resulting procedure is easy to implement and only relies on
arm-specific summary statistics, which may be useful when individual-level data are
not directly available.

Third, when an additional binary post-treatment variable  is observed before the
final outcome, we extend the framework to principal stratification. We first develop
new transportability-based identification routes for the joint distribution of
the potential surrogates and outcomes, with and without additional monotonicity restrictions. These
routes can accommodate trial-specific principal stratification average
causal effects (PSACEs). We then show how these results complement the  monotonicity-based route of \citet{Jiang-etal2016}. In this sense, PSACEs are treated here as an important application of a broader joint-distribution
framework.}

\bcol{Throughout the main text, we focus on a covariate-free, nonparametric multi-trial identification framework. We make this choice deliberately rather than incidentally, and it aligns with our empirical application, in which no individual-level baseline covariates are available. In Supplementary Material \ref{app:covariates}, we discuss how to extend the proposed framework to accommodate baseline covariates.}

The idea of combining two or more datasets to identify causal effects and enhance estimation efficiency has garnered significant attention in the field of causal inference~\citep{Colnet-etal2023, Degtiar-Rose2023, Wu-etal-2024-Compare,athey-etal2019, Yang-Ding2020, imbens2024long, kallus2024role, hu2023longterm, Humermund-Bareinboim2025}. However, most existing methods consider causal effects involving the marginal distribution of potential outcomes. In contrast, we focus on identifying and estimating causal estimands that involve the joint distribution of potential outcomes, thus extending and complementing this class of methodologies. 



The rest of this article is organized as follows. In Section \ref{sec2}, we outline the basic setup. In Section \ref{sec3}, we present the identification assumptions and establish the identification for the joint distribution of potential outcomes. In Section \ref{sec4}, we propose the least-squares-based method for estimating the transition probabilities and present a method for testing the key assumption of transportability of state transition probabilities. Section \ref{sec6} extends the proposed method to identify and estimate PSACEs. Section \ref{sec7} evaluates the finite-sample performance of the proposed method through extensive simulations. Section \ref{sec8} demonstrates the proposed methods in a real-world application, and Section \ref{sec9} concludes the paper. 


\section{Setup}    \label{sec2}

Let $A \in \mathcal{A} = \{0, 1\}$ represent a binary treatment, where $A=1$  indicates the treatment condition and $A = 0$ indicates the control condition.  
Let $Y\in \mathcal{Y}$ be the outcome of interest.  
To define causal estimands, we adopt the potential outcome framework~\citep{neyman1923, Rubin1974} and denote $Y^1$ and $Y^0$ as the potential outcomes under treatment and control, respectively.  We make the stable unit treatment value assumption (i.e., no multiple  treatment versions and no interference across units), which ensures the well-definedness of potential outcome $Y^a$~\citep{Imbens-Rubin2015}. Accordingly, the observed outcome is linked to the potential outcomes through $Y = (1-A)Y^0 + A Y^1$. 

Suppose we have access to data from individuals in a collection of $m$  trials, indexed by $\cG = \{1, ...,  m\}$.   
For each trial $g \in \cG$, the observed data consist of realizations of independent random tuples $\{(G_i = g, X_i, A_i, Y_i), i=1, ..., n_g\}$, where $n_g$ denotes the number of individuals in trial $g$, and we let $n = \sum_{g=1}^{m} n_g$ represent the total sample size. We adopt a non-nested design \citep{dahabreh2021study} in which the observed samples from different trials are independent. Without loss of generality, under a superpopulation model $\P$,  
 we assume that the observed data in trial $g$ form an \emph{i.i.d.} sample from $\P(\cdot \mid G=g)$, which implies that all observed data are obtained by stratified sampling from a population that is stratified by $G$. 
   




In this paper, our primary objective is to identify and estimate the joint distribution of potential outcomes $Y^0, Y^1$ from the data of multiple trials. Given that there might be potential distribution discrepancies across trials, we are particularly interested in the distribution $\P(Y^0, Y^1 \mid G = g)$ for each trial $g \in \mathcal{G}$. 
Notably, this task diverges from and is more challenging than what most existing data-combination-based causal inference methods can handle. The latter mainly focuses on estimating causal effects defined in terms of the marginal distribution of potential outcomes \citep[e.g.,][]{Rosenman2023,Colnet-etal2023, Wu-etal-2024-Compare}.


We mainly consider the case of a binary outcome $Y \in \{0, 1\}$ and number of trials $m \geq 3$ (i.e., at least three trials). This case is of particular interest because it leads to overidentification and offers an opportunity to test the key identification assumption. Additionally, for identifying the principal causal effects, we require $m \geq 4$ due to the extra complexity introduced by principal stratification \bcol{and relax this to $m \ge 2$ under additional monotonicity conditions}, see Section \ref{sec6} for details.    


\section{Nonparametric Identification}  \label{sec3}


\bcol{In this section, we establish the nonparametric identification of the joint distribution of potential outcomes. We do not consider any covariates in the main text and defer the discussion of covariates to Supplementary Material \ref{app:covariates}. }  


\subsection{Basic Assumptions}  \label{sec3-1}

We begin by presenting the basic assumptions required to identify $\P(Y^0, Y^1 \mid G=g)$. 

\begin{assumption}[Within-trial Randomization]  \label{assump1}  For all $g \in \cG = \{1, ..., m\}$, 
 (i) $A \indep (Y^0, Y^1) \mid G = g$, and  
  (ii) $0 <  \P(A=1\mid G=g) < 1$.
\end{assumption}

Assumption \ref{assump1} means that the treatment $A$ is randomized within each trial and both treatment arms are assigned with a positive probability. This assumption trivially holds for data from randomized trials, and it identifies the marginal potential outcome distributions within each trial. However, it is insufficient to identify the joint distribution $\P(Y^0, Y^1 \mid G=g)$, \bcol{because the observed data only provide information about single-world marginals rather than the cross-world dependence between $Y^0$ and $Y^1$. Therefore, if our goal is to identify the joint distribution of potential outcomes, some cross-world structure is unavoidable. Moreover, because we aim to combine information across multiple trials, some cross-trial structure is also needed; otherwise there is no basis for borrowing information across trials. We therefore introduce the following invariance assumption.}


\begin{assumption}[Transportability of state transition probabilities] \label{assump2} $Y^1\indep G\mid Y^0$.   
\end{assumption} 

\bcol{
Assumption \ref{assump2} requires that the cross-world conditional distribution of $Y^1$ given $Y^0$ is invariant across trials, i.e.,
$   \P(Y^1 = 1  \mid Y^0, G=1) =  \P(Y^1 = 1  \mid Y^0, G=2)   = \cdots =   \P(Y^1 = 1 \mid Y^0, G=m)$.
This means that the transition probabilities from the untreated state $Y^0$ to the treated state $Y^1$ are the same across trials.
}

\bcol{A useful way to interpret Assumption \ref{assump2} is through \emph{untreated prognosis}. In many applications, $Y^0$ can be viewed as a latent summary of an individual’s untreated prognosis, or natural outcome tendency under control, whereas $Y^1$ represents the outcome after treatment acts on this prognosis. Under this interpretation, Assumption \ref{assump2} states that, once untreated prognosis is fixed, the treatment-response mechanism is stable across trials, even though the marginal distribution of untreated prognosis may differ across trials. In Section \ref{sec8}, our application focuses on the Adjuvant Colon Clinical Trials (ACCTs), where $Y^1$ and $Y^0$ refer to cancer survival with and without treatment, respectively. In this setting, Assumption \ref{assump2} means that, given the untreated prognosis summarized by $Y^0$, the survival response under fluorouracil-based chemotherapy is approximately the same across trials. The same logic may also arise in other contexts, where populations may differ in baseline natural outcome tendency across sites while the incremental response mechanism conditional on the natural tendency is comparatively stable.}

\bcol{Assumption \ref{assump2} can also be understood from the perspective of effect modifiers. In particular, it implies that any effect modification by $G$ operates through the untreated potential outcome $Y^0$. That is, once we condition on $Y^0$, the trial indicator $G$ does not further modify the treatment response. Thus, Assumption \ref{assump2} still allows substantial heterogeneity across trials in the marginal distributions of potential outcomes and in average treatment effects, but it requires the conditional transition law from $Y^0$ to $Y^1$ to remain stable.
This is weaker than the full transportability assumption $G \indep (Y^0, Y^1)$ that requires identical effects across trials. 
This stronger assumption is ruled out in our analysis of the ACCT data, as shown in Figure \ref{fig1} in Section \ref{sec8}, where the average treatment effects vary across trials.
}

\bcol{Assumption \ref{assump2} has both strengths and limitations. On the one hand, it has a clear scientific interpretation as a stable transition mechanism from untreated prognosis to treated outcome, while allowing substantial cross-trial heterogeneity through differences in $\P(Y^0 \mid G=g)$. On the other hand, it is a substantive cross-world assumption and is not guaranteed by randomized design alone. It may fail when treatment versions, adherence, supportive care, endpoint definitions, or latent effect modifiers differ across trials in ways that still affect treatment response after conditioning on $Y^0$.
}

\bcol{Importantly, Assumption \ref{assump2} has observable implications in an overidentified setting, namely when the number of trials is strictly larger than the number of outcome categories. For a binary outcome, this means $m \ge 3$. In this case, the same transition probabilities must simultaneously explain the observed arm-specific outcome distributions across all trials. In Section \ref{sec-test}, we develop a chi-squared overidentification test for the restrictions implied by Assumption \ref{assump2}. Such a test assesses whether the observable implications of Assumption \ref{assump2} are compatible with the data. A rejection indicates that these implied restrictions are violated by the observed data, whereas a failure to reject should only be interpreted as a lack of evidence against Assumption \ref{assump2}. We also provide visual diagnostics in Section \ref{sec8}.
}

\subsection{Identifiability for Joint Distribution of Potential Outcomes}   \label{sec3-2}

An important implication of Assumptions \ref{assump1}--\ref{assump2} is that the joint distributions of potential outcomes are identifiable from the data of multiple trials. Assumption \ref{assump2} imposes invariant potential outcome state transition probabilities across trials:
\begin{equation*}
    \pi_{b|a} :=   \P(Y^1 = b \mid  Y^0 = a)  =  \P(Y^1 = b \mid  Y^0 = a, G=g),  \quad a,b=0,1; g \in \mathcal{G}.    
\end{equation*}
By law of total probability, the following holds for all $g \in \mathcal{G}$:
\begin{equation}\label{eq1} 
    \P(Y^1 = 1 \mid G=g) =     \P(Y^0 = 0 \mid  G=g) \cdot \pi_{1|0} +    \P(Y^0 = 1 \mid  G=g) \cdot \pi_{1|1},
\end{equation}
where $\P(Y^1 =1 \mid G=g)$ and $\P(Y^0 = a \mid G=g)$ for $a = 0, 1$ are all identifiable under Assumption \ref{assump1}. 
\bcol{Therefore, identifying $\pi_{1|0}$ and $\pi_{1|1}$ amounts to solving the system of $m$ linear equations in \eqref{eq1}.
The solution is unique, and hence point identification holds, under the following full-rank condition on the coefficient matrix in \eqref{eq1}.}

\begin{condition}[Full-column rank] \label{cond1}  The matrix  $(\P(Y^0 = 0 \mid G= g),  \P(Y^0 = 1  \mid G= g )_{m \times 2}$ 
     has a full-column rank (i.e., rank $2$).  
\end{condition}

\bcol{For a binary outcome, Condition \ref{cond1} is equivalent to requiring that the untreated outcome distribution differs across at least two trials, namely that the untreated base rate $\P(Y^0=1 \mid G=g)$ is not the same for all $g$. 
Thus, for point identification, the condition does not require the untreated base rates to be \emph{very} different; any nonzero cross-trial variation is sufficient in principle. At the same time, larger variation generally improves numerical stability and finite-sample precision, whereas near collinearity in the design matrix can make estimation more variable. 
}

\bcol{At first glance, Assumption \ref{assump2} and Condition \ref{cond1} may appear to pull in different directions. Assumption \ref{assump2} requires the conditional law $\P(Y^1 \mid Y^0, G=g)$ to be invariant across trials, whereas Condition \ref{cond1} requires the marginal untreated outcome distribution $\P(Y^0 \mid G=g)$ to vary across trials. We do not view these two requirements as being in any necessary tension. Rather, they concern different aspects of the data-generating structure and play complementary roles in identification. Condition \ref{cond1} provides the identifying leverage through cross-trial variation in untreated prognosis, whereas Assumption \ref{assump2} imposes a common conditional transition law that links these heterogeneous trial populations. Different trials may therefore have different untreated base rates, and even different treated marginal risks, while still sharing the same conditional response mechanism once $Y^0$ is fixed. The key substantive issue is whether the factors driving cross-trial variation in $Y^0$ are adequately summarized by $Y^0$ itself. If they are, then marginal heterogeneity in $Y^0$ and conditional invariance of $Y^1 \mid Y^0$ are compatible; if they are not, then Assumption \ref{assump2} becomes less plausible.}


\bcol{Condition \ref{cond1} is satisfied only if $m \ge 2$ for a binary outcome. If we only have access to data from a single trial, one might imagine splitting the data into two independent subsamples to create two pseudo-trials. However, this would not generate new identifying variation: although the two subsamples would be independent, their untreated outcome distributions would be the same at the population level, so Condition \ref{cond1} would fail.
}


\begin{theorem}[Binary outcome] \label{prop1} Under Assumptions \ref{assump1}--\ref{assump2} and Condition \ref{cond1}, the distributions $\P(Y^1 \mid Y^0, G=g)$ and $\P(Y^1, Y^0 \mid G=g)$ for $g\in \cG$ are identifiable.  
\end{theorem}

\bcol{Theorem \ref{prop1} shows that the conditional distribution $\P(Y^1 \mid Y^0, G=g)$ and the joint distribution $\P(Y^1, Y^0 \mid G=g)$ are identified for each trial $g \in \cG$. Intuitively, Assumption \ref{assump2} implies that the same transition probabilities $(\pi_{1|0}, \pi_{1|1})$ must fit all trials simultaneously, and Condition \ref{cond1} ensures that the resulting linear system has a unique solution. Once $\P(Y^1 \mid Y^0, G=g)$ is identified, the joint distribution follows immediately from $\P(Y^1=b, Y^0=a \mid G=g)
=
\P(Y^1=b \mid Y^0=a, G=g)\P(Y^0=a \mid G=g)$ for $a,b \in \{0,1\}$.}


 The identification Assumption \ref{assump2} and Condition \ref{cond1} resemble the standard assumptions in the context of instrumental variable (IV) analysis~\citep{Angrist-etal1996, Imbens2004}. We may view the dataset indicator $G$ as an ``IV", the potential outcome $Y^0$ as the ``treatment", and the potential outcome $Y^1$ as the ``outcome". Then  Assumption \ref{assump2} corresponds to the IV exclusion restriction assumption, and Condition \ref{cond1} corresponds to the full rank assumption of IV on treatment (or IV relevance), which requires a strong enough association between $G$ and $Y^0$.

\bcol{We note that one may instead impose the symmetric assumption $Y^0 \indep G \mid Y^1$. Then, by the same argument as above, $\P(Y^0, Y^1 \mid G=g)$ remains identifiable provided that the marginal distribution of $Y^1$ varies sufficiently across trials, namely that the matrix $\bigl(\P(Y^1 = 0 \mid G=g),, \P(Y^1 = 1 \mid G=g)\bigr)_{m\times 2}$ has full column rank. We focus on Assumption~\ref{assump2} because it is more natural scientifically: in many applications, $Y^0$ represents untreated prognosis, and $\P(Y^1 \mid Y^0, G)$ then describes how treatment acts on that prognosis. By contrast, $\P(Y^0 \mid Y^1, G)$ has backward-looking interpretation. Thus, although the two formulations are mathematically symmetric, Assumption~\ref{assump2} provides the more natural primary formulation.}

  

Following Theorem \ref{prop1}, we can identify various causal estimands involving the joint distribution of potential outcomes. Below are some examples. 

\begin{example}[Probability of causation]
    Causal inference involves not only evaluating the effects of causes but also deducing the causes of given effects~\citep{Dawid2022}, where the latter is also referred to as attribution analysis~\citep{pearl2009causality, Pearl-etal2016-primer, Pearl-Mackenzie2018}.   
The probability of sufficient causation (PS) and the probability of necessary causation (PN) are two standard quantities for attribution analysis. They are defined by  $\textup{PS}(A \Rightarrow Y)    = \P( Y^1= 1 \mid A = 0, Y = 0 )$  and $\textup{PN}(A \Rightarrow Y)  = \P(  Y^0 = 0 \mid A=1, Y=1 )$ respectively.   
For example, the quantity $\textup{PN}(A \Rightarrow Y)$ can be written as $ \P(  Y^0 = 0, Y^1 = 1 \mid A=1 )/\P(Y^1=1 \mid A=1)$, 
where the denominator is an identifiable quantity and the numerator equals to $\P(Y^0 = 0, Y^1 = 1)$ for randomized treatment assignment. 
\end{example}

\begin{example}[Effect of persuasion]
Let $A \in \{0, 1\}$ be a binary indicator for an individual's exposure to certain persuasive information, and let $Y$ be a binary indicator representing the individual's behavior, \bcol{where $Y=0$ denotes a negative response to the information and $Y=1$ denotes a positive response}.   
\citet{Jun-Lee2023} defined the persuasion rate as $\P(Y^1 = 1 \mid Y^0 = 0)$, which quantifies the proportion of individuals in the subpopulation with \bcol{$Y^0 = 0$ who would change  from negative to positive behavior as a result of the exposure to persuasive information}. 
\end{example}

\begin{example}[Treatment benefit and harm rates] 
 Denote $Y=1$ as a favorable outcome (e.g., survival) and $Y=0$ as an unfavorable outcome (e.g., death). 
The average treatment harm rate~\citep{shen2013treatment, Wu-etal-2024-Harm} is defined as $\textup{THR} = \P(Y^0 = 1, Y^1 = 0),$ 
 which quantifies the percentage of individuals experiencing worse outcomes under treatment than under control. 
Similarly, the average treatment benefit rate is defined as $\textup{TBR} = \P(Y^0 = 0, Y^1 = 1).$
\end{example}

\bcol{In some scenarios, the target data may contain only control units~\citep{li2023improving}. 
For example, this could arise if the target population has only received an old drug ($A=0$) and researchers want to assess the efficacy of a new drug ($A=1$) unavailable in this population.   
Our identification result can be readily generalized to such a control-only target population.} 

\bcol{\begin{example}[Generalization to a control-only target population]
Consider a simple random sample from the target population with only control units (denoted by $G = 0$). 
 In the control-only dataset, we can straightforwardly identify $\P(Y^0 = a \mid  G=0)$. Moreover,applying Theorem \ref{prop1} allows us to identify $\pi_{b \mid a}$ for $a,b=0, 1$ from the $m$ trials. Taken together, these results yield identification of the joint distribution in the control-only   target population: $ \P(Y^0 = a, Y^1=b \mid G=0) =  \pi_{b|a} \cdot \P(Y^0 = a \mid  G=0)$. 
\end{example}}

\subsection{Extension to Categorical Outcomes}

The identifiability results can be also extended to general categorical outcomes. Specifically, for a categorical outcome with cardinality $k$,  let $\P(Y^0  \mid G) = ( \P(\bcol{Y^0} = i  \mid G=g) )_{m \times k}$ be a $m \times k$ matrix whose $(g, i)$-th element is $\P(\bcol{Y^0} = i  \mid G=g)$. 

\begin{condition}  \label{cond2}
The matrix  $\P(Y^0 \mid G)$ has full-column rank.
\end{condition} 
Similar to Condition \ref{cond1}, Condition \ref{cond2} holds only when $m \geq k$. 

\begin{theorem}[Categorical outcome] \label{thm1}  Let $Y$ be a categorical outcome with $k$ possible values. Under Assumptions \ref{assump1}--\ref{assump2} and Condition \ref{cond2}, the  joint distributions $\P(Y^1, Y^0  \mid  G=g)$ for $g \in \cG$ are identifiable.  
\end{theorem}

Theorem \ref{thm1} extends the result of Theorem \ref{prop1}. \bcol{We also discuss extensions to more general treatment settings; see Supplementary Material \ref{app-multivalued-treatment} for details.}
After establishing identifiability, we then consider the estimation of the joint distributions and the testing of the key Assumption \ref{assump2}.  

\section{Estimation and Inference}  \label{sec4}
In this section, we first present an estimator of $\pi_{b|a}$ for $a, b=0,1$, which are key invariant parameters for identifying the joint distribution $\P(Y^1, Y^0 \mid G=g)$.
Then, we provide a method to test   Assumption \ref{assump2} when $m = |\mathcal{G}| \geq 3$.  

\subsection{Estimation}  \label{sec4-1}

We now present a simple least-squares-based estimation method for $\pi_{b|a}$, $a, b \in \{0,1\}$. Under Assumptions \ref{assump1}--\ref{assump2},  (\ref{eq1}) gives the following for $g \in \mathcal{G}$: 
  {\small\begin{equation}   \label{eq2} 
		\P(Y = 1   \mid G=g, A=1)   =  \pi_{1|0}   \P(Y = 0  \mid  G=g, A=0)  + \pi_{1|1}   \P(Y = 1  \mid  G=g, A=0)   
\end{equation}}	 

When $m = 2$, the quantities $\pi_{b|a}$ for $a, b \in \{0,1\}$ are \emph{just-identified} as the number of parameters to be identified equals the number of equations. For example, when $m = 2$, 
solving the two equations in (\ref{eq2}) identifies $\pi_{1|0}$ and $\pi_{1|1}$. Then $\pi_{0|0}$ and  $\pi_{0|1}$ are identified through $\pi_{0|0}= 1 -  \pi_{1|0}$ and $\pi_{0|1} = 1 - \pi_{1|1}$, respectively.  In contrast, if $m > 2$,  (\ref{eq2}) includes more equations than parameters. In such a case, the parameters $\pi_{b|a}$ for $a, b \in \{0,1\}$ are \emph{overidentified}.  


In both just-identified and overidentified cases,  
we can estimate $\theta := (\pi_{1|0}, \pi_{1|1})$ via a linear least-squares estimator.
Formally,  we let 
  $\tilde Y_g = \P(Y=1\mid G=g, A=1),$ $\tilde X_{1g} = \P(Y=0\mid G=g, A=0),$ and $\tilde X_{2g} = \P(Y=1\mid G=g, A=0)$, and denote $\tilde X_g = (\tilde X_{1g}, \tilde X_{2g})^\intercal$.   
By \eqref{eq2}, we have $\tilde Y_g = \tilde X_g^\top \theta$ for $g\in\mathcal{G}$. Under the full rank condition (Condition \ref{cond1}), we can further write $\theta$ as
$$\theta =  \left ( \frac{1}{m} \sum_{g=1}^m \tilde X_{g} \tilde X_{g}^\intercal \right )^{-1} \cdot \frac{1}{m} \sum_{g=1}^m \tilde X_g \tilde Y_g.$$ 
Although the conditional probabilities $\tilde X_g$ and $\tilde Y_g$ are unknown, they can be easily estimated by the corresponding sample frequencies, e.g.,  $\hat Y_g = \sum_{i=1}^n \mathbb{I}(Y_i = 1, G_i = g, A_i = 1) \big / \sum_{i=1}^n \mathbb{I}(G_i = g, A_i = 1) $. Given estimators  $\hat X_g$ and $\hat Y_g$, the resulting estimator for $\hat\theta$ is  
   \[  \hat \theta = \left ( \frac{1}{m} \sum_{g=1}^m \hat X_{g} \hat X_{g}^\intercal \right )^{-1} \cdot \frac{1}{m} \sum_{g=1}^m \hat X_g \hat Y_g.  \] 
This amounts to the ordinary least squares (OLS) coefficient estimator with $\hat Y_g$ as the response and $\hat X_g$ as the covariates. It is noteworthy that while $\hat \theta$ resembles the OLS estimator in a linear model, \bcol{it differs in that} the  ``effective sample size'' $m$ is fixed. 
 A potential strength of this estimator is that it does not need the individual-level data. Instead, it only involves some sample summary statistics $\hat X_g$ and $\hat Y_g$. This may be helpful for privacy protection~\citep{Han-etal-2023}.    
Next, we show that estimator $\hat \theta$ is consistent and asymptotically normal. \bcol{
In our theoretical analysis, for each trial $g$, the sample size $n_g \to \infty$, and the proportion $n_g / n$ converges to a strictly positive constant in $(0,1)$.}


 \begin{theorem}  \label{thm2}
 The estimator $\hat \theta$ is consistent and asymptotically normal,  satisfying $ \sqrt{n}(\hat \theta - \theta) \xrightarrow{d} N(0, \Sigma),$ 
where $\Sigma = C^{-1} V C^{-1}$, $C = m^{-1} \sum_{g=1}^m \tilde X_{g} \tilde X_{g}^\intercal$, and 
   \begin{align*}
    V =   &  \frac{1}{m^2}  \sum_{g=1}^m  \mathrm{Var}\Biggl \{  \tilde Y_g  \begin{pmatrix}
        \dfrac{   \mathbb{I}(Y = 0, G = g, A = 0)-
        \P(Y=0, G=g, A=0) }{\P(G=g, A=0)}   \\
        \dfrac{   \mathbb{I}(Y = 1, G = g, A = 0)-
        \P(Y=1, G=g, A=0)  }{\P(G=g, A=0)}  
    \end{pmatrix}  \\
     {}& +      \dfrac{  \mathbb{I}(Y = 1, G = g, A = 1)- \P(Y=1, G=g, A=1)}{\P(G=g, A=1)}\tilde X_g  \Biggr \}. 
   \end{align*}
 \end{theorem}
\bcol{From Theorem \ref{thm2}, we could use the bootstrap method to estimate the asymptotic variance of the proposed estimator, as detailed in Supplementary Material \ref{appendix-s3.3}.} 

\subsection{Testing Assumption 2 under Overidentification}  \label{sec-test} 
If the parameter $\theta$ is overidentified, we can further test for the observable implications of Assumption \ref{assump2}---the key identification assumption.
 Specifically, given that Assumption \ref{assump1} holds, Assumption  \ref{assump2} leads to the null hypothesis  $H_0: \tilde Y_g = \tilde X_g^\intercal \theta,  \text{ for all } g = 1, ..., m,$ 
 that is, there is a linear relationship between $\tilde Y_g$ and $\tilde X_g$ for all $g\in \cG$. 

Let $\boldsymbol{\hat \epsilon}  = ( \hat Y_1 - \hat X_1^\intercal \hat \theta, ... , \hat Y_m - \hat X_m^\intercal \hat \theta)^\intercal$ be the residual vector. We use the following statistic to test Assumption~\ref{assump2}: 
\[
J = n\;\hat{\boldsymbol{\epsilon}}^\intercal\,\boldsymbol{\Sigma}_*^+\,\hat{\boldsymbol{\epsilon}},
\]
 where $\boldsymbol{\Sigma}_*^+$ denotes the Moore-Penrose pseudo-inverse of $\boldsymbol{\Sigma}_*$, and  $\boldsymbol{\Sigma}_*$ is the asymptotic covariance matrix of $\sqrt{n}\boldsymbol{\hat{\epsilon}}$, which can be estimated via the bootstrap.    
  In Supplementary Material \ref{test-statistics-proof}, we 
  show that the test statistic is asymptotically chi-squared, i.e., $J  \xrightarrow{d} \chi^2_{m - 2}$, 
  where $m$ is the number of trials and $2$ is the number of outcome categories. In addition, we provide the procedure for implementing the hypothesis test in Supplementary Material \ref{implement-test-statistic}.  
Intuitively, if Assumption \ref{assump2} is significantly violated, the $J$-statistic will be far from 0, leading to a rejection of the null hypothesis $H_0$. However, it should be noted that the failure to reject $H_0$ does not necessarily imply that Assumption \ref{assump2} holds, because this may simply result from the test's lack of power or \bcol{insufficient evidence to reject Assumption \ref{assump2}}. Nevertheless, the test could be useful for detecting the violation of Assumption \ref{assump2}.

   \bcol{We emphasize that the overidentification test is used here as a diagnostic tool for the observable implications of Assumption~\ref{assump2}, not as a pretest that determines whether the estimation and inference result is reported. The asymptotic results are therefore derived under the maintained identifying assumptions and do not condition on the test outcome. If the test were instead used as a pretest for model selection, additional analysis would be needed to address the resulting post-selection inference, which is beyond the scope of this paper.}

    The test on Assumption \ref{assump2} is analogous to the test of exclusion restriction assumption in the IV  setting, such as \citet{Frank2019, Kiviet2020, Carrasco-Doukali2022}, and overidentified testing method in the GMM framework \citep{Newey1985,NEWEY19942111}.


\section{Extension to Principal Causal Effects}  \label{sec6}



\bcol{In this section, we extend the framework developed in Sections~\ref{sec3}--\ref{sec4} to the setting with an additional binary post-treatment variable $S$, and study the identification of principal stratification average causal effects (PSACEs). As before, we focus on the binary outcome setting.} 
\bcol{In addition to variables $(G,A,Y)$, we now observe a binary
post-treatment variable $S$ (e.g., a surrogate endpoint) measured before the outcome $Y$. Let $S^0$ and $S^1$ denote the potential
outcomes of $S$ under treatment arms $0$ and $1$, respectively. Our goal is to
identify   
$\P(Y^a,S^0,S^1 \mid G=g)$ for $g \in \cG$ and
$a=0,1$. Once these joint distributions are identified, we can identify the PSACEs:}
                   \[     \text{PSACE}_{ab|g} = \E[ Y^1 - Y^0 \mid S^0 = a, S^1 = b, G=g], \quad a = 0, 1; b = 0, 1.   \]
     These quantities have many applications, such as noncompliance~\citep{Imbens-Angrist1994}, truncation by death~\citep{Rubin2006}, and surrogate endpoint  evaluation~\citep{Jiang-etal2016}. See \citet{Frangakis-Rubin2002} for more examples.  
 
     \bcol{
      We consider two conceptually distinct identification routes in this setting. The
      first route, developed in Section~\ref{sec6-2}, extends our transportability of
      state transition probabilities from Sections~\ref{sec3}--\ref{sec4} to the pair
      of states $(S^0,Y^0)$ and $(S^1,Y^1)$. This route yields two versions:
      Corollary~\ref{coro3} without monotonicity, and Theorem~\ref{thm-add} with
      additional monotonicity, where monotonicity weakens the rank requirement. The
      second route, discussed in Section~\ref{sec6-3}, is the monotonicity-based route
      of \citet{Jiang-etal2016}. 
      We focus on the first route and the new transportability-based results. Then we compare these two routes in detail. 
     }


 \subsection{Our Transportability-Based Route}    \label{sec6-2}


\bcol{Following the spirit of Sections~\ref{sec3}--\ref{sec4}, and in parallel to
Assumptions~\ref{assump1}--\ref{assump2} and Condition~\ref{cond1}, we introduce
the following assumptions and rank condition.}

\begin{assumption} \label{assump5}   
   $A \indep (S^0, S^1, Y^0, Y^1) \mid G = g$  and $0 < \P\left(A = 1 \mid G = g \right) < 1$ for all $g \in \mathcal{G}$. 
 \end{assumption}

\begin{assumption} \label{assump6}  $G \indep (S^1, Y^1) \mid S^0, Y^0$. 
\end{assumption}

\begin{condition} \label{cond3} The number of trials $m \geq 4$, and the matrix $(  \P(S^0 = 0, Y^0 = 0 \mid G=g), \P(S^0 = 0, Y^0 = 1 \mid G=g), \P(S^0 = 1, Y^0 = 0 \mid G=g), \P(S^0 = 1, Y^0 = 1 \mid G=g)   )_{m\times 4} $    
         has full-column rank.     
\end{condition}


Assumption~\ref{assump5} is analogous to Assumption~\ref{assump1} and holds when
all trials are randomized experiments. \bcol{Assumption~\ref{assump6} is the
principal-stratification analogue of Assumption~\ref{assump2}: it imposes
transportability of the state transition from the untreated state $(S^0,Y^0)$ to
the treated state $(S^1,Y^1)$ across trials. In this setting, we may again regard
$(S^0,Y^0)$ as a latent summary of untreated prognosis, so that
Assumption~\ref{assump6} describes a stable transition mechanism across trials.}

\begin{corollary} \label{coro3} Under Assumptions \ref{assump5}, \ref{assump6} and Condition \ref{cond3},  the joint distributions $\P(S^0, S^1, Y^0, Y^1 \mid G=g)$ for all  $g \in \mathcal{G}$ are identifiable, and therefore $\text{PSACE}_{ab|g}$ for $a,b \in \{0,1\}$ and $g \in \mathcal{G}$ are also identifiable. 
\end{corollary} 

\bcol{
  Corollary~\ref{coro3} is the baseline transportability-based route in this
section. It does not impose monotonicity, but requires the stronger four-column
rank condition (Condition~\ref{cond3}), and hence at least four trials. Essentially, we can view Corollary~\ref{coro3} as an instantiation of Theorem~\ref{thm1}, treating $(Y^1, S^1)$ and $(Y^0, S^0)$ as two categorical variables, each with four possible values. 
}

\bcol{In addition, we can relax Condition~\ref{cond3} by additionally imposing monotonicity.}



\begin{condition} \label{cond-add}
(i) The matrix 
             $  (  \P(S^0 = 0, Y^0 = 0 \mid G=g), \P(S^0 = 0, Y^0 = 1 \mid G=g)   )_{m\times 2} 
             $      
         has full column rank;    
(ii) the matrix 
             $  (  \P(S^0 = 0, Y^0 = 0 \mid G=g), \P(S^0 = 1, Y^0 = 0 \mid G=g)   )_{m\times 2} 
             $      
         has full column rank.    
 \end{condition}        
    

 \begin{theorem}  \label{thm-add}
 If $Y^1 \geq Y^0$ and $S^1 \geq S^0$, then under Assumptions \ref{assump5}, \ref{assump6} and Condition \ref{cond-add},  the joint distributions $\P(S^0, S^1, Y^0, Y^1 \mid G=g)$ for $g \in \mathcal{G}$ are identifiable, and therefore $\text{PSACE}_{ab|g}$ for $a,b \in \{0,1\}$ and $g \in \mathcal{G}$ are also identifiable. 
 \end{theorem}

 \bcol{
  Theorem~\ref{thm-add} should be read as a monotonicity-strengthened version of
Corollary~\ref{coro3}. By imposing the additional monotonicity conditions
$S^1 \geq S^0$ and $Y^1 \geq Y^0$, it relaxes the identifying-variation
requirement from Condition~\ref{cond3} to Condition~\ref{cond-add}, and hence
from $m \ge 4$ to $m \ge 2$. In this sense, Corollary~\ref{coro3} and
Theorem~\ref{thm-add} are best viewed as two versions of the same
transportability-based route: Corollary~\ref{coro3} is the no-monotonicity
version with a stronger rank condition, whereas Theorem~\ref{thm-add} is the
monotonicity-strengthened version with a weaker rank condition.
 }

 \bcol{
  In addition, under the identification assumptions in Corollary~\ref{coro3} and
Theorem~\ref{thm-add}, we can also construct least-squares-based estimators for
the invariant transition parameters and the induced PSACEs. To keep the main text
concise, we defer these estimation details to Supplementary Material \ref{appendix-s3}.
 }

  If only partial monotonicity holds (i.e., $S^1 \geq S^0$ or $Y^1 \geq Y^0$, but
not both), then under Assumptions~\ref{assump5}, \ref{assump6} and
Condition~\ref{cond-add}, only half of the quantities in $\{\P(S^0=a, S^1=b, Y^0=c, Y^1=d \mid G=g): a,b,c,d = 0,1;\ g \in \mathcal{G}\}$ are identifiable (see Supplementary Material \ref{appendix-s2}). \bcol{Thus, partial monotonicity
alone cannot replace Condition~\ref{cond3} by
Condition~\ref{cond-add} for the purpose of identifying the whole joint
distribution. Nevertheless, partial monotonicity can still simplify estimation
when Condition~\ref{cond3} holds; see Supplementary Material \ref{appendix-s2-3}. Moreover, if we
relax Assumption~\ref{assump6} to partial but not joint conditional independence
$G \indep S^1 \mid (S^0, Y^0)$ and $G \indep Y^1 \mid (S^0, Y^0)$ in
Corollary~\ref{coro3}, then we can identify $\P(S^0, S^1, Y^0 \mid G=g)$ and
$\P(S^0, Y^0, Y^1 \mid G=g)$ but not the full joint distribution needed to
identify the PSACEs; see Supplementary Material \ref{appendix-s4} for details.
 }


\subsection{Comparison with Monotonicity-based Route in \citet{Jiang-etal2016}} \label{sec6-3}
\bcol{Our paper is closely related to \citet{Jiang-etal2016}, as both works leverage
multiple trials to identify causal quantities that are not identifiable from a
single trial. The two papers, however, have different targets.
\citet{Jiang-etal2016} focus on the PSACEs for surrogate endpoint evaluation. By contrast, our primary target is the
joint distribution of potential outcomes itself, with PSACEs
treated as one important extension. This broader target covers causal estimands
beyond PSACEs (see the examples in Section~\ref{sec3-2}).

A second key difference is in the cross-trial invariance assumption.
\citet{Jiang-etal2016} impose outcome homogeneity within principal strata, namely
$Y^a \indep G \mid (S^0,S^1)$ for $a=0,1$, which implies that the PSACEs are invariant across trials. In contrast, our framework is based on
transportability of state transition probabilities. This allows the marginal distributions of $(S^0,Y^0)$ and, in general,
the PSACEs themselves to vary across trials. Conceptually,
\citet{Jiang-etal2016} assume invariance of outcomes within principal strata,
whereas we assume invariance of the transition mechanism from untreated states to
treated states.

In addition to the cross-trial invariance assumption, \citet{Jiang-etal2016} crucially relies on the monotonicity condition $S^1 \ge S^0$ for the identification of PSACEs. For brevity, we do not
restate their theorem and detailed assumptions here. Instead, Supplementary Material \ref{app-s7} provides
a formal restatement of the monotonicity-based route of \citet{Jiang-etal2016}. There we also provide an alternative proof for their identification theorem, which motivates a least-squares estimator. These complement the identification proof and Bayesian estimator in \citet{Jiang-etal2016}. 
We remark that  this monotonicity-based route is a  
 benchmark for comparison,  whereas
Corollary~\ref{coro3} and Theorem~\ref{thm-add} are our new identification
results.}

\bcol{
  Without the monotonicity condition $S^1 \geq S^0$, 
\citet{Jiang-etal2016} also shows that a necessary condition for \emph{local
identifiability} of the principal effects is $m \ge 3$. In contrast, our 
Corollary~\ref{coro3}, which does not impose monotonicity, establishes sufficient
conditions for global identifiability when $m \ge 4$. In this sense, our new
identification results provide alternative conditions leading to stronger global
identification, which complement the findings of \citet{Jiang-etal2016}.
}

\bcol{In summary, our paper complements \citet{Jiang-etal2016} by developing new transportability-based identification routes from the broader perspective of joint potential-outcome distributions, while their monotonicity-based route serves as a useful benchmark. In Sections \ref{sec7} and \ref{sec8}, we compare these two routes in both simulation and real-data analyses.}


\section{Simulation}  \label{sec7}

We perform simulation studies to evaluate the finite-sample performance of the proposed method.  We consider both scenarios, with and without the post-treatment surrogate $S$. 
The replication code for both simulation and application is available at \url{https://github.com/pengwu1224/The-Promises-of-Multiple-Experiments}.    
  Throughout this simulation, the number of trials is set to 10, and for each trial $g$, the binary treatment $A$ is randomly assigned with probability $\P(A=1 \mid G = g) = 0.5$, the sample size is set to 100, 200, and 500, respectively. 

\bigskip 
 {\bf Study I} (without the post-treatment variable $S$). We first  examine the methods developed in Sections \ref{sec3}--\ref{sec4} by considering two data-generating processes for $(Y^1, Y^0)$.  
 
\begin{itemize} 

\item[(C1)]  $\P(Y^1=1\mid Y^0, G=g) = \text{expit}(Y^0 - 0.5)$ for $g = 1, ..., 10$,  where $\text{expit}(x) = \exp(x)/\{1 + \exp(x)\}$ is the standard logistic function.  For each trial $g = 1, 2, ..., 10$, the potential outcome $Y^0$ follows from a Bernoulli distribution with $\P(Y^0 = 1  \mid G=g) = 0.5 + (g-1)/30$, i.e., 
    taking evenly spaced values at equal intervals from 0.5 to 0.8.  
    
\item[(C2)]   $\P(Y^1=1\mid Y^0, G=g) = \text{expit}(Y^0 + 0.5)$   for $g = 1, ..., 10$ and $Y^0$ is generated according to the process described in  (C1).
\end{itemize}

Assumptions \ref{assump1}-\ref{assump2} and Condition \ref{cond1} hold for both cases (C1) and (C2). The true values of $\theta = ( \pi_{1|0},  \pi_{1|1} ) = ( \P(Y^1=1\mid Y^0=0), \P(Y^1=1 \mid Y^0=1) )$ are  $(0.378, 0.622)$ and $(0.622, 0.818)$ for cases (C1) and (C2), respectively.  
We replicate each simulation case 1,000 times and calculate the Bias, SD, ESE, and CP95 as evaluation
metrics, where Bias and SD represent the Monte Carlo bias and standard deviation of the point estimates over the 1,000 replicates, ESE denotes the square root of the average of the estimated asymptotic variances, and CP95 refers to the coverage proportion of the 95\% confidence intervals. Both ESE and CP95 are calculated using the estimated asymptotic variance based on 100 bootstraps.

Table \ref{tab2} summarizes the numerical results for the proposed estimator of $\theta$ for cases (C1)--(C2).  From the table, the Bias is small and decreases as the sample size increases, demonstrating the consistency of the proposed estimator. As the sample size grows, ESE approaches SD, and CP95 converges to its nominal value of 0.95, indicating the asymptotic normality of the proposed estimator and validating the bootstrap method for estimating asymptotic variance.

\begin{table*}[t]
 \centering  
\caption{\centering Simulation results for cases (C1)--(C2).  \label{tab2}}
\resizebox{1\textwidth}{!}{\begin{tabular}{cc | rrrr | rrrr | rrrr}
  \hline
      & &     \multicolumn{4}{c|}{$n_g =100$}    & \multicolumn{4}{c|}{$n_g =200$} &  \multicolumn{4}{c}{$n_g =500$}  \\
Case & $\theta$ &  Bias & SD & ESE & CP95 & Bias & SD & ESE & CP95 & Bias & SD & ESE & CP95 \\ 
  \hline
\multirow{2}{*}{(C1)}  & $\pi_{1|0}$ & 0.041 & 0.136 & 0.141 & 0.938 & 0.017 & 0.104 & 0.107 & 0.946 & 0.009 & 0.068 & 0.070 & 0.952 \\ 
      &  $\pi_{1|1}$ & -0.024 & 0.074 & 0.077 & 0.943 & -0.009 & 0.057 & 0.058 & 0.945 & -0.006 & 0.037 & 0.038 & 0.956 \\  
   \hline  
\multirow{2}{*}{(C2)}  & $\pi_{1|0}$ & 0.037 & 0.123 & 0.122 & 0.937 & 0.024 & 0.092 & 0.092 & 0.944 & 0.007 & 0.060 & 0.060 & 0.951 \\ 
 & $\pi_{1|1}$ &  -0.021 & 0.067 & 0.067 & 0.935 & -0.013 & 0.049 & 0.050 & 0.939 & -0.003 & 0.032 & 0.033 & 0.947 \\ 
   \hline 
\end{tabular}}
\begin{flushleft} \footnotesize 
Note: Bias and SD are the Monte Carlo bias and standard deviation over the 1000 simulations of the point estimates,  ESE and CP95 are the estimated asymptotic variances and coverage proportions of the 95\% confidence intervals based on 100 bootstraps, respectively. 
\end{flushleft}  
\end{table*}

 \bigskip 
 {\bf Study II} (with the post-treatment variable $S$).  We then explore the proposed method in the presence of a post-treatment variable $S$. Two additional data-generating processes for the potential outcomes $(S^0, S^1, Y^0, Y^1)$ are considered. 
\begin{itemize} 
	\item[(C3)]   $\P(Y^1=1\mid Y^0, S^0, G=g) = \text{expit}\{ (S^0 + Y^0 + 1)/2 \}$ and   $\P(S^1=1\mid Y^0, S^0, G=g) = \text{expit}\{ (S^0 + Y^0 - 1)/2 \}$, and $S^0$ and $Y^0$ are independent, both drawn from the Bernoulli distribution with success probability $0.5 + (g-1)/30$.   
	
		\item[(C4)]  $\P(Y^1=1\mid S^0, S^1, G=g) = \text{expit}\{ (S^0 + S^1 + 1)/2 \}$ and   $\P(Y^0=1\mid S^0, S^1, G=g) = \text{expit}\{ (S^0 + S^1 - 1)/2 \}$, and 
   $S^0$ and $S^1$ are independent, both drawn from the Bernoulli distribution, with the success probabilities $0.3 + (g-1)/30$ and $0.5 + (g-1)/30$, respectively. 
 After generating $(S^0, S^1)$, we
further adjust the value of $S^0$, setting $S^0$ to 0 if $S^1 = 0$ to ensure monotonicity. 
\end{itemize}

Assumptions \ref{assump5}--\ref{assump6} and Condition \ref{cond3} hold for case (C3), \bcol{while Assumptions \ref{assump5}, \ref{assump7}--\ref{assump8} and Condition \ref{cond4} in Supplementary Material \ref{app-s7} hold for case (C4)}. 
 In case (C3),  we denote 
  $\pi_{s|ab} = \P(S^1 = 1  \mid S^0 = a, Y^0 = b)$ and  $\pi_{y|ab} = \P(Y^1 = 1 \mid S^0 = a, Y^0 = b)$ for $a, b = 0, 1$. 
The parameters of interest are
 $\theta = (\pi_{s|00}, \pi_{s|01}, \pi_{s|10}, \pi_{s|11},  \pi_{y|00}, \pi_{y|01}, \pi_{y|10}, \pi_{y|11}),$ 
which are the key invariant parameters for estimating the joint distributions $\P(S^0, S^1, Y^0 \mid G=g)$ and $\P(S^0, S^1, Y^1 \mid G=g)$ for $g = 1, ..., 10$. 
In case (C4), we define 
  $\pi_{1 |ab} = \P(Y^0 = 1 \mid S^0 = a, S^1= b)$ and  $\tilde \pi_{1 |ab} = \P(Y^1 = 1 \mid S^0 = a, S^1 = b)$ for $a, b = 0, 1$.  
  By the monotonicity condition $S^1 \geq S^0$, $\pi_{1| 1 0}$ and $\tilde \pi_{1 | 1 0}$ are undefined. 
  Thus, the key invariance parameters are 
$\theta = (\pi_{1|00}, \pi_{1|01}, \pi_{1|11},  \tilde \pi_{1|00},  \tilde \pi_{1|01}, \tilde \pi_{1|11}).$ 
\bcol{It is noteworthy that cases (C3) and (C4) correspond to two different identification routes in Section~\ref{sec6}. Case (C3) illustrates our new transportability-based identification route in Corollary~\ref{coro3}. By contrast, case (C4) corresponds to the monotonicity-based route of \citet{Jiang-etal2016}, which is restated in Supplementary Material \ref{app-s7}.}


\begin{table*}[!h]
 \centering 
\caption{\centering Simulation results for case (C3).  \label{tab3}}
\resizebox{1\textwidth}{!}{\begin{tabular}{cc | rrrr | rrrr | rrrr}
  \hline
      & &     \multicolumn{4}{c|}{$n_g =100$}    & \multicolumn{4}{c|}{$n_g =200$} &  \multicolumn{4}{c}{$n_g =500$}  \\
Case & $\theta$ &  Bias & SD & ESE & CP95 & Bias & SD & ESE & CP95 & Bias & SD & ESE & CP95 \\ 
  \hline
\multirow{8}{*}{(C3)}  & $\pi_{s|00}$ & 0.032 & 0.431 & 0.420 & 0.951 & 0.012 & 0.382 & 0.388 & 0.964 & 0.017 & 0.357 & 0.354 & 0.952 \\
      &  $\pi_{s|01}$ &  0.005 & 0.427 & 0.375 & 0.933 & -0.008 & 0.393 & 0.378 & 0.942 & -0.010 & 0.381 & 0.365 & 0.945 \\ 
  & $\pi_{s|10}$ & -0.005 & 0.396 & 0.379 & 0.944 & 0.020 & 0.416 & 0.374 & 0.939 & -0.003 & 0.375 & 0.362 & 0.948 \\  
 & $\pi_{s|11}$ &  -0.011 & 0.140 & 0.141 & 0.955 & -0.011 & 0.118 & 0.121 & 0.952 & 0.000 & 0.101 & 0.102 & 0.963 \\  
 & $\pi_{y|00}$ &  0.026 & 0.396 & 0.363 & 0.940 & 0.013 & 0.337 & 0.334 & 0.956 & 0.014 & 0.310 & 0.308 & 0.961  \\ 
 & $\pi_{y|01}$ & 0.007 & 0.363 & 0.324 & 0.937 & -0.013 & 0.346 & 0.323 & 0.945 & -0.021 & 0.349 & 0.320 & 0.943 \\ 
  & $\pi_{y|10}$ & -0.002 & 0.342 & 0.327 & 0.944 & 0.006 & 0.340 & 0.322 & 0.948 & 0.007 & 0.339 & 0.314 & 0.940  \\ 
 & $\pi_{y|11}$ & -0.009 & 0.127 & 0.120 & 0.937 & -0.001 & 0.102 & 0.104 & 0.962 & 0.003 & 0.089 & 0.088 & 0.957 \\ 
   \hline 
\end{tabular}}
\end{table*}

\begin{table*}[!h]
 \centering 
 \small 
\caption{\centering Simulation results for case (C4).  \label{tab4}}
\resizebox{1\textwidth}{!}{\begin{tabular}{cc | rrrr | rrrr | rrrr}
  \hline
      & &     \multicolumn{4}{c|}{$n_g =100$}    & \multicolumn{4}{c|}{$n_g =200$} &  \multicolumn{4}{c}{$n_g =500$}  \\
Case & $\theta$ &  Bias & SD & ESE & CP95 & Bias & SD & ESE & CP95 & Bias & SD & ESE & CP95 \\ 
  \hline
\multirow{6}{*}{(C4)}  & $\pi_{1|00}$ &  0.040 & 0.155 & 0.142 & 0.931 & 0.033 & 0.130 & 0.128 & 0.938 & 0.020 & 0.094 & 0.101 & 0.954  \\
      &  $\pi_{1 |01}$ &  -0.043 & 0.160 & 0.144 & 0.923 & -0.037 & 0.131 & 0.131 & 0.943 & -0.021 & 0.097 & 0.103 & 0.946 \\ 
  & $\pi_{1 |11}$ &  -0.000 & 0.039 & 0.039 & 0.943 & 0.001 & 0.028 & 0.028 & 0.949 & -0.001 & 0.018 & 0.018 & 0.940 \\  
 & $\tilde \pi_{1 |00}$ &  0.001 & 0.037 & 0.037 & 0.942 & 0.000 & 0.026 & 0.026 & 0.952 & -0.000 & 0.017 & 0.016 & 0.948  \\  
 & $\tilde \pi_{1|01}$ &   0.019 & 0.109 & 0.113 & 0.953 & 0.012 & 0.092 & 0.094 & 0.962 & 0.004 & 0.064 & 0.065 & 0.947 \\ 
 & $\tilde \pi_{1|11}$ &  -0.022 & 0.125 & 0.127 & 0.954 & -0.013 & 0.106 & 0.106 & 0.956 & -0.004 & 0.073 & 0.074 & 0.940 \\
   \hline 
\end{tabular}}
\end{table*}

Tables \ref{tab3} and \ref{tab4} present the numerical results for cases (C3) and (C4), respectively. The simulation results are similar to those in cases (C1)–(C2), indicating that the extension of the proposed method  presented in Section \ref{sec6} performs well.
\bcol{Taken together, these two cases show that the proposed least-squares framework performs well both under our new transportability-based route and under the monotonicity-based route corresponding to \citet{Jiang-etal2016}.
To further evaluate the robustness of the proposed method, we present additional simulations for the just-identified case and for the violation of Assumption \ref{assump2} in Supplementary Material \ref{app-s9}.}


\section{Application to the Adjuvant Colon Clinical Trials}  \label{sec8}

We demonstrate the proposed methodology using the data from phase III adjuvant colon clinical trials (ACCTs).  The initial ACCTs data consist of 20,898 patients from 18 randomized phase III clinical trials, with the enrollment period spanning from 1977 to 1999~\citep{Sargent-etal2005}.  
Among these 18 trials, the data from 10 trials are publicly available in \citet{Baker-etal2012}, including a total of 9,102 patients. \bcol{We base our analysis on these 10 trials.} 
    In each trial, we have a contingency table of observed frequencies for three \emph{binary} variables:  treatment $A$,  surrogate $S$,  and outcome $Y$. 
   Here $S=0$ indicates cancer recurrence within 3 years and $S=1$ otherwise; $Y=0$ denotes mortality within 5 years and $Y=1$ indicates survival beyond 5 years; $A =1$ and $A=0$ denote receiving treatment (\bcol{fluorouracil-based chemotherapy}) and not, respectively.   \bcol{The public data do not contain individual-level baseline covariates. Accordingly,
   as in \citet{Baker-etal2012} and \citet{Jiang-etal2016}, our empirical analysis
   here does not adjust for covariates.}

   
The goal of the ACCTs is to determine whether cancer recurrence within 3 years ($S$) can serve as an appropriate surrogate for overall survival with a 5-year follow-up ($Y$). 
 Using the ACCTs data, \citet{Sargent-etal2005} identified a strong correlation between $S$ and $Y$.  
From the perspective of principal stratification in causal inference, \citet{Jiang-etal2016} investigated this question by first estimating PSACEs and then evaluating the surrogate using the causal necessity and causal sufficiency criteria~\citep{Gilbert-Hudgens2008}: for all $g \in \cG$,  causal necessity requires $\text{PSACE}_{11 |g}  = 0$ and $\text{PSACE}_{00 |g}  = 0$, and causal sufficiency requires  $\text{PSACE}_{10 |g}  \neq 0$ and $\text{PSACE}_{01 |g}  \neq 0$. 
\bcol{Causal necessity means that if $A$ has no effect on $S$ (i.e., $S^1=S^0$), then it should also have no effect on $Y$, whereas causal sufficiency means that if $A$ affects $S$ (i.e., $S^1\neq S^0$), then it should also affect $Y$.}


 In this article, we further explore the problem by estimating both the joint distributions of $\P(S^0, S^1\mid G=g)$ and $\P(Y^0, Y^1 \mid G=g)$, as well as the PSACEs under different sets of identification assumptions.
Notably, for the  ACCTs data, Assumptions \ref{assump1} and \ref{assump5} naturally hold due to randomization. 
   

\subsection{State Transition Probabilities and Joint Distribution of  Potential Surrogates and Potential Outcomes}

\begin{figure}[t]
\centering
\vspace{2pt} 
\begin{minipage}[t]{0.8\linewidth}
\centering
\includegraphics[width=1\textwidth]{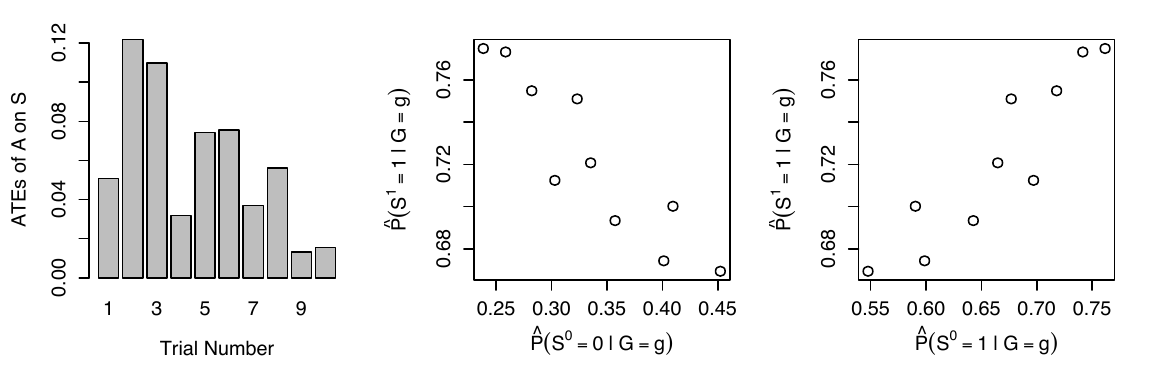}
\end{minipage}%
\par\medskip

\begin{minipage}[t]{0.8\linewidth}
\centering
\includegraphics[width=1\textwidth]{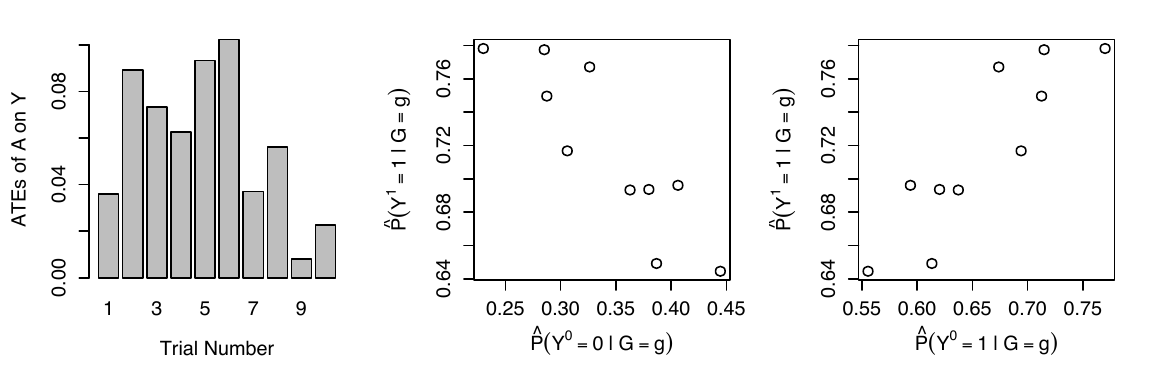}
\end{minipage}%
\caption{\bcol{(a) Estimated ATEs of $A$ on $S$, and scatter plots of $\hat \P(S^1=1\mid G=g)$
against $\hat \P(S^0=0\mid G=g)$ and $\hat \P(S^0=1\mid G=g)$, respectively. 
The scatter plots show strong linear correlations, providing informal visual
evidence for the linear restrictions implied by $S^1 \indep G \mid S^0$. A linear regression yields an $R^2$ of 0.9996; the formal  over-identification test assessment is reported in Table~\ref{tab5}. 
 (b) Estimated ATEs of $A$ on $Y$, and
scatter plots of $\hat \P(Y^1=1\mid G=g)$ against $\hat \P(Y^0=0\mid G=g)$ and
$\hat \P(Y^0=1\mid G=g)$, respectively. The scatter plots likewise provide
informal visual evidence for the linear restrictions implied by
$Y^1 \indep G \mid Y^0$. A linear regression yields an $R^2$ of $0.9991$; the formal assessment is again reported in Table~\ref{tab5}.}}
\label{fig1}
\end{figure}

To get an intuitive understanding of the heterogeneity across different trials, we estimate the average treatment effects (ATEs) of  $A$ on $S$ (or $Y$) for each trial $g \in \cG$ by computing  the average contrasts of $S$ (or $Y$) between the groups $(A=1, G=g)$ and $(A=0, G=g)$.   
\bcol{As illustrated in the barplots of Figure~\ref{fig1},
the ATEs of $A$ on $S$ and $Y$ vary substantially across trials, indicating strong
cross-trial heterogeneity and ruling out stronger homogeneity assumptions that require common ATEs across trials.}
 In addition, the ATEs of $A$ on $S$ and $Y$ show a similar pattern, indicating that $S$ may have the potential to serve as an appropriate surrogate for $Y$.  

By employing the proposed method outlined in Sections \ref{sec3}--\ref{sec4}, and under Assumption \ref{assump2} (i.e., $Y^1 \indep G \mid Y^0$), we can estimate the state transition probabilities from $Y^0$ to $Y^1$, i.e., $\P(Y^1 \mid Y^0 )$.  This allows us to subsequently estimate the joint distributions $\P(Y^0, Y^1\mid G=g)$ for $g \in \cG$. Similarly, if we assume $S^1 \indep G \mid S^0$, then we can apply the same methods to estimate the  transition probabilities $\P(S^1 \mid S^0)$ and the joint distributions $\P(S^0, S^1\mid G=g)$.

\bcol{
  The key assumptions $Y^1 \indep G \mid Y^0$ and $S^1 \indep G \mid S^0$ imply
  a linear relationship between $\P(Y^1 = 1\mid G=g)$ on $\P(Y^0=1\mid G=g)$ and $\P(Y^0=0\mid G=g))$,  as well as between   $\P(S^1 = 1\mid G=g)$ on $\P(S^0=1\mid G=g)$ and $\P(S^0=0\mid G=g))$. We
illustrate these possible linear relationships through the scatter plots in
Figure~\ref{fig1}, where probabilities are replaced by observed frequencies. These
plots provide informal visual evidence for the implied restrictions. Moreover, a linear regression of $\hat \P(S^1=1\mid G=g)$ on  $(\hat \P(S^0=0\mid G=g), \hat \P(S^0=1\mid G=g))$  yields an $R^2$ of 0.9996.  Similarly, A linear regression of  $\hat \P(Y^1=1\mid G=g)$ on $(\hat \P(Y^0=0\mid G=g), \hat \P(Y^0=1\mid G=g))$ yields an $R^2$ of $0.9991$. These results indicate a strong linear relationship.
We then 
estimate the invariant parameters
$(\P(S^1=1\mid S^0=0),\P(S^1=1\mid S^0=1))$ and
$(\P(Y^1=1\mid Y^0=0),\P(Y^1=1\mid Y^0=1))$, with the results reported
in Table~4. Table~4 also reports the overidentification tests for $S^1 \indep G \mid S^0$ and $Y^1 \indep G \mid Y^0$. The corresponding $p$-values are 0.846 and 0.325, respectively, both
of which are very large. So we do not reject the observable restrictions implied by these two
assumptions.
}



\begin{table*}[t]
 \centering 
\caption{\centering State transition probabilities and tests for $S^1 \indep G \mid S^0$ and $Y^1 \indep G \mid Y^0$.  \label{tab5}}
\resizebox{1\textwidth}{!}{\begin{tabular}{c | cccc c } 
  \hline
  Parameters &  Estimate (ESE) &  95\% CI & $J$-Statistic & $p$-value  &  null hypothesis $H_0$  \\
  \hline
  $\P(S^1=1\mid S^0 = 0)$  & 0.379 (0.107)  &  (0.170, 0.590)  &   \multirow{4}{*}{\bcol{4.119}}   & \multirow{4}{*}{\bcol{0.846}}  & \multirow{4}{*}{$S^1 \indep G \mid S^0$}   \\  
    $\P(S^1=1\mid S^0 = 1)$  & 0.896 (0.049)  & (0.800, 0.992)   &    &      \\ 
    $\P(S^1=0 \mid S^0 = 0)$  & 0.621 (0.107)  & (0.411, 0.830) \\
    $\P(S^1=0 \mid S^0 = 1)$  &  0.104 (0.488) & (0.008, 0.200)  \\ \hline 
          $\P(Y^1=1\mid Y^0 = 0)$  & 0.275 (0.101) & (0.062, 0.488)  &  \multirow{4}{*}{\bcol{9.208}}   & \multirow{4}{*}{\bcol{0.325}}  & \multirow{4}{*}{$Y^1 \indep G \mid Y^0$  }   \\  
         $\P(Y^1=1\mid Y^0 = 1)$  & 0.946 (0.051)  & (0.846, 1.044) &   & \\  
          $\P(Y^1=0 \mid Y^0 = 0)$  &  0.725 (0.101)   & (0.512, 0.938)    &   \\  
         $\P(Y^1=0 \mid Y^0 = 1)$  & 0.054 (0.051)    &  (-0.045, 0.154)   &  & &     \\  
   \hline 
\end{tabular}}
\begin{flushleft} \footnotesize 
Note: ESE is the estimated asymptotic standard error based on 500 bootstraps, 95\% CI represents 95\% confidence interval, and the $J$-Statistic and $p$-value correspond to the null hypothesis.    
\end{flushleft} 
\end{table*}


\begin{figure}[t]
\centering
\includegraphics[width=0.9 \linewidth]{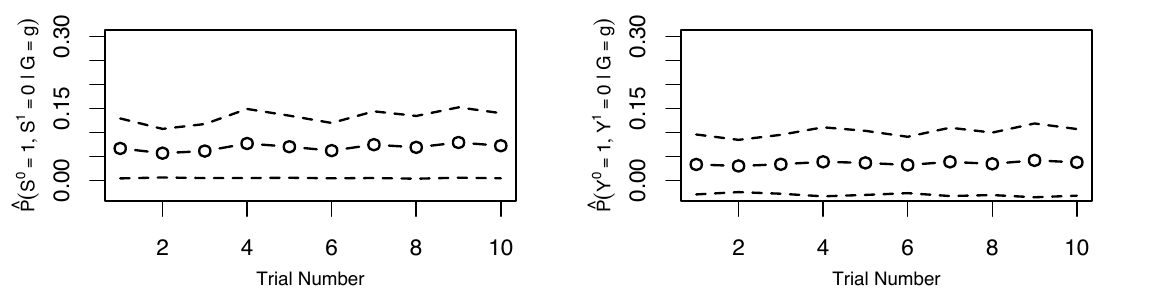} 
\caption{\bcol{Estimated joint probabilities $\hat \P(S^0=1,S^1=0\mid G=g)$ and
$\hat \P(Y^0=1,Y^1=0\mid G=g)$ for all $g\in\mathcal{G}$. Since these quantities
would be zero under the monotonicity conditions $S^1\geq S^0$ and $Y^1\geq Y^0$,
respectively, the figure provides direct evidence for assessing the plausibility of
the two monotonicity assumptions.}}
\label{fig2} 
\end{figure}


\bcol{
Another useful by-product of our framework is that it allows us to assess whether the monotonicity assumptions are plausible in the ACCT data. \citet{Jiang-etal2016} also examined monotonicity by comparing models with and without monotonicity. By contrast, once our framework identifies the joint distributions $\P(S_0,S_1 \mid G=g)$ and $\P(Y_0,Y_1 \mid G=g)$, we can directly examine quantities such as $\P(S_0=1,S_1=0 \mid G=g)$ and $\P(Y_0=1,Y_1=0 \mid G=g)$, which would be zero under the corresponding monotonicity assumptions.
} 
Figure \ref{fig2} displays the point estimates along with the corresponding 95\% confidence intervals (CIs).  The lower bounds of the 95\% CIs for $\hat \P(S^0 = 1, S^1 = 0 \mid G=g)$ for all trials are strictly positive, suggesting that the monotonicity condition (Assumption \ref{assump8}, $S^1 \geq S^0$) may not hold.  In contrast, the 95\% CIs for $\hat \P(Y^0 = 1, Y^1 = 0 \mid G=g)$ cover 0 in all trials.  The other estimated values of $\P(S^0=a, S^1=b \mid G=g)$ and $\P(Y^0=a, Y^1=b \mid G=g)$ for $a, b \in \{0, 1\}$ are provided in Supplementary Material \ref{app-s9}. 


\subsection{Evaluation of Principal Surrogate}
\begin{figure}[th]
\centering
\vspace{2pt}  
\begin{minipage}[t]{0.24\linewidth} 
\centering
\textbf{Method 1}\par
\vspace{2pt}
\includegraphics[width=1\textwidth]{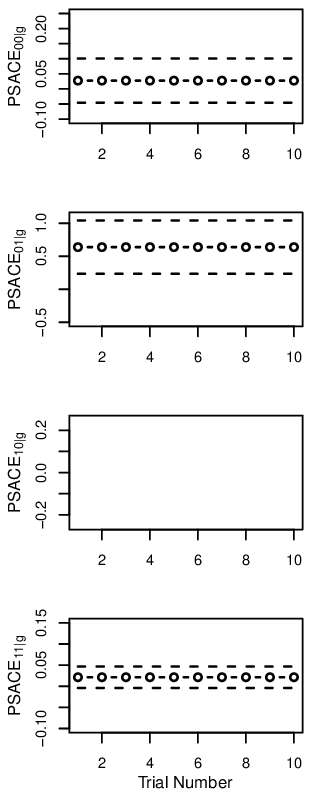}
\end{minipage}%
\hfill
\begin{minipage}[t]{0.24\linewidth}
\centering
\textbf{Method 2}\par
\vspace{2pt}
\includegraphics[width=1\textwidth]{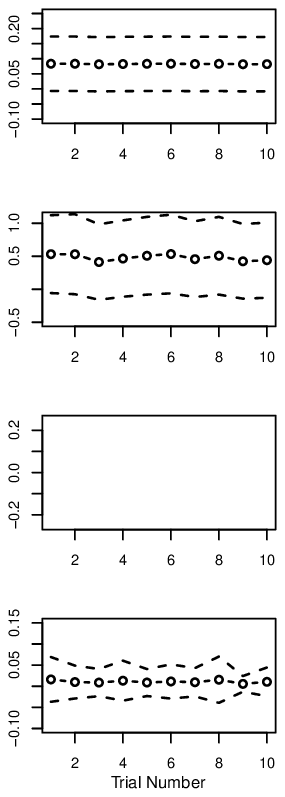}
\end{minipage} 
\hfill
\begin{minipage}[t]{0.24\linewidth}
\centering
\textbf{Method 3}\par
\vspace{2pt}
\includegraphics[width=1\textwidth]{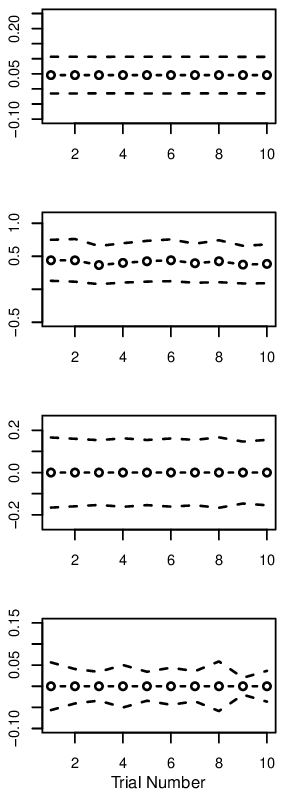}
\end{minipage} 
\hfill
\begin{minipage}[t]{0.24\linewidth}
\centering
\textbf{Method 4}\par
\vspace{2pt}
\includegraphics[width=1\textwidth]{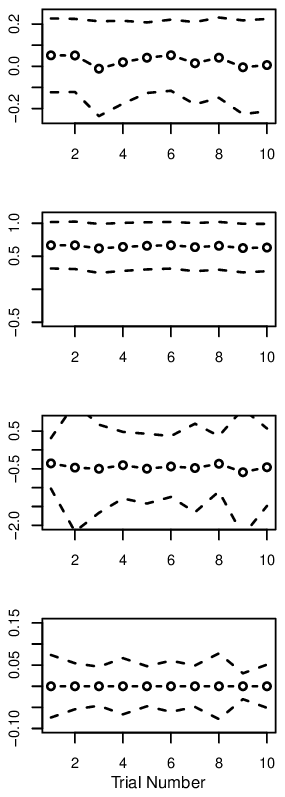}
\end{minipage}%
\caption{\bcol{Point estimates and 95\% confidence intervals of $\textup{PSACE}_{ab\mid g}$
across trials under Methods~1--4. From top to bottom, the four rows correspond to
$(a,b)=(0,0),(0,1),(1,0),(1,1)$. Methods~1 and 2 assume monotonicity
$S^1\geq S^0$, so $\textup{PSACE}_{10\mid g}$ is undefined. The vertical scale in
Method~4 differs from that in the other columns because its confidence intervals
are substantially wider.}}
\label{fig3} 
\end{figure}

To evaluate the surrogate, we estimate the principal stratification average causal effects $\text{PSACE}_{ab|g}$ for $g \in \cG$. \bcol{Based on the methods proposed in Section \ref{sec6}, we consider the following four approaches under different
sets of assumptions. The four methods serve both as a sensitivity analysis and as
a way to distinguish the monotonicity-based route most closely associated with
\citet{Jiang-etal2016} from our transportability-based routes.}  
  \begin{itemize}
   \item \bcol{Method 1: Based on the monotonicity-based identification route of \citet{Jiang-etal2016}, relying on Assumptions \ref{assump5}, \ref{assump7}, and Condition \ref{cond4}, with monotonicity $S^1 \geq S^0$. 
   See Supplementary Material \ref{app-s7} for details on identification and estimation.}
   
   \item Method 2: Based on Theorem \ref{thm-add},  relying on  Assumptions \ref{assump5}, \ref{assump6}, and Condition \ref{cond-add}, with monotonicity $Y^1 \geq Y^0$ and $S^1 \geq S^0$. See Supplementary Material \ref{appendix-s3} for estimation details.
   
   \item Method 3: Based on  Corollary \ref{coro3},  relying on Assumptions \ref{assump5}, \ref{assump6},  and Condition \ref{cond3}.
   \bcol{We additionally impose the partial monotonicity $Y^1 \geq Y^0$ to simplify the identification and estimation of a subset of parameters}. See Supplementary Material \ref{appendix-s2} for estimation details.  
   \bcol{We include this method for comparison primarily because Figure~\ref{fig2} suggests
   that $S^1\geq S^0$ may fail.} 
   
   \item Method 4: Based on Corollary \ref{coro3}, relying on Assumptions \ref{assump5}, \ref{assump6},   and Condition \ref{cond3}, \emph{without} any monotonicity. See Supplementary Material \ref{appendix-s3} for estimation details.

\end{itemize}
  Figure~\ref{fig3} presents the point estimates and the associated 95\% confidence
intervals of $\text{PSACE}_{ab\mid g}$ across trials, with the four columns
corresponding to Methods~1--4, respectively. When $S^1\geq S^0$ is assumed
(Methods~1 and 2), $\text{PSACE}_{10\mid g}$ is undefined and is therefore left
blank. \bcol{In addition, under Assumption~\ref{assump7} ($G \indep Y^a \mid (S^0, S^1)$, Method~1), $\text{PSACE}_{ab\mid g}$ is
invariant across trials for fixed $a,b\in\{0,1\}$. 
}

\bcol{
  Figure~\ref{fig3} suggests that causal necessity may hold, as the 95\% confidence
intervals of $\text{PSACE}_{00\mid g}$ and $\text{PSACE}_{11\mid g}$ all include
zero, indicating no clear evidence that treatment affects survival when it does not
materially affect cancer recurrence. By contrast, the evidence for causal
sufficiency is mixed. The point estimates of $\text{PSACE}_{01\mid g}$ are
uniformly positive, suggesting that patients whose cancer recurrence would be
prevented by treatment are also likely to benefit in survival. However, the 95\%
confidence intervals of $\text{PSACE}_{10\mid g}$ generally include zero,
indicating ambiguous survival responses among patients whose cancer recurrence
would be worsened by treatment. So these results provide only partial
support for causal sufficiency.
}

\bcol{Method~1 corresponds most closely to the monotonicity-based identification route
in \citet{Jiang-etal2016}, although our estimation procedure here is least-squares-based
rather than Bayesian. In their preferred empirical analysis without monotonicity,
\citet{Jiang-etal2016} conclude that both causal necessity and causal sufficiency
are supported. Our Methods~3 and 4, which do not rely on the monotonicity condition $S^1\geq S^0$ and are
therefore more plausible in light of Figure~\ref{fig2}, likewise support causal
necessity and also show positive evidence for the causal sufficiency within the latent stratum with $S^0=0$ and
$S^1=1$. However, because the confidence intervals for $\text{PSACE}_{10\mid g}$
generally cover zero, our analysis yields a more cautious conclusion
about full causal sufficiency. Thus, the two analyses agree that monotonicity is
questionable for the ACCT data and that causal necessity is supported, but our
approach yields a more cautious conclusion about full causal sufficiency.}

\section{Conclusion}  \label{sec9} 

In this work, we propose a novel framework that leverages multiple experimental datasets to identify and estimate the joint distribution of potential outcomes. First, we established the identifiability of joint distributions for binary and categorical outcomes under the assumption of transportability of state transition probabilities (Assumption \ref{assump2}) and a rank condition (Conditions \ref{cond1} and \ref{cond2}). Second, we introduce a simple least-squares-based method for estimating the joint distribution of potential outcomes. Finally, we extended our framework to identify and estimate principal stratification causal effects. These extend and complement existing methods in causal inference that integrate information from multiple datasets.  



\bibliographystyle{plainnat}
\bibliography{ref}

@article{imbens2024long,
  title={Long-term causal inference under persistent confounding via data combination},
  author={Imbens, Guido and Kallus, Nathan and Mao, Xiaojie and Wang, Yuhao},
  journal={Journal of the Royal Statistical Society Series B: Statistical Methodology},
  pages={qkae095},
  year={2024},
  publisher={Oxford University Press UK}
}

@article{kallus2024role,
  title={On the role of surrogates in the efficient estimation of treatment effects with limited outcome data},
  author={Kallus, Nathan and Mao, Xiaojie},
  journal={Journal of the Royal Statistical Society Series B: Statistical Methodology},
  pages={qkae099},
  year={2024},
  publisher={Oxford University Press UK}
}

@manual{Wang-etal2020,
	Author = {Yong Wang and Charles L. Lawson and  Richard J. Hanson},
	Title = {lsei: Solving Least Squares or Quadratic Programming Problems under Equality/Inequality Constraints},
	Url = {https://CRAN.R-project.org/package=lsei},
	Year = {2020}}

@article{Yin-etal2018,
	author = {Yunjian Yin and Lan Liu and Zhi Geng},
	journal = {Statistica Sinica},
	pages = {115-135},
	title = {Assessing the Treatment Effect Heterogeneity with a Latent Variable},
	volume = {28},
	year = {2018}}

@article{Zhang-etal2013,
	author = {Zhiwei Zhang and Chenguang Wang and Lei Nie and Guoxing Soon},
	journal = {Journal of the Royal Statistical Society: Series C (Applied Statistics)},
	pages = {687--704},
	title = {Assessing the heterogeneity of treatment effects via potential outcomes of individual patients},
	volume = {62},
	year = {2013}}

@article{Gilbert-Hudgens2008,
author = {Peter B Gilbert and Michael G Hudgens},
title = {Evaluating candidate principal surrogate endpoints},
volume = {64},
journal = {Biometrics},
number = {4},
 year = {2008}, 
pages = {1146--1154}}

@article{Rubin2006,
author = {Donald B. Rubin},
title = {Causal Inference Through Potential Outcomes and Principal Stratification: Application to Studies with “Censoring” Due to Death},
volume = {21},
journal = {Statistical Science},
number = {3},
 year = {2006}, 
pages = {299 -- 309}}

@article{Imbens-Angrist1994,
 author = {Guido W. Imbens and Joshua D. Angrist},
 journal = {Econometrica},
 number = {2},
 pages = {467--475},
 title = {Identification and Estimation of Local Average Treatment Effects},
 urldate = {2025-01-22},
 volume = {62},
 year = {1994}
}

@article{Imbens2004,
    author = {Imbens, Guido W.},
    title = {Nonparametric Estimation of Average Treatment Effects Under Exogeneity: A Review},
    journal = {The Review of Economics and Statistics},
    volume = {86},
    number = {1},
    pages = {4-29},
    year = {2004}}

@article{Angrist-etal1996,
author = {Joshua D. Angrist and Guido W. Imbens and Donald B. Rubin},
title = {Identification of Causal Effects Using Instrumental Variables},
journal = {Journal of the American Statistical Association},
volume = {91},
number = {434},
pages = {444--455},
year = {1996}
}

@article{Rosenman2023,
  title={Combining observational and experimental datasets using shrinkage estimators},
  author={Evan T.R. Rosenman and Guillaume Basse and Art B. Owen and Mike Baiocchi},
  journal={Biometrics},
  volume={79},
  number={4},
  year={2023},
    pages={2961--2973}
}

@article{Frangakis-Rubin2002,
  title={Principal Stratification in Causal Inference},
  author={Constantine E. Frangakis and Donald B. Rubin},
  journal={Biometrics},
  volume={58},
  number={1},
  pages={21--29},
  year={2002},
  publisher={International Biometric Society}
}

@article{Carrasco-Doukali2022,
    author = {Carrasco, Marine and Doukali, Mohamed},
    title = "{Testing overidentifying restrictions with many instruments and heteroscedasticity using regularised jackknife IV}",
    journal = {The Econometrics Journal},
    volume = {25},
    number = {1},
    pages = {71-97},
    year = {2022}
}

@article{Frank2019,
    author = {Windmeijer, Frank},
    title = "{Two-stage least squares as minimum distance}",
    journal = {The Econometrics Journal},
    volume = {22},
    number = {1},
    pages = {1--9},
    year = {2019}}

@article{Newey1985,
title = {Generalized method of moments specification testing},
journal = {Journal of Econometrics},
volume = {29},
number = {3},
pages = {229-256},
year = {1985},
issn = {0304-4076},
author = {Whitney K. Newey}
}

@article{Han-etal-2023,
	Author = {Larry Han and Jue Hou and Kelly Cho and Rui Duan and Tianxi Cai},
	Journal = {arXiv:2112.09313},
	Title = {Federated Adaptive Causal Estimation (FACE) of Target Treatment Effects},
	Year = {2023}}

@article{2022nathan,
	author = {Kallus, Nathan},
	journal = {arXiv preprint arXiv:2205.10327},
	title = {What's the Harm? Sharp Bounds on the Fraction Negatively Affected by Treatment},
	year = {2022}}

@inproceedings{2022NathanFacct,
	address = {New York, NY, USA},
	author = {Kallus, Nathan},
	booktitle = {2022 ACM Conference on Fairness, Accountability, and Transparency},
	numpages = {1},
	pages = {213},
	publisher = {Association for Computing Machinery},
	series = {FAccT '22},
	title = {Treatment Effect Risk: Bounds and Inference},
	year = {2022}}

@inproceedings{li2023trustworthy,
  title={Trustworthy policy learning under the counterfactual no-harm criterion},
  author={Li, Haoxuan and Zheng, Chunyuan and Cao, Yixiao and Geng, Zhi and Liu, Yue and Wu, Peng},
  booktitle={International Conference on Machine Learning},
  pages={20575--20598},
  year={2023},
  organization={PMLR}
}

@article{Wu-etal-2024-Harm,
	Author = {Peng Wu and Peng Ding and Zhi Geng and Yue Li},
	Journal = {arXiv preprint arXiv:2402.10537},
	Title = {Quantifying Individual Risk for Binary Outcome},
	Year = {2024}}

@book{pearl2009causality,
	author = {Pearl, Judea},
	publisher = {Cambridge university press},
	title = {Causality},
	year = {2009}}

@book{Pearl-Mackenzie2018,
	author = {Judea Pearl and Dana Mackenzie},
	publisher = {Hachette Book Group},
	title = {The Book of Why: The New Science of Cause and Effect},
	year = {2018}}

@book{Pearl-etal2016-primer,
	author = {Judea Pearl and Madelyn Glymour and Nicholas P. Jewell},
	publisher = {John Wiley \& Sons},
	title = {Causal Inference in Statistics: A Primer},
	year = {2016}}

@book{Imbens-Rubin2015,
	author = {Guido W. Imbens and Donald B. Rubin},
	publisher = {Cambridge University Press},
	title = {Causal Inference For Statistics Social and Biomedical Science},
	year = {2015}}

@book{Hernan-Robins2020,
	author = {Miguel A. Hern{\'a}n and James M. Robins},
	publisher = {Boca Raton: Chapman and Hall/CRC},
	title = {Causal Inference: What If},
	year = {2020}}

@article{Yin2018-second,
	author = {Yunjian Yin and Zheng Cai and Xiao-Hua Zhou},
	journal = {Biometrical Journal},
	pages = {879--892},
	title = {Using secondary outcome to sharpen bounds for treatment harm rate in characterizing heterogeneity},
	volume = {60},
	year = {2018}}

@article{Jun-Lee2023,
  title={Identifying the effect of persuasion},
  author={Sung Jae Jun and Sokbae Lee},
  journal={Journal of Political Economy},
  volume={131},
  number={8},
  pages={2032--2058},
  year={2023}
}

@article{Jun-Lee2024,
  title={Learning the Effect of Persuasion via Difference-In-Differences},
  author={Sung Jae Jun and Sokbae Lee},
  journal={arXiv preprint arXiv:2410.14871},
  year={2024}
}

@article{tian2000probabilities,
  title={Probabilities of causation: Bounds and identification},
  author={Tian, Jin and Pearl, Judea},
  journal={Annals of Mathematics and Artificial Intelligence},
  volume={28},
  number={1},
  pages={287--313},
  year={2000},
  publisher={Springer}
}

@article{li2022probabilities,
  title={Probabilities of causation: Adequate size of experimental and observational samples},
  author={Li, Ang and Mao, Ruirui and Pearl, Judea},
  journal={arXiv preprint arXiv:2210.05027},
  year={2022}
}

@article{shen2013treatment,
  title={Treatment benefit and treatment harm rate to characterize heterogeneity in treatment effect},
  author={Shen, Changyu and Jeong, Jaesik and Li, Xiaochun and Chen, Peng-Sheng and Buxton, Alfred},
  journal={Biometrics},
  volume={69},
  number={3},
  pages={724--731},
  year={2013},
  publisher={Oxford University Press}
}

@article{lu2023evaluating,
  title={Evaluating causes of effects by posterior effects of causes},
  author={Lu, Zitong and Geng, Zhi and Li, Wei and Zhu, Shengyu and Jia, Jinzhu},
  journal={Biometrika},
  volume={110},
  number={2},
  pages={449--465},
  year={2023},
  publisher={Oxford University Press}
}

@article{Dawid2022,
  title={Effects of causes and causes of effects},
  author={Dawid, A Philip and Musio, Monica},
  journal={Annual Review of Statistics and Its Application},
  volume={9},
  number={1},
  pages={261--287},
  year={2022},
  publisher={Annual Reviews}
}

@article{pearl1999,
  author={Judea Pearl},
  title={Probabilities of causation: three counterfactual interpretations and their identification},
journal={Synthese},
  volume={121},
  pages={93--149},
  year={1999}
}

@article{Wu-etal-2024-Compare,
	Author = {Peng Wu and Shanshan Luo and Zhi Geng},
	Journal = {Journal of the American Statistical Association},
	Title = {On the Comparative Analysis of Average Treatment Effects Estimation via Data Combination},
	Year = {2024}}

@article{Baker-etal2012,
	author = {Stuart G. Baker and Daniel J. Sargent and Marc Buyse and Tomasz Burzykowski},
	journal = {Biometrics},
	number = {68},
	pages = {248--257},
	title = {Predicting Treatment Effect from Surrogate Endpoints and Historical Trials: An Extrapolation Involving Probabilities of a Binary Outcome or Survival to a Specific Time},
	volume = {1},
	year = {2012}}

@article{Sargent-etal2005,
	author = {DJ Sargent  and HS Wieand  and DG Haller  and R Gray and JK Benedetti  and M Buyse and R Labianca and JF Seitz  and CJ O'Callaghan  and G Francini  and A Grothey  and M O'Connell  and PJ Catalano and CD Blanke  and D Kerr  and E Green  and N Wolmark  and T Andre  and RM  Goldberg and A De Gramont },
	journal = {Journal of Clinical Oncology},
	number = {34},
	pages = {8664--8670},
	title = {Disease-free survival versus overall survival as a primary end point for adjuvant colon cancer studies: individual patient data from 20,898 patients on 18 randomized trials},
	volume = {23},
	year = {2005}}

@article{Shahn-Madigan2025,
	author = {Zach Shahn and David Madigan},
	journal = {arXiv preprint arXiv:2509.20506},
	title = {Identification and Estimation of Joint Potential Outcome Distributions from a Single Study},
	year = {2025}}

@article{Jiang-etal2016,
	author = {Jiang, Zhichao and Ding, Peng and Geng, Zhi},
	journal = {Journal of the Royal Statistical Society Series B: Statistical Methodology},
	month = {11},
	number = {4},
	pages = {829-848},
	title = {{Principal Causal Effect Identification and Surrogate end point Evaluation by Multiple Trials}},
	volume = {78},
	year = {2016}}

@incollection{NEWEY19942111,
	author = {Whitney K. Newey and Daniel McFadden},
	booktitle = {Handbook of Econometrics},
	issn = {1573-4412},
	pages = {2111-2245},
	publisher = {Elsevier},
	title = {Chapter 36 Large sample estimation and hypothesis testing},
	volume = {4},
	year = {1994}}

@article{Kiviet2020,
	author = {Jan F. Kiviet},
	issn = {0304-4076},
	journal = {Journal of Econometrics},
	number = {2},
	pages = {294-316},
	title = {Testing the impossible: Identifying exclusion restrictions},
	volume = {218},
	year = {2020}}

@article{neyman1923,
	author = {Neyman, Jerzy},
	journal = {Statistical Science},
	pages = {465--472},
	title = {On the Application of Probability Theory to Agricultural Experiments},
	volume = {5},
	year = {1923}}

@article{li2023improving,
	author = {Li, Xinyu and Miao, Wang and Lu, Fang and Zhou, Xiao-Hua},
	journal = {Biometrics},
	number = {1},
	pages = {394--403},
	publisher = {Wiley Online Library},
	title = {Improving efficiency of inference in clinical trials with external control data},
	volume = {79},
	year = {2023}}

@article{Rubin1974,
	author = {Rubin, Donald B.},
	journal = {Journal of educational psychology},
	pages = {688-701},
	title = {Estimating causal effects of treatments in randomized and nonrandomized studies},
	volume = {66},
	year = {1974}}

@book{Rosenbaum2020,
	author = {Paul R. Rosenbaum},
	publisher = {Springer},
	title = {Design of Observational Studies},
	year = {2020}}

@article{dahabreh2021study,
	author = {Dahabreh, Issa J and Haneuse, Sebastien JP A and Robins, James M and Robertson, Sarah E and Buchanan, Ashley L and Stuart, Elizabeth A and Hern{\'a}n, Miguel A},
	journal = {American journal of epidemiology},
	number = {8},
	pages = {1632-1642},
	publisher = {Oxford University Press},
	title = {Study designs for extending causal inferences from a randomized trial to a target population},
	volume = {190},
	year = {2021}}

@article{Degtiar-Rose2023,
	author = {Irina Degtiar and Sherri Rose},
	journal = {Annual Review of Statistics and Its Application},
	pages = {501-524},
	title = {A Review of Generalizability and Transportability},
	volume = {10},
	year = {2023}}

@techreport{athey-etal2019,
	author = {Athey, Susan and Chetty, Raj and Imbens, Guido and Kang, Hyunseung},
	institution = {National Bureau of Economic Research},
	number = {26463},
	title = {The Surrogate Index: Combining Short-Term Proxies to Estimate Long-Term Treatment Effects More Rapidly and Precisely},
	type = {Working Paper},
	year = {2019}}

@article{hu2023longterm,
  title={Identification and estimation of treatment effects on long-term outcomes in clinical trials with external observational data},
  author={Wenjie Hu and Xiao-Hua Zhou and Peng Wu},
  journal={Statistica Sinica},
	pages = {1--22},
	volume = {35},
  year={2025}
}

@article{Humermund-Bareinboim2025,
	author = {Paul Hünermund and Elias Bareinboim},
	journal = {The Econometrics Journal},
	number = {1},
	pages = {41--82},
	title = {Causal inference and data fusion in econometrics},
	volume = {28},
	year = {2025}}

@article{Yang-Ding2020,
	author = {Shu Yang and Peng Ding},
	journal = {Journal of the American Statistical Association},
	number = {531},
	pages = {1540--1554},
	title = {Combining Multiple Observational Data Sources to Estimate Causal Effects},
	volume = {115},
	year = {2020}}

@article{Colnet-etal2023,
	author = {B{\'e}n{\'e}dicte Colnet and Imke Mayer and Guanhua Chen and Awa Dieng and Ruohong Li and Ga{\"e}l Varoquaux and Jean-Philippe Vert and Julie Josse and Shu Yang},
	journal = {Statistical Science},
	title = {Causal inference methods for combining randomized trials and observational studies: a review},
	volume = {To Appear},
	year = {2023}}

@article{Holland1986,
	author = {Paul W. Holland},
	journal = {Journal of the American Statistical Association},
	pages = {945-960},
	title = {Statistics and Causal Inference},
	volume = {81},
	year = {1986}}
 


%

 \newpage

  \begin{center}
\bf \Large 
Supplementary Material
\end{center}

\hypersetup{linkcolor=blue}
 


%

\setcounter{equation}{0}
\setcounter{section}{0}
\setcounter{figure}{0}
\setcounter{example}{0}
\setcounter{proposition}{0}
\setcounter{corollary}{0}
\setcounter{theorem}{0}
\setcounter{table}{0}
\setcounter{condition}{0}
\setcounter{lemma}{0}
\setcounter{remark}{0}

\renewcommand {\theproposition} {S\arabic{proposition}}
\renewcommand {\theexample} {S\arabic{example}}
\renewcommand {\thefigure} {S\arabic{figure}}
\renewcommand {\thetable} {S\arabic{table}}
\renewcommand {\theequation} {S\arabic{equation}}
\renewcommand {\thelemma} {S\arabic{lemma}}
\renewcommand {\thesection} {S\arabic{section}}
\renewcommand {\thetheorem} {S\arabic{theorem}}
\renewcommand {\thecorollary} {S\arabic{corollary}}
\renewcommand {\thecondition} {S\arabic{condition}}
\renewcommand {\thepage} {S\arabic{page}}
\renewcommand {\theremark} {S\arabic{remark}} 
  \renewcommand{\theassumption}{S\arabic{assumption}}

\setcounter{page}{1}

  \setcounter{equation}{0}
\renewcommand {\theequation} {S\arabic{equation}}
  \setcounter{lemma}{0}
\renewcommand {\thelemma} {S\arabic{lemma}}
   \setcounter{definition}{0}
\renewcommand {\thedefinition} {S\arabic{definition}}
   \setcounter{example}{0}
\renewcommand {\theexample} {S\arabic{example}}
   \setcounter{proposition}{0}
\renewcommand {\theproposition} {S\arabic{proposition}}
   \setcounter{corollary}{0}
\renewcommand {\thecorollary} {S\arabic{corollary}}
\setcounter{assumption}{0}
  \renewcommand{\theassumption}{S\arabic{assumption}}

 \section{Technical Proofs}
 Throughout the proofs,  for events $B_1$ and $B_2$, we set $\P( B_1 \mid B_2 ) = 0$   whenever  $\P(B_2) = 0$ to ensure well-definedness.

 \subsection{Proof of Theorem \ref{prop1}}
 \begin{proof}[Proof of  Theorem \ref{prop1}] 
  Recall that $ \pi_{b|a} =  \P(Y^1 = b \mid Y^0 = a)$ for $a,b=0,1$ are invariant across trials under Assumption \ref{assump2}.  Note that 
 \begin{equation*} 
 	\begin{cases} 
		\P(Y^1 = 1 \mid G=1) =  \pi_{1|0}  \cdot  \P(Y^0 = 0  \mid  G=1)  + \pi_{1|1}  \cdot  \P(Y^0 = 1  \mid  G=1)   	\\
		\P(Y^1 = 1  \mid G=2) =  \pi_{1|0}  \cdot  \P(Y^0 = 0  \mid  G=2)  +  \pi_{1|1} \cdot  \P(Y^0 = 1  \mid  G=2)   \\ 
		\vdots \\
		\P(Y^1 = 1 \mid G=m) =  \pi_{1|0}  \cdot  \P(Y^0 = 0  \mid \bcol{G=m})  +  \pi_{1|1} \cdot  \P(Y^0 = 1  \mid  G=m)  
	\end{cases} 
\end{equation*}	 
	and $\P(Y^1 =b  \mid G=g)$ and $\P(Y^0 = a  \mid G=g)$ are identifiable under Assumption \ref{assump1}. Then, the above system of equations contains a total of $m$ equations and $2$ unknown free parameters ($\pi_{1|0}$ and $\pi_{1|1}$). Thus, under Condition \ref{cond1} $\pi_{1|0}$ and $\pi_{1|1}$ are identifiable by solving the system of equations. In addition, $\pi_{0|0} = 1- \pi_{1|0} $ and $\pi_{0|1} = 1- \pi_{1|1}$ are identifiable. 
	
The identifiability of $\P(Y^1=b, Y^0 = a \mid G=g)$ follows immediately from the fact that 
	\begin{align*}
	       \P(Y^1=b, Y^0 = a \mid G=g) ={}&  \P(Y^1 = b \mid Y^0 = a) \cdot \P(Y^0 = a \mid  G=g) \\
	       = {}& \pi_{b|a} \cdot \P(Y = a \mid A = 0, G=g) ~ \text{for } g = \bcol{1, 2, ..., m}; a,b=0, 1. 
	       \end{align*}
\end{proof}



\subsection{Proof of Theorem \ref{thm1}}
 \begin{proof}[Proof of Theorem \ref{thm1}]  
Under Assumption  \ref{assump2}, $\pi_{i | j } :=   \P(Y^1 = i  \mid Y^0 = j)$ for $i,j=1, 2, ..., k$ are  invariant across trials.  Thus, we have that 
                  \begin{equation*}  
	             \P(Y^1 = i \mid G = g)  = \sum_{j=1}^k \pi_{i | j } \P(Y^0 = j \mid G=g), \quad  i=1, ..., k-1; g = 1, ..., m, 
\end{equation*}	 
where $ \P(Y^1 = i \mid G = g) $ and $\P(Y^0 = j \mid G=g)$ are identifiable under Assumption \ref{assump1}. Under Condition \ref{cond2}, 
the above system of equations include a total of $m \times (k-1)$ equations, however, the quantities $\{\pi_{i|j}, i= 1, ..., k; j = 1, ..., k\}$ contains $k\times (k-1)$ free parameters due to the truth that $\sum_{i=1}^k \pi_{i|j} = 1$ for each $j$.  
Thus, when $m \geq k$, the quantities $\{\pi_{i|j}, i= 1, ..., k; j = 1, ..., k\}$ are identifiable, and the conclusion of  Theorem \ref{thm1} holds immediately from the fact that 
	\[       \P(Y^1=i, Y^0 = j \mid G=g) =  \pi_{i | j }   \cdot \P(Y^0 = j \mid  G=g)~ \text{for }  i, j = 1, ..., k.  \]
This finishes the proof. 
    
\end{proof}

 \subsection{Proof of Theorem \ref{thm2}}
 
\begin{proof}[Proof of Theorem \ref{thm2}]  Recall that $m$ is the number of trials, which is a fixed number; and $n_g$ is the sample size of observed data in the $g$-th trial, which converges to infinity when we discussing the large sample properties, and $n = \sum_{g=1}^m n_g$ is the total sample size of the observed data.

For ease of presentation, we let 
  $\tilde Y_g = \P(Y=1 \mid G=g, A=1),$ $\tilde X_{1g} = \P(Y=0 \mid  G=g, A=0),$ and $\tilde X_{2g} = \P(Y=1 \mid  G=g, A=0)$. 
Denote $\tilde X_g = (\tilde X_{1g}, \tilde X_{2g})^\intercal$.   
Then by equation \eqref{eq2}, the true value of $\theta = ( \pi_{1|0},  \pi_{1|1} )$ can be written as 
   \[  \theta =  \left ( \frac{1}{m} \sum_{g=1}^m \tilde X_{g} \tilde X_{g}^\intercal \right )^{-1} \cdot \frac{1}{m} \sum_{g=1}^m \tilde X_g \tilde Y_g.   \]
However, if we use linear regression for estimating $\theta$, its estimator is 
     \[  \hat \theta = \left ( \frac{1}{m} \sum_{g=1}^m \hat X_{g} \hat X_{g}^\intercal \right )^{-1} \cdot \frac{1}{m} \sum_{g=1}^m \hat X_g \hat Y_g,   \]
where $\hat X_g$ and $\hat Y_g$ are the estimators of $\tilde X_g$ and $\tilde Y_g$, respectively, by replacing probabilities with frequencies.  
For example, 
  \[    \hat Y_g = \frac{ n^{-1} \sum_{i=1}^n \mathbb{I}(Y_i = 1, G_i = g, A_i = 1)  }{ n^{-1} \sum_{i=1}^n \mathbb{I}(G_i = g, A_i = 1)   }.   \]
For $\hat Y_g$, we have that 
   \begin{align*}
       \sqrt{n} (\hat Y_g - \tilde Y_g) ={}& 
       \frac{ n^{-1/2} \sum_{i=1}^n \mathbb{I}(Y_i = 1, G_i = g, A_i = 1)   }{ \P(G=g, A=1) } - \frac{ n^{-1/2} \sum_{i=1}^n \P(Y=1, G=g, A=1)  }{ \P(G=g, A=1) } + o_{\P}(1) \\
       ={}& \frac{1}{\P(G=g, A=1)} \cdot  \frac{1}{\sqrt{n}} \sum_{i=1}^n [  \mathbb{I}(Y_i = 1, G_i = g, A_i = 1)-
        \P(Y=1, G=g, A=1)]  + o_{\P}(1). 
   \end{align*}
Similarly, 
 \begin{align*}
    \sqrt{n} (\hat X_g - \tilde X_g) ={}&   \frac{1}{\sqrt{n}} \sum_{i=1}^n  \begin{pmatrix}
        \dfrac{  \mathbb{I}(Y_i = 0, G_i = g, A_i = 0)-
        \P(Y=0, G=g, A=0)}{\P(G=g, A=0)}   \\
        \dfrac{  \mathbb{I}(Y_i = 1, G_i = g, A_i = 0)-
        \P(Y=1, G=g, A=0)}{\P(G=g, A=0)}  
    \end{pmatrix}  + o_{\P}(1).    
 \end{align*}

By the strong law of large numbers, $\hat X_g$ converges to $\tilde X_g$ almost surely, which implies that $m^{-1} \sum_{g=1}^m \hat X_{g} \hat X_{g}^\intercal$ converges to 
$m^{-1} \sum_{g=1}^m \tilde X_{g} \tilde X_{g}^\intercal $ almost surely.  
Thus, $\hat \theta$
 has the same asymptotical distribution as 
     \[  \bar \theta = \left ( \frac{1}{m} \sum_{g=1}^m \tilde X_{g} \tilde X_{g}^\intercal \right )^{-1} \cdot \frac{1}{m} \sum_{g=1}^m \hat X_g \hat Y_g.   \]
Next, we focus on analyzing $\tilde \theta$. Let 
$C := m^{-1} \sum_{g=1}^m \tilde X_{g} \tilde X_{g}^\intercal$, then 
  \begin{align*}
      \sqrt{n}(\bar \theta - \theta)
={}& C^{-1} \cdot  \frac{\sqrt{n}}{m}  \sum_{g=1}^m 
 (\hat X_g \hat Y_g - \tilde X_g \tilde Y_g ). \\
 ={}&  C^{-1} \cdot  \frac{1}{m}  \sum_{g=1}^m 
 \{ \sqrt{n}(\hat X_g - \tilde X_g) \hat Y_g +  \sqrt{n}(\hat Y_g - \tilde Y_g)  \tilde X_g\}  \\
 ={}&   C^{-1} \cdot  \frac{1}{m}  \sum_{g=1}^m 
 \{ \sqrt{n}(\hat X_g - \tilde X_g) \tilde Y_g +  \sqrt{n}(\hat Y_g - \tilde Y_g)  \tilde X_g\}  + o_{\P}(1) \\
 ={}&  C^{-1} \cdot \left \{ \frac{1}{m}  \sum_{g=1}^m  \tilde Y_g  \cdot   \frac{1}{\sqrt{n}} \sum_{i=1}^n  \begin{pmatrix}
        \dfrac{  \mathbb{I}(Y_i = 0, G_i = g, A_i = 0)-
        \P(Y=0, G=g, A=0)}{\P(G=g, A=0)}   \\
        \dfrac{  \mathbb{I}(Y_i = 1, G_i = g, A_i = 0)-
        \P(Y=1, G=g, A=0)}{\P(G=g, A=0)}  
    \end{pmatrix} \right \}  \\
    +{}& ~  C^{-1} \cdot \left \{ \frac{1}{m}  \sum_{g=1}^m  \tilde X_g  \cdot \frac{1}{\sqrt{n}} \sum_{i=1}^n  \dfrac{  \mathbb{I}(Y_i = 1, G_i = g, A_i = 1)-
        \P(Y=1, G=g, A=1)}{\P(G=g, A=1)}\right \}  + o_{\P}(1) \\
         ={}&  C^{-1} \cdot  \frac{1}{\sqrt{n}} \sum_{i=1}^n  \left \{  \frac{1}{m}  \sum_{g=1}^m  \tilde Y_g  \begin{pmatrix}
        \dfrac{   \mathbb{I}(Y_i = 0, G_i = g, A_i = 0)-
        \P(Y=0, G=g, A=0) }{\P(G=g, A=0)}   \\
        \dfrac{   \mathbb{I}(Y_i = 1, G_i = g, A_i = 0)-
        \P(Y=1, G=g, A=0)  }{\P(G=g, A=0)}  
    \end{pmatrix}  \right \} \\
    +{}& ~  C^{-1} \cdot \frac{1}{\sqrt{n}} \sum_{i=1}^n\left ( \frac{1}{m}  \sum_{g=1}^m  \tilde X_g  \cdot   \dfrac{  \mathbb{I}(Y_i = 1, G_i = g, A_i = 1)- \P(Y=1, G=g, A=1)}{\P(G=g, A=1)}\right )  + o_{\P}(1). 
  \end{align*} 
Therefore, we have that 
 \[
   \sqrt{n}(\bar \theta - \theta) \xrightarrow{d} N(0, \sigma^2),
 \]
where $\sigma^2 = C^{-1} V C^{-1}$, $V$ equals to the variance of 
   \begin{align*}
      &  \frac{1}{m}  \sum_{g=1}^m  \Biggl \{  \tilde Y_g  \begin{pmatrix}
        \dfrac{   \mathbb{I}(Y_i = 0, G_i = g, A_i = 0)-
        \P(Y=0, G=g, A=0) }{\P(G=g, A=0)}   \\
        \dfrac{   \mathbb{I}(Y_i = 1, G_i = g, A_i = 0)-
        \P(Y=1, G=g, A=0)  }{\P(G=g, A=0)}  
    \end{pmatrix}  \\
     {}& + \tilde X_g  \cdot   \dfrac{  \mathbb{I}(Y_i = 1, G_i = g, A_i = 1)- \P(Y=1, G=g, A=1)}{\P(G=g, A=1)}  \Biggr \}. 
   \end{align*}
This finishes the proof. 

\end{proof}

 \subsection{Proof of Corollary \ref{coro3}}

%
%
%

\begin{proof}[Proof of Corollary \ref{coro3}]  
Define $\pi_{cd|ab} = \P( S^1 = c, Y^1  = d  \mid S^0 = a, Y^0 =b)$ for $a,b,c,d=0,1$, which are invariant parameters across trials by Assumption \ref{assump6}.    
Taking $(S^1, Y^1)$ and $(Y^0, S^0)$ as two category variables taking four different values, then following the proof of Theorem \ref{thm1}, under Assumption \ref{assump6}, we have the following system of equations 
    \begin{equation*}  
 	\begin{cases} 
	\P(S^1 = c, Y^1 = d  \mid G=g) = \sum_{a, b}   \pi_{cd | ab}   \P(S^0 =a, Y^0 = b  \mid  G=g),  ~ \text{for } c,d=0,1; g = 1, ..., m  \\
       \sum_{c, d} \P(S^1 = c, Y^1 = d  \mid G=g) = 1,   \quad g = 1, ..., m \\
        \sum_{a, b} \P(S^0 = a, Y^0 = b  \mid G=g) = 1, \quad g = 1, ..., m
	\end{cases} 
\end{equation*}	  
where $\P(S^1 = c, Y^1 = d  \mid G=g)$ and $ \P(S^0 =a, Y^0 = b \mid  G=g)$ are identifiable quantities under Assumption \ref{assump5}.

The above system of equations contains a total of $m \times 3$ linearly independent equations, 
and $3 \times 4 = 12$ unknown free parameters (i.e., $\{ (\pi_{11|ab}, \pi_{10|ab}, \pi_{01|ab}): a, b = 0, 1\}$) due to $\sum_{c, d} \pi_{cd | ab} = 1$ for $c=0,1; d = 0, 1$.  Then under Condition \ref{cond3}, $ \pi_{cd | ab} $ for $c=0,1; d = 0, 1$ are  identifiable. 
This also implies the identifiability of  $\P(S^0, S^1, Y^0, Y^1 \mid G=g)$ by noting that 
         \[   \P(S^0 =a, S^1=c, Y^0 = b, Y^1 = d \mid G=g) =   \pi_{cd | ab} \cdot \P(S^0 =a, Y^0 = b\mid G=g).     \]
Therefore, $\text{PSACE}_{ab|g}$ for $a,b\in\{0, 1\}$ and $g \in \cG$ are also identifiable. 

\end{proof}

 \subsection{Proof of Theorem \ref{thm-add}} 
 
\begin{proof}[Proof of Theorem \ref{thm-add}]  
Under Assumption \ref{assump6},   
$$\pi_{cd|ab} = \P( S^1 = c, Y^1  = d  \mid S^0 = a, Y^0 =b), \quad a,b,c,d=0,1,$$
 are invariant parameters across trials. When the monotonicity condition $S^1 \geq S^0$ and $Y^1 \geq Y^0$ holds, it has 9 possible values of $(a, b, c, d)$ with 5 free parameters in $\{ \pi_{cd|ab}, a,b,c,d=0,1 \}$, as shown in Table \ref{tab-s1}.    
 
\begin{table}[h!]\centering
\caption{Parameters under Assumption \ref{assump6} and the monotonicity condition $S^1\geq S^0$ and $Y^1\geq Y^0$}
\begin{tabular}{cccc | c|c}
\hline
$S^0=a$ & $Y^0=b$ & $S^1=c$  & $Y^1=d$  &  Restriction &  Number of Free Parameters                                  \\\hline
0 & 0 & 0  & 0  &    \multirow{4}{*}{$\sum_{c=0}^1 \sum_{d=0}^1 \pi_{cd|00}= 1$} & \multirow{4}{*}{3}          \\
0 & 0 & 0  & 1  &        \\
0 & 0 & 1  & 0  &        \\
0 & 0 & 1  & 1  &      \\ \hline 
0 & 1 & 0  & 1  &  \multirow{2}{*}{$\sum_{c=0}^1 \pi_{c1|01}= 1$} & \multirow{2}{*}{1}              \\
0 & 1 & 1  & 1  &        \\  \hline 
1 & 0 & 1  & 0  &     \multirow{2}{*}{$\sum_{d=0}^1 \pi_{1d|10}= 1$} & \multirow{2}{*}{1}       \\
1 & 0 & 1  & 1  &        \\ \hline 
1 & 1 & 1  & 1  &     $\pi_{11|11}= 1 $  & 0      \\
\hline 
\end{tabular}
\label{tab-s1}
\end{table}

On the other hand, under Assumption \ref{assump6}, we have the following system of equations 
    \begin{equation*}  
 	\begin{cases} 
	\P(S^1 = 0, Y^1 = 0  | G=g) ={}& \pi_{00 | 00}  \cdot  \P(S^0 =0, Y^0 = 0 |  G=g),  \\
    \P(S^1 = 0, Y^1 = 1  | G=g) ={}& \pi_{01 | 00}  \cdot  \P(S^0 =0, Y^0 = 0 |  G=g) + \pi_{01 | 01}  \cdot  \P(S^0 =0, Y^0 = 1 |  G=g),    \\
    \P(S^1 = 1, Y^1 = 0  | G=g) ={}& \pi_{10 | 00}  \cdot  \P(S^0 =0, Y^0 = 0 |  G=g)   + \pi_{10 | 10}  \cdot  \P(S^0 =1, Y^0 = 0 |  G=g) ,   \\
        \P(S^1 = 1, Y^1 = 1  | G=g) ={}& \pi_{11 | 00}  \cdot  \P(S^0 =0, Y^0 = 0 |  G=g) + \pi_{11 | 01}  \cdot  \P(S^0 =0, Y^0 = 1 |  G=g) \\
    \qquad \qquad \qquad \qquad \qquad \quad  +{}& \pi_{11 | 10}  \cdot  \P(S^0 =1, Y^0 = 0 |  G=g) + \pi_{11 | 11}  \cdot  \P(S^0 =1, Y^0 = 1 |  G=g), \\
      \quad \text{for }  g = 1, ..., m. 
	\end{cases} 
\end{equation*}	  
where $\P(S^1 = c, Y^1 = d  \mid G=g)$ and $ \P(S^0 =a, Y^0 = b  \mid  G=g)$ are identifiable quantities under Assumption \ref{assump5}.  
Clearly, $\pi_{00|00}$ is identifiable from the first row of the above system of equations. When Condition \ref{cond-add} holds, $\{\pi_{01|00}, \pi_{01|01},  \pi_{10|00},  \pi_{10|10} \}$ are identifiable from the second and third rows of the above system of equations. 
Then by the restriction (the second column) in Table \ref{tab-s1},  all the parameters in   $\{ \pi_{cd|ab}, a,b,c,d=0,1 \}$ are identifiable.  
 Then,  $\P(S^0, S^1, Y^0, Y^1 \mid G=g)$ is identifiable due to 
         \[   \P(S^0 =a, S^1=c, Y^0 = b, Y^1 = d \mid G=g) =   \pi_{cd | ab} \cdot \P(S^0 =a, Y^0 = b \mid G=g).     \] 
Therefore, $\text{PSACE}_{ab|g}$ for $a,b\in\{0, 1\}$ and $g \in \cG$ are also identifiable. 
         
\end{proof}

\subsection{Proof of Theorem \ref{thm3}} \label{app-s1.6}

\begin{proof}[Proof of Theorem \ref{thm3}]  As discussed below Assumption \ref{assump8} in the main text,  under Assumptions \ref{assump5} and \ref{assump8}, the principal scores $\delta_{ab|g} = \P(S^0 = a, S^1 = b \mid G=g)$ for $a = 0, 1$ and $b = 0, 1$ are identifiable.

We first prove Theorem \ref{thm3}(a).
Define  
       \[        \pi_{1|ab} = \P(Y^1 = 1 \mid S^0 =a , S^1 = b), \quad a = 0, 1; b = 0, 1,       \]
    which are invariant parameters across trials by Assumption \ref{assump7}.     
Note that by Assumptions \ref{assump5} and \ref{assump8},  $\pi_{1| 1 0} =0$ and 
       $$\pi_{1| 0 0} =  \P(Y^1 = 1 \mid S^0 =0 , S^1 = 0) =  \P(Y^1 = 1  \mid S^1 = 0)  =  \P(Y = 1  \mid S = 0, A  =1 )$$
 is identifiable. Thus, the set $\{ \pi_{1|ab}, a = 0, 1; b =0,1\}$ contains only two free parameters $\pi_{1| 01}$ and $\pi_{1| 11}$.  On the other hand, we have the following decomposition 
      \begin{align*} 
       \P(Y^1 =1  \mid G = g) 
      ={}& \P(Y^1 =1  \mid S^0 = 0, S^1 = 0, G = g) \P(S^0 = 0, S^1 = 0  \mid G = g) \\
      {}& +  \P(Y^1 =1  \mid S^0 = 0, S^1 = 1, G = g) \P(S^0 = 0, S^1 = 1  \mid G = g)  \\
      {}& + \P(Y^1 =1  \mid S^0 = 1, S^1 = 1, G = g) \P(S^0 = 1, S^1 = 1  \mid G = g)    \\
      ={}& \pi_{1| 0 0}  \delta_{00|g} +  \pi_{1| 0 1}  \delta_{01|g} + \pi_{1| 1 1}  \delta_{11|g},  \quad g = 1, ..., m,     
      \end{align*} 
  which has $m$ equations. Clearly, when Condition \ref{cond4}(i) holds, $\pi_{1| 01}$ and $\pi_{1| 11}$ are identifiable by solving the above equations.   Therefore, $\P(Y^1 \mid S^0, S^1, G=g)$ is identifiable. 

  In addition, the joint distribution  $\P(Y^1, S^0, S^1| G=g)$ for $g \in \cG$ are also identifiable by noting that   
           \[   \P(Y^1 = d, S^0 =a, S^1=b \mid G=g) =   \pi_{d | ab} \cdot \delta_{ab|g}.     \]
and both $\pi_{d | ab}$ and $\delta_{ab|g}$ are identifiable. 

Next, we prove Theorem \ref{thm3}(b). 
  Define  
       \[        \tilde \pi_{1|ab} = \P(Y^0 = 1 \mid S^0 =a , S^1 = b), \quad a = 0, 1; b = 0, 1,       \]
    which are invariant across trials by Assumption \ref{assump7}.     
By Assumptions \ref{assump5} and \ref{assump8},  $\tilde \pi_{1| 1 0} =0$ and  
       $$\tilde \pi_{1| 11} =  \P(Y^0 = 1 | S^0 =1 , S^1 = 1) =  \P(Y^0 = 1 | S^0 = 1)  =  \P(Y = 1 | S = 1, A  =0 )$$
 is identifiable. Thus, the set $\{ \tilde \pi_{1|ab}, a = 0, 1; b =0,1\}$ contains only two free parameters $\pi_{1| 00}$ and $\pi_{1| 01}$.               
 In addition, by Assumption \ref{assump7}, we have the following decomposition 
      \begin{align*} 
       \P(Y^0 =1 | G = g) 
      ={}& \tilde \pi_{1| 0 0}  \delta_{00|g} +   \tilde \pi_{1| 0 1}  \delta_{01|g} +  \tilde  \pi_{1| 1 1}  \delta_{11|g},  \quad g = 1, ..., m,     
      \end{align*} 
  which has $m$ equations. When Condition \ref{cond4}(b) holds, $\tilde \pi_{1| 00}$ and $\tilde \pi_{1| 01}$ are identifiable by solving the above equations. 
  Therefore, $\P(Y^0 \mid S^0, S^1, G=g)$ is identifiable. 
Also, the joint distribution of $(Y^0, S^0, S^1| G=g)$ for $g in \cG$ are also identifiable by noting that   
           \[   \P(Y^0 = c, S^0 =a, S^1=b | G=g) =  \tilde \pi_{c | ab} \cdot \delta_{ab|g}.     \]
and both $\tilde \pi_{c | ab}$ and $\delta_{ab|g}$ are identifiable. 

Theorem \ref{thm3}(c) is a direct corollary of Theorems \ref{thm3}(a) and \ref{thm3}(b). This finishes the proof. 

\end{proof}

\subsection{Proof of Theorem \ref{thm-s2}} \label{app-s1.7}

Before presenting the detailed proof of Theorem~\ref{thm-s2}, we first give a general result (Lemma \ref{lemma-s1}) that will be used in the proof of Theorem~\ref{thm-s2}.

\begin{lemma} \label{lemma-s1}
Consider a general estimator of $\theta \in \mathbb{R}^p$. Suppose that the true parameter $\theta$ and its estimator $\hat{\theta}$ take the forms 
 \[  \theta =  \left ( \frac{1}{m} \sum_{g=1}^m  X_{g} X_{g}^\intercal \right )^{-1} \cdot \frac{1}{m} \sum_{g=1}^m  X_g  Y_g,   \]
     \[  \hat \theta = \left ( \frac{1}{m} \sum_{g=1}^m \hat X_{g} \hat X_{g}^\intercal \right )^{-1} \cdot \frac{1}{m} \sum_{g=1}^m \hat X_g \hat Y_g,   \]
where $m$ is a fixed constant, $X_g$ and $Y_g$ are constants indexed by $g$,   
$\hat X_g$ and $\hat Y_g$ are estimators of $X_g \in \mathbb{R}^p$ and $Y_g \in \mathbb{R}$, respectively. Assume that they admit asymptotic linear representations: 
    \[ \sqrt{n}(\hat X_g - X_g) = \frac{1}{\sqrt{n}} \sum_{i=1}^n \phi_g(X_i, A_i, S_i, Y_i) + o_{\P}(1),\]
\[ \sqrt{n}(\hat Y_g - Y_g) = \frac{1}{\sqrt{n}} \sum_{i=1}^n \varphi_g(X_i, A_i, S_i, Y_i) + o_{\P}(1),\]
where $\phi_g(X_i, A_i, S_i, Y_i) \in \mathbb{R}^p$ and $\varphi_g(X_i, A_i, S_i, Y_i) \in \mathbb{R}$ have mean zero and finite variance, $o_{\P}(1)$ converges to zero in probability uniformly as $n \to \infty$. Then we have
 $ \sqrt{n}(\hat \theta - \theta) \xrightarrow{d} N(0, \sigma^2),$ 
where $\sigma^2 = C^{-1} V C^{-1}$, $C = m^{-1} \sum_{g=1}^m  X_{g} X_{g}^\intercal$, and  
   \begin{align*}
    V =   & 
   \mathrm{Var} \Biggl \{ \frac{1}{m}  \sum_{g=1}^m  \Big( \phi_g(X_i, A_i, S_i, Y_i) Y_g +  
 \varphi_g(X_i, A_i, S_i, Y_i) X_g \Big) \Biggr \}.
   \end{align*}
\end{lemma}

\begin{proof}[Proof of Lemma \ref{lemma-s1}]
By the strong law of large numbers, 
$\hat \theta$ 
 has the same asymptotical distribution as 
     \[  \bar \theta = \left ( \frac{1}{m} \sum_{g=1}^m  X_{g}  X_{g}^\intercal \right )^{-1} \cdot \frac{1}{m} \sum_{g=1}^m \hat X_g \hat Y_g = C^{-1} \frac{1}{m} \sum_{g=1}^m \hat X_g \hat Y_g.   \]
Thus, it is sufficient to analyzing $\tilde \theta$. Note that 
  \begin{align*}
      \sqrt{n}(\bar \theta - \theta)
={}& C^{-1} \cdot  \frac{\sqrt{n}}{m}  \sum_{g=1}^m 
 (\hat X_g \hat Y_g - X_g  Y_g ). \\
 ={}&  C^{-1} \cdot  \frac{1}{m}  \sum_{g=1}^m 
 \{ \sqrt{n}(\hat X_g - X_g) \hat Y_g +  \sqrt{n}(\hat Y_g -  Y_g)  X_g\}  \\
 ={}&   C^{-1} \cdot  \frac{1}{m}  \sum_{g=1}^m 
 \{ \sqrt{n}(\hat X_g -  X_g)  Y_g +  \sqrt{n}(\hat Y_g -  Y_g) X_g\}  + o_{\P}(1) \\
 ={}&  C^{-1} \cdot  \frac{1}{m}  \sum_{g=1}^m 
 \left\{ \frac{1}{\sqrt{n}} \sum_{i=1}^n \Big( \phi_g(X_i, A_i, S_i, Y_i) Y_g +  
 \varphi_g(X_i, A_i, S_i, Y_i) X_g \Big) \right\}  + o_{\P}(1) \\
 ={}&  C^{-1} \cdot  \frac{1}{\sqrt{n}} \sum_{i=1}^n 
 \left\{  \frac{1}{m}  \sum_{g=1}^m  \Big( \phi_g(X_i, A_i, S_i, Y_i) Y_g +  
 \varphi_g(X_i, A_i, S_i, Y_i) X_g \Big) \right\}  + o_{\P}(1).
 \end{align*}
This finishes the proof. 

\end{proof}

Next, we prove Theorem  \ref{thm-s2}. 

\begin{proof}[Proof of Theorem \ref{thm-s2}]

Recall that 
      \[        \pi_{1|ab} = \P(Y^1 = 1 \mid S^0 =a , S^1 = b),~    \tilde \pi_{1|ab} = \P(Y^0 = 1 \mid S^0 =a , S^1 = b).      \]
    for $a,b \in \{0, 1\}$. 
In addition, under the monotonicity condition $S^1 \geq S^0$, $\pi_{1|10}$ and $\tilde \pi_{1|10}$ are undefined and do not need estimation.  Let $\beta :=  (\pi_{1|00}, \pi_{1|01}, \pi_{1|11})$ and $\gamma := (\tilde \pi_{1|00}, \tilde  \pi_{1|01}, \tilde  \pi_{1|11})$.  
The estimation of $\beta$ and $\gamma$ relies on estimating $\delta_{ab\mid g} := \P(S^0 = a, S^1 = b \mid G = g)$ for $a,b \in \{0,1\}$. We therefore first establish the asymptotic linear representation of these estimators.

\smallskip 
{\bf First, we focus on $\delta_{ab\mid g}$ for $a, b \in \{0,1\}$}. 
 For any given $g$, $\hat \delta_{10 |g} = 0$  by Assumption \ref{assump8}, and 
the estimator of $\delta_{11|g}, \delta_{01|g}$, and $\delta_{00|g}$ are given as 
 \begin{align*}
     \hat \delta_{11|g} ={}& \frac{n^{-1}\sum_{i=1}^n \mathbb{I}(S_i = 1, A_i = 0, G_i = g) }{ n^{-1}\sum_{i=1}^n \mathbb{I}(A_i = 0, G_i = g) } \\
     \hat \delta_{01|g} ={}&  \frac{n^{-1}\sum_{i=1}^n \mathbb{I}(S_i = 1, A_i = 1, G_i = g) }{ n^{-1}\sum_{i=1}^n \mathbb{I}(A_i = 1, G_i = g) } -  \hat \delta_{11|g}\\
    \hat \delta_{00|g} ={}& \frac{n^{-1}\sum_{i=1}^n \mathbb{I}(S_i = 0, A_i = 1, G_i = g) }{ n^{-1}\sum_{i=1}^n \mathbb{I}(A_i = 1, G_i = g) }, 
 \end{align*}
where the estimate of $\delta_{00|g}$ is based on the observation that $\delta_{00|g} = \P(S^0 = 0, S^1 = 0\mid G=g) =  \P(S^1 = 0\mid G=g)  = \P(S=0\mid A=1, G=g)$. It is easy to show that 
 \begin{align*}
    \sqrt{n}( \hat \delta_{11|g} - \delta_{11|g}) ={}&    \frac{1}{\P(A=0, G=g)} \cdot  \frac{1}{\sqrt{n}} \sum_{i=1}^n [  \mathbb{I}(S_i = 1, A_i = 0, G_i = g) - 
        \P(S=1, A=0, G=g)] + o_{\P}(1), \\
  \sqrt{n}( \hat \delta_{01|g} - \delta_{01|g}) ={}&  \frac{1}{\P(A=1, G=g)} \cdot  \frac{1}{\sqrt{n}} \sum_{i=1}^n [  \mathbb{I}(S_i = 1, A_i = 1, G_i = g) - 
        \P(S=1, A=1, G=g)]  \\ 
   -{}&   \frac{1}{\P(A=0, G=g)} \cdot  \frac{1}{\sqrt{n}} \sum_{i=1}^n [  \mathbb{I}(S_i = 1, A_i = 0, G_i = g) -  \P(S=1, A=0, G=g)] + o_{\P}(1), \\
\sqrt{n}( \hat \delta_{00|g} - \delta_{00|g}) ={}&    \frac{1}{\P(A=1, G=g)} \cdot  \frac{1}{\sqrt{n}} \sum_{i=1}^n [  \mathbb{I}(S_i = 0, A_i = 1, G_i = g) - 
        \P(S=0, A=1, G=g)] + o_{\P}(1). 
 \end{align*}

\smallskip 
{\bf Second, we focus on  $\hat \pi_{1| 0 0} $ and $\hat{\tilde \pi}_{1| 11}$.} 
   \begin{align*}
       \hat \pi_{1| 0 0}  = {}& \frac{n^{-1}\sum_{i=1}^n \mathbb{I}(Y_i = 1, S_i = 0, A_i = 1) }{ n^{-1}\sum_{i=1}^n \mathbb{I}(S_i = 0, A_i = 1) }   \\
       \hat{\tilde \pi}_{1| 11} ={}& \frac{n^{-1}\sum_{i=1}^n \mathbb{I}(Y_i = 1, S_i = 1, A_i = 0) }{ n^{-1}\sum_{i=1}^n \mathbb{I}(S_i = 1, A_i = 0)}.
   \end{align*}
We could show that 
\begin{align*}
   & \sqrt{n}( \hat \pi_{1| 0 0}  -\pi_{1| 0 0}) \\
   ={}& \frac{1}{\P(S=0, A=1)} \cdot  \frac{1}{\sqrt{n}} \sum_{i=1}^n [  \mathbb{I}(Y_i = 1, S_i = 0, A_i = 1) - 
        \P(Y=1, S=0, A=1)] + o_{\P}(1), \\
         &  \sqrt{n}( \hat{\tilde \pi}_{1| 11}  - \tilde \pi_{1| 11} ) \\
         ={}&    \frac{1}{\P(S=1, A=0)} \cdot  \frac{1}{\sqrt{n}} \sum_{i=1}^n [  \mathbb{I}(Y_i = 1, S_i = 1, A_i = 0) - 
        \P(Y=1, S=1, A=0)] + o_{\P}(1).
\end{align*}

\smallskip 
{\bf Third, we focus on $(\hat \pi_{1|01}, \hat \pi_{1|11})$ and $(\hat{\tilde \pi}_{1|00},  \hat{\tilde \pi}_{1|01})$.}
For $(\hat \pi_{1\mid 01}, \hat \pi_{1\mid 11})$, for ease of presentation, we let 
    \begin{align*}
        Y_g :={}&   \P(Y =1 \mid G = g, A=1) - \pi_{1| 0 0}  \delta_{00|g}, \\
        X_g :={}& (X_{1g}, X_{2g})^\intercal = (\delta_{01|g}, \delta_{11|g})^\intercal.
    \end{align*}
Then the true value of $(\pi_{1|01},  \pi_{1|11})^\intercal$ can be written as 
   \[  (\pi_{1|01},  \pi_{1|11})^\intercal =  \left ( \frac{1}{m} \sum_{g=1}^m  X_{g}  X_{g}^\intercal \right )^{-1} \cdot \frac{1}{m} \sum_{g=1}^m  X_g  Y_g,  \]
  and the estimator is
\[ (\hat \pi_{1\mid 01}, \hat \pi_{1\mid 11})^\intercal =  \left ( \frac{1}{m} \sum_{g=1}^m \hat X_{g} \hat X_{g}^\intercal \right )^{-1} \cdot \frac{1}{m} \sum_{g=1}^m \hat X_g \hat Y_g,  \]  
  where $\hat X_g$ and $\hat Y_g$ are estimates of $X_g$ and $Y_g$, respectively. 
Note that \begin{align*} 
 & \sqrt{n} (\hat \pi_{1| 0 0} \hat  \delta_{00|g} -  \pi_{1| 0 0}  \delta_{00|g}  ) \\
 ={}& 
\sqrt{n} (\hat \pi_{1| 0 0}  \hat  \delta_{00|g} - \hat \pi_{1| 0 0}  \delta_{00|g})  + \sqrt{n} ( \hat \pi_{1| 0 0}  \delta_{00|g}  -   \pi_{1| 0 0}  \delta_{00|g}  )  \\
={}&  \pi_{1| 0 0}  \sqrt{n} ( \hat  \delta_{00|g} -   \delta_{00|g})  +  \delta_{00|g}  \sqrt{n} ( \hat \pi_{1| 0 0}  -   \pi_{1| 0 0}   )   + o_{\P}(1) \\
={}&     \frac{1}{\sqrt{n}}  \sum_{i=1}^n \pi_{1|00} \frac{  \mathbb{I}(S_i = 0, A_i = 1, G_i = g)-     \P(S=0, A=1, G=g)}{\P(A=1, G=g)} \\
{}& +  \frac{1}{\sqrt{n}} \sum_{i=1}^n  \delta_{00|g} \frac{ \mathbb{I}(Y_i = 1, S_i = 0, A_i = 1) - 
        \P(Y=1, S=0, A=1)}{\P(S=0, A=1)} + o_{\P}(1), 
\end{align*} 
we have  
  \begin{align*}
  	\sqrt{n} (\hat Y_g - Y_g) ={}&  \frac{1}{\sqrt{n}} \sum_{i=1}^n \frac{ \mathbb{I}(Y_i =1, G_i = g, A_i = 1) - \P(Y=1, G=g, A=1) }{ \P(A=1, G=g) } \\
	-{}&   \frac{1}{\sqrt{n}}  \sum_{i=1}^n \pi_{1|00} \frac{  \mathbb{I}(S_i = 0, A_i = 1, G_i = g)-    \P(S=0, A=1, G=g)}{\P(A=1, G=g)} \\
-{}&  \frac{1}{\sqrt{n}} \sum_{i=1}^n  \delta_{00|g} \frac{ \mathbb{I}(Y_i = 1, S_i = 0, A_i = 1) - 
        \P(Y=1, S=0, A=1)}{\P(S=0, A=1)} + o_{\P}(1). 
  \end{align*}
In addition, 
   \begin{align*}
   & \sqrt{n}(\hat X_g - X_g) = o_{\P}(1)  +  \frac{1}{\sqrt{n}} \sum_{i=1}^n    \\
   {}&    
   \begin{pmatrix}    
    \dfrac{\mathbb{I}(S_i = 1, A_i = 1, G_i = g) - 
        \P(S=1, A=1, G=g)}{\P(A=1, G=g)}  - \dfrac{\mathbb{I}(S_i = 1, A_i = 0, G_i = g) - 
        \P(S=1, A=0, G=g)}{\P(A=0, G=g)}   \\
        \dfrac{\mathbb{I}(S_i = 1, A_i = 0, G_i = g) - 
        \P(S=1, A=0, G=g)}{\P(A=0, G=g)} 
        \end{pmatrix}.  
   \end{align*}
Then, by Lemma \ref{lemma-s1}, Theorem \ref{thm-s2}(a) holds.

Similarly, we could show Theorem \ref{thm-s2}(b) by setting 
        \begin{align*}
        Y_g :={}&  
             \P(Y=1 \mid A=0, G = g) - \tilde  \pi_{1| 1 1}  \delta_{11|g}, \\
        X_g :={}& (X_{1g}, X_{2g})^\intercal = (\delta_{00|g}, \delta_{01|g})^\intercal, 
    \end{align*} 
 observing that the true value of
$({\tilde \pi}_{1|00},  {\tilde \pi}_{1|01})$ can be written as 
   \[  (\pi_{1|01},  \pi_{1|11})^\intercal =  \left ( \frac{1}{m} \sum_{g=1}^m  X_{g}  X_{g}^\intercal \right )^{-1} \cdot \frac{1}{m} \sum_{g=1}^m  X_g  Y_g,  \]
  and the estimator is
\[ (\hat{\tilde \pi}_{1|00},  \hat{\tilde \pi}_{1|01})^\intercal =  \left ( \frac{1}{m} \sum_{g=1}^m \hat X_{g} \hat X_{g}^\intercal \right )^{-1} \cdot \frac{1}{m} \sum_{g=1}^m \hat X_g \hat Y_g,  \]  
and 
   \begin{align*}
   & \sqrt{n}(\hat X_g - X_g) = o_{\P}(1)  +  \frac{1}{\sqrt{n}} \sum_{i=1}^n    \\
   {}&    
   \begin{pmatrix}    
        \dfrac{ \mathbb{I}(S_i = 0, A_i = 1, G_i = g) - 
        \P(S=0, A=1, G=g)}{\P(A=1, G=g)} \\
    \dfrac{\mathbb{I}(S_i = 1, A_i = 1, G_i = g) - 
        \P(S=1, A=1, G=g)}{\P(A=1, G=g)}  - \dfrac{\mathbb{I}(S_i = 1, A_i = 0, G_i = g) - 
        \P(S=1, A=0, G=g)}{\P(A=0, G=g)} 
        \end{pmatrix} 
   \end{align*} 
  \begin{align*}
  	\sqrt{n} (\hat Y_g - Y_g) ={}&  \frac{1}{\sqrt{n}} \sum_{i=1}^n \frac{ \mathbb{I}(Y_i =1, A_i = 0, G_i = g) - \P(Y=1, A=0,  G=g) }{ \P(A=0, G=g) } \\
	-{}&   \frac{1}{\sqrt{n}}  \sum_{i=1}^n \tilde \pi_{1|11}\frac{ \mathbb{I}(S_i = 1, A_i = 0, G_i = g) - 
        \P(S=1, A=0, G=g)}{\P(A=0, G=g)} \\
-{}&  \frac{1}{\sqrt{n}} \sum_{i=1}^n  \delta_{11|g} \frac{ \mathbb{I}(Y_i = 1, S_i = 1, A_i = 0) - 
        \P(Y=1, S=1, A=0)}{\P(S=1, A=0)} + o_{\P}(1). 
  \end{align*}

\end{proof}

\subsection{\bcol{Proof of the Asymptotic Distribution of the Test Statistic}}  \label{test-statistics-proof}

Recall that for each trial $g = 1,\dots,m$, $\tilde X_g = (\tilde X_{1g}, \tilde X_{2g})^\intercal$, $\hat X_g = (\hat X_{1g}, \hat X_{2g})^\intercal$, $\theta = (\pi_{1|0}, \pi_{1|1})^\intercal$ is the true value of the parameter. Under the null hypothesis, 
\[
\tilde Y_g = \tilde X_g^\intercal \theta \qquad\text{for all }g.
\]
Define the residual evaluated at $\theta$ as $\hat\epsilon_{\theta,g} = \hat Y_g - \hat  X_g^\intercal \theta$, and let 
\[ \hat{\boldsymbol{\epsilon}}_\theta = (\hat\epsilon_{\theta,1},\dots,\hat\epsilon_{\theta,m})^\intercal.
\]

We note that $\hat{Y}_g$ and $\hat{X}_g$ are asymptotically normal with convergence rate of order $1/\sqrt{n_g}$. It then follows that $\hat{\boldsymbol{\epsilon}}_\theta$ is asymptotically normal. When the proportion $n_g / n$ converges to a strictly positive constant in $(0,1)$, we can use CLT to easily prove that  
\[
\sqrt{n}\,\hat{\boldsymbol{\epsilon}}_\theta \xrightarrow{d} N(\mathbf{0},\boldsymbol{\Omega}),
\]
where $\boldsymbol{\Omega}$ is a diagonal matrix of full rank $m$. This follows from the independence of samples across trials, which implies that $\operatorname{cov}(\hat{\epsilon}_{\theta, g}, \hat{\epsilon}_{\theta, g'}) = 0$ for $g \neq g'$.

Let $\hat\epsilon_g = \hat Y_g - \hat  X_g^\intercal \hat \theta$ be the residual evaluated at $\hat \theta$, and 
  $\hat{\boldsymbol{\epsilon}} = (\hat\epsilon_{1},\dots,\hat\epsilon_{m})^\intercal$, then 
\[
J =(\sqrt{n}\,\hat{\boldsymbol{\epsilon}})^\intercal\,\boldsymbol{\Sigma}_*^+\,(\sqrt{n}\,\hat{\boldsymbol{\epsilon}})
= n\;\hat{\boldsymbol{\epsilon}}^\intercal\,\boldsymbol{\Sigma}_*^+\,\hat{\boldsymbol{\epsilon}},
\]
where $\boldsymbol{\Sigma}_*^+$ denotes the Moore–Penrose pseudo‑inverse of $\boldsymbol{\Sigma}_*$, 
  \[
\boldsymbol{\Sigma}_* = (\mathbf{I}_m - \tilde{\mathbf{P}})\,\boldsymbol{\Omega}\,(\mathbf{I}_m - \tilde{\mathbf{P}})^\intercal, 
\]
$\tilde{\mathbf{P}} = \tilde{\mathbf{X}}(\tilde{\mathbf{X}}^\intercal\tilde{\mathbf{X}})^{-1}\tilde{\mathbf{X}}^\intercal$ is the orthogonal projection matrix onto the column space of $\tilde{\mathbf{X}}$ (rank $2$), and 
 $\tilde{\mathbf{X}}$ be the $m\times 2$ matrix whose $g$-th row is $\tilde X_g^\intercal = (\tilde X_{1g},\tilde X_{2g})$.

Next, we focus on showing that $J \xrightarrow{d} \chi^2_{m-2}$.

\bigskip 
{\bf OLS Estimation and Residuals}. 
Set $\hat{\mathbf{Y}} = (\hat Y_1,\dots,\hat Y_m)^\intercal$ and let $\hat{\mathbf{X}}$ be the $m\times 2$ matrix whose $g$-th row is $\hat X_{g}^\intercal$.  The OLS estimator is 
\[
\hat{\theta} = (\hat{\mathbf{X}}^\intercal\hat{\mathbf{X}})^{-1}\hat{\mathbf{X}}^\intercal \hat{\mathbf{Y}}.
\]
By the definition of $\hat{\boldsymbol{\epsilon}}_\theta$, we could write $\hat{\mathbf{Y}} = \hat{\mathbf{X}}\theta + \hat{\boldsymbol{\epsilon}}_\theta$. Then
\[
\hat{\theta} - \theta = (\hat{\mathbf{X}}^\intercal\hat{\mathbf{X}})^{-1}\hat{\mathbf{X}}^\intercal\hat{\boldsymbol{\epsilon}}_\theta
= (\tilde{\mathbf{X}}^\intercal\tilde{\mathbf{X}})^{-1}\tilde{\mathbf{X}}^\intercal\hat{\boldsymbol{\epsilon}}_\theta + o_{\P}(1/\sqrt{n}),
\]
where the last equality follows from $(\hat{\mathbf{X}}^\intercal\hat{\mathbf{X}})^{-1} - (\tilde{\mathbf{X}}^\intercal\tilde{\mathbf{X}})^{-1} = O_{\P}(1)$, $\hat{\mathbf{X}} - \tilde{\mathbf{X}} = o_{\P}(1)$, and $\hat{\boldsymbol{\epsilon}}_\theta = O_{\P}(n^{-1/2})$.

The OLS residual vector $\hat{\boldsymbol{\epsilon}}$ is 
\[
\hat{\boldsymbol{\epsilon}} = \hat{\mathbf{Y}} - \hat{\mathbf{X}}\hat{\theta} 
= \hat{\boldsymbol{\epsilon}}_\theta - \hat{\mathbf{X}}(\hat{\theta}-\theta).
\]
Substituting the expansion for $\hat{\theta}-\theta$ gives 
\[
\hat{\boldsymbol{\epsilon}} = \hat{\boldsymbol{\epsilon}}_\theta - \tilde{\mathbf{X}}(\tilde{\mathbf{X}}^\intercal\tilde{\mathbf{X}})^{-1}\tilde{\mathbf{X}}^\intercal\hat{\boldsymbol{\epsilon}}_\theta + o_{\P}(1/\sqrt{n})
= (\mathbf{I}_m - \tilde{\mathbf{P}})\hat{\boldsymbol{\epsilon}}_\theta + o_{\P}(1/\sqrt{n}).
\]
Next, multiplying both sides by $\sqrt{n}$, we obtain 
\[
\sqrt{n}\,\hat{\boldsymbol{\epsilon}} = (\mathbf{I}_m - \tilde{\mathbf{P}})\sqrt{n}\,\hat{\boldsymbol{\epsilon}}_\theta + o_{\P}(1)
\xrightarrow{d} N(\mathbf{0},\boldsymbol{\Sigma}_*). 
\]
The matrix $\boldsymbol{\Sigma}_*$ is singular of rank $m-2$.

\bigskip 
{\bf Asymptotic distribution of $J$-Statistic}. Let $\mathbf{B} = \mathbf{I}_m - \tilde{\mathbf{P}}$, then $\hat{\boldsymbol{\epsilon}} = \mathbf{B}\hat{\boldsymbol{\epsilon}}_\theta + o_{\P}(1/\sqrt{n})$. Since $\mathbf{B}$ is symmetric idempotent with $\operatorname{rank}(\mathbf{B}) = m-2$, there exists an $m\times(m-2)$ matrix $\mathbf{V}$ whose columns form an orthonormal basis for the column space of $\mathbf{B}$, i.e.,  
\[
\mathbf{V}^\intercal\mathbf{V} = \mathbf{I}_{m-2},\quad \mathbf{V}\mathbf{V}^\intercal = \mathbf{B}.
\]
Project the scaled residuals onto this subspace: 
\[
\mathbf{z} = \mathbf{V}^\intercal(\sqrt{n}\,\hat{\boldsymbol{\epsilon}})
= \mathbf{V}^\intercal\mathbf{B}(\sqrt{n}\,\hat{\boldsymbol{\epsilon}}_\theta) + o_{\P}(1)
= \mathbf{V}^\intercal (\mathbf{V}\mathbf{V}^\intercal)   (\sqrt{n}\,\hat{\boldsymbol{\epsilon}}_\theta) + o_{\P}(1) 
= \mathbf{V}^\intercal  (\sqrt{n}\,\hat{\boldsymbol{\epsilon}}_\theta) + o_{\P}(1).
\]
Hence, we have 
\[
\mathbf{z} \xrightarrow{d} N(\mathbf{0},\,\mathbf{V}^\intercal\boldsymbol{\Omega}\mathbf{V}),
\] 
and $\mathbf{V}^\intercal\boldsymbol{\Omega}\mathbf{V}$ is a positive definite $(m-2)\times(m-2)$ matrix.

In addition, observe that 
\[
(\sqrt{n}\,\hat{\boldsymbol{\epsilon}})^\intercal\mathbf{V}(\mathbf{V}^\intercal\boldsymbol{\Omega}\mathbf{V})^{-1}\mathbf{V}^\intercal(\sqrt{n}\,\hat{\boldsymbol{\epsilon}})
= \mathbf{z}^\intercal(\mathbf{V}^\intercal\boldsymbol{\Omega}\mathbf{V})^{-1}\mathbf{z} + o_{\P}(1),
\]
and since $\mathbf{z} \xrightarrow{d} N(\mathbf{0},\mathbf{V}^\intercal\boldsymbol{\Omega}\mathbf{V})$, the standard quadratic form theory yields
\[
(\sqrt{n}\,\hat{\boldsymbol{\epsilon}})^\intercal\mathbf{V}(\mathbf{V}^\intercal\boldsymbol{\Omega}\mathbf{V})^{-1}\mathbf{V}^\intercal(\sqrt{n}\,\hat{\boldsymbol{\epsilon}}) \xrightarrow{d} \chi^2_{m-2}.
\]
Therefore, to establish that $J \xrightarrow{d} \chi^2_{m-2}$, it suffices to verify that 
\[
\boldsymbol{\Sigma}_*^+ = \mathbf{V}\,(\mathbf{V}^\intercal\boldsymbol{\Omega}\mathbf{V})^{-1}\,\mathbf{V}^\intercal.
\]
This can be shown by the following verification: 
\begin{itemize}

    \item Recall that $\boldsymbol{\Sigma}_* = \mathbf{B}\boldsymbol{\Omega}\mathbf{B}= \mathbf{V}\mathbf{V}^\intercal\boldsymbol{\Omega}\mathbf{V}\mathbf{V}^\intercal$, $\mathbf{V}^\intercal\mathbf{V} = \mathbf{I}_{m-2}$. 
    By calculation, 
     $$\boldsymbol{\Sigma}_*^+ \boldsymbol{\Sigma}_* =  \mathbf{V}\,(\mathbf{V}^\intercal\boldsymbol{\Omega}\mathbf{V})^{-1}\,\mathbf{V}^\intercal \mathbf{V}\mathbf{V}^\intercal\boldsymbol{\Omega}\mathbf{V}\mathbf{V}^\intercal =  \mathbf{V}\,(\mathbf{V}^\intercal\boldsymbol{\Omega}\mathbf{V})^{-1}\mathbf{V}^\intercal\boldsymbol{\Omega}\mathbf{V}\mathbf{V}^\intercal = \mathbf{V}\mathbf{V}^\intercal = \mathbf{B}.$$ 
     $$\boldsymbol{\Sigma}_* \boldsymbol{\Sigma}_*^+ =  \mathbf{V}\mathbf{V}^\intercal\boldsymbol{\Omega}\mathbf{V}\mathbf{V}^\intercal \mathbf{V}\,(\mathbf{V}^\intercal\boldsymbol{\Omega}\mathbf{V})^{-1}\,\mathbf{V}^\intercal = \mathbf{V}\mathbf{V}^\intercal\boldsymbol{\Omega}\mathbf{V} (\mathbf{V}^\intercal\boldsymbol{\Omega}\mathbf{V})^{-1}\,\mathbf{V}^\intercal =  \mathbf{V}\mathbf{V}^\intercal = \mathbf{B}.$$

\item Verify  $\boldsymbol{\Sigma}_*^+ \boldsymbol{\Sigma}_* \boldsymbol{\Sigma}_*^+ =\boldsymbol{\Sigma}_*^+$.  It holds by noting that 
 $\boldsymbol{\Sigma}_*^+ \boldsymbol{\Sigma}_* \boldsymbol{\Sigma}_*^+ = \mathbf{B} \boldsymbol{\Sigma}_*^+ = \mathbf{B} \mathbf{V}\,(\mathbf{V}^\intercal\boldsymbol{\Omega}\mathbf{V})^{-1}\,\mathbf{V}^\intercal = \mathbf{V}\mathbf{V}^\intercal  \mathbf{V}\,(\mathbf{V}^\intercal\boldsymbol{\Omega}\mathbf{V})^{-1}\,\mathbf{V}^\intercal  = \mathbf{V}(\mathbf{V}^\intercal\boldsymbol{\Omega}\mathbf{V})^{-1}\,\mathbf{V}^\intercal = \boldsymbol{\Sigma}_*^+$.

\item Verify $\boldsymbol{\Sigma}_* \boldsymbol{\Sigma}_*^+ \boldsymbol{\Sigma}_*= \boldsymbol{\Sigma}_*$. It holds by noting that 
  $\boldsymbol{\Sigma}_* \boldsymbol{\Sigma}_*^+ \boldsymbol{\Sigma}_* =\mathbf{B}  \boldsymbol{\Sigma}_* = \mathbf{B} \mathbf{B}\boldsymbol{\Omega}\mathbf{B} = \mathbf{B}\boldsymbol{\Omega}\mathbf{B} = \boldsymbol{\Sigma}_*$. ($\mathbf{B}$ is idempotent matrix)

\item Verify $(\boldsymbol{\Sigma}_* \boldsymbol{\Sigma}_*^+)^\intercal = \boldsymbol{\Sigma}_* \boldsymbol{\Sigma}_*^+$ and
  $(\boldsymbol{\Sigma}_*^+ \boldsymbol{\Sigma}_* )^\intercal = \boldsymbol{\Sigma}_*^+ \boldsymbol{\Sigma}_*$. They hold by noting that 
  $\boldsymbol{\Sigma}_* \boldsymbol{\Sigma}_*^+ = \mathbf{B} = \boldsymbol{\Sigma}_*^+ \boldsymbol{\Sigma}_*$ and $\mathbf{B}$ is symmetrical. 
\end{itemize}

\section{\bcol{Discussion: Incorporation of Covariates}}\label{app:covariates}

In the main text, we focus on the non-covariate, nonparametric multi-trial
identification framework. This is  
consistent with the ACCT application in Section~7 where no 
individual-level baseline covariates are available. At the same time, incorporating baseline
covariates is important. Covariates may improve the plausibility of the
transportability assumption by absorbing observed case-mix differences, and in many
settings they are also needed either for confounding adjustment or for capturing
effect heterogeneity that is not adequately represented in the marginal
untreated potential outcomes. This importance is also reflected in subsequent methodological
development: \citet{Shahn-Madigan2025} extend our
framework to settings with covariates. In this subsection, we provide a brief discussion on the implications of incorporating covariates, and refer to \citet{Shahn-Madigan2025} for more details. 

To clarify the technical implications of incorporating covariates, it is helpful
to distinguish two roles of covariates.

\paragraph{Role 1: Confounding adjustment.}
If treatment assignment is randomized only conditional on baseline covariates, or
if one considers an observational analogue, then the arm-specific counterfactual
risks within each trial (e.g., $\P(Y^0 \mid G = g)$ and $\P(Y^1 \mid G = g)$) must first be estimated after adjusting for covariates,
for example by standardization, weighting, or doubly robust procedures. Our
identifying equations then apply after replacing simple sample proportions by
these covariate-adjusted risk estimates.

\paragraph{Role 2: Effect-heterogeneity adjustment.}
A second and more direct role is that the key transportability assumption itself
may only hold after conditioning on additional baseline covariates. In this case,
Assumption~2 becomes a conditional invariance assumption.

We formalize this setting below for the binary-outcome case.

\begin{assumption}[Conditional unconfoundedness of trials]
\label{assump-cov1}
For all $g \in \cG = \{1,\ldots,m\}$,
(i) $A \indep (Y^0,Y^1) \mid G=g,X$, and
(ii) $0 < \P(A=1 \mid G=g,X) < 1$ almost surely.
\end{assumption}

\begin{assumption}[Conditional transportability of state transition probability]
\label{assump-cov2}
$Y^1 \indep G \mid Y^0, X$.
\end{assumption}

For each covariate value $x$ in the common support of $X$, define
\[
p_{1g}(x) := \P(Y=1 \mid A=1,G=g,X=x),
\qquad
p_{0g}(x) := \P(Y=1 \mid A=0,G=g,X=x).
\]
Under Assumption~\ref{assump-cov1},
\[
p_{1g}(x)=\P(Y^1=1 \mid G=g,X=x),
\qquad
p_{0g}(x)=\P(Y^0=1 \mid G=g,X=x).
\]
Moreover, Assumption~\ref{assump-cov2} implies that the following transitional probabilities do not depend on $G$:
\[
\pi_{1\mid 0}(x):=\P(Y^1=1 \mid Y^0=0,X=x),
\qquad
\pi_{1\mid 1}(x):=\P(Y^1=1 \mid Y^0=1,X=x),
\]
Hence, for every $g \in \cG$ and every $x$,
\begin{equation}
\label{eq-cov-main}
p_{1g}(x)
=
\pi_{1\mid 0}(x)\{1-p_{0g}(x)\}
+
\pi_{1\mid 1}(x)p_{0g}(x).
\end{equation}

In addition to Assumptions~\ref{assump-cov1}--\ref{assump-cov2}, one also needs
common support of $X$ across the relevant trials. Otherwise, the conditional
transition law at a given covariate value cannot be learned from all trials. Moreover, we also need to extend the rank condition to incorporate the covariates.

\begin{condition}[Overlap Condition]
  \label{cond-cov1}
  For every $g = 1, \dots, m$, $\P(G = g \mid X) > 0$ almost surely. 
  \end{condition}

\begin{condition}[Conditional full-column rank]
\label{cond-cov2}
For every $x$ in the common support of $X$, the matrix
$
\bigl(
\P(Y^0=0 \mid G=g,X=x),\,
\P(Y^0=1 \mid G=g,X=x)
\bigr)_{m\times 2}$ 
has full column rank.
\end{condition}

\begin{theorem}[Covariate-adjusted binary outcome]
\label{thm-cov1}
Under Assumptions~\ref{assump-cov1}--\ref{assump-cov2} and
Conditions~\ref{cond-cov1}--\ref{cond-cov2}, the conditional distributions
$\P(Y^1 \mid Y^0,G=g,X=x)$ and $\P(Y^1,Y^0 \mid G=g,X=x)$ are identifiable for all
$g \in \cG$. Consequently, the marginal
joint distribution for trial $g$ is identifiable through
\[
\P(Y^1,Y^0 \mid G=g)
=
\int
\P(Y^1,Y^0 \mid G=g,X=x)\,dF_{X\mid G=g}(x).
\]
\end{theorem}

Theorem~\ref{thm-cov1} shows that the law-of-total-probability argument in the
main text extends pointwise in $X$. At the same time, it also makes clear how the
required conditions change once covariates are introduced. In particular, the
positivity and rank conditions now become conditional on $X$, and the identifying
variation must remain after conditioning on covariates.

\paragraph{Low-dimensional discrete covariates.}
If $X$ takes only a small number of discrete values, then estimation and inference
are conceptually straightforward. One may proceed stratum by stratum. For each
covariate stratum $X=x$, estimate $p_{0g}(x)$ and $p_{1g}(x)$, solve the same
least-squares problem as in Section~4 of the main text to recover
$\pi_{1\mid 0}(x)$ and $\pi_{1\mid 1}(x)$, reconstruct
$\P(Y^1,Y^0 \mid G=g,X=x)$, and then standardize over $X \mid G=g$ to obtain
$\P(Y^1,Y^0 \mid G=g)$.

\paragraph{Continuous or high-dimensional covariates.}
When the relevant covariates are continuous or high-dimensional, the stratum-by-stratum
strategy is no longer viable. Exact stratification leads to sparsity, the functions
$\pi_{1\mid 0}(x)$ and $\pi_{1\mid 1}(x)$ become unknown infinite-dimensional
objects, and the conditional arm-specific risks $p_{0g}(x)$ and $p_{1g}(x)$ must
also be estimated as functions of $x$. As a result, one typically needs additional
structure.

A natural approach is to impose finite-dimensional parametric models such as
\[
\pi_{1\mid 0}(x)=f(x;\beta^\star),
\qquad
\pi_{1\mid 1}(x)=h(x;\lambda^\star),
\]
where $\beta^\star$ and $\lambda^\star$ are finite-dimensional parameters and
$f(\cdot;\beta)$ and $h(\cdot;\lambda)$ are known functions, for example generalized
linear models. Under these models, equation~\eqref{eq-cov-main} becomes
\[
p_{1g}(x)
=
f(x;\beta^\star)\{1-p_{0g}(x)\}
+
h(x;\lambda^\star)p_{0g}(x).
\]
In this parametric regime, one no longer needs to impose the previous pointwise
rank condition for every $x$; rather, it suffices that the observed data provide
enough variation across trials to identify the finite-dimensional parameters
$(\beta^\star,\lambda^\star)$. In other words, parametric identification requires weaker identifying variations across trials. 

One possible estimation strategy is to first construct nuisance estimators
$\hat p_{1g}(x)$ and $\hat p_{0g}(x)$, and then solve
\[
(\hat\beta,\hat\lambda)
\in
\arg\min_{\beta,\lambda}
\frac{1}{n}
\sum_{g=1}^m \sum_{i=1}^n
I(G_i=g)
\Bigl[
\hat p_{1g}(X_i)
-
f(X_i;\beta)\{1-\hat p_{0g}(X_i)\}
-
h(X_i;\lambda)\hat p_{0g}(X_i)
\Bigr]^2.
\]
In this regime, however, the problem is no longer a direct extension of the
nonparametric least-squares framework in the main text. The main challenge shifts
from solving a finite-dimensional linear system to learning covariate-dependent
nuisance objects and covariate-dependent transition laws. In particular,
high-dimensional or flexible nuisance estimation typically calls for more structured
semiparametric or debiased approaches. See \citet{Shahn-Madigan2025} for one such
development based on Neyman-orthogonal moments and sample splitting.

Overall, conditioning on measured baseline covariates may improve the plausibility
of the transportability assumption, because some between-trial heterogeneity that
would otherwise appear as violations of $Y^1 \indep G \mid Y^0$ may instead be
explained by observed covariates. On the other hand, this comes at a
cost: the relevant overlap and rank conditions become conditional on $X$, and
continuous or high-dimensional covariates require additional modeling and inference
machinery. Analogous covariate-adjusted extensions in the principal-stratification
setting are possible but would be substantially more involved, and we do not pursue
them here.

\section{\bcol{Implementation of the Test for Assumption \ref{assump2}}}
\label{implement-test-statistic}

In this section, we describe a feasible procedure for conducting the test of Assumption \ref{assump2} using the $J$-statistic. The main practical challenge is estimating the asymptotic covariance matrix $\boldsymbol{\Sigma}_*$ of the infeasible sample moment vector $\sqrt{n}\hat{\boldsymbol{\epsilon}}$; see Section \ref{test-statistics-proof} for details. We use the bootstrap to obtain a consistent estimator $\hat{\boldsymbol{\Sigma}}_*^{\text{boot}}$, from which we compute the required pseudo-inverse. The steps below outline the procedure.

\begin{enumerate}
    \item Obtain the OLS estimator $\hat{\theta}$, and compute the OLS residuals $\hat{\boldsymbol{\epsilon}} = \hat{\mathbf{Y}} - \hat{\mathbf{X}}\hat{\theta}$.
   
    \item Use the bootstrap to estimate $\boldsymbol{\Sigma}_*$. This gives $\hat{\boldsymbol{\Sigma}}_*^{\text{boot}}$. The detailed procedure is as follows: 
        \begin{enumerate}
            \item For $b = 1, \dots, B$ (e.g., $B = 500$):
            \begin{enumerate}
                \item Generate a bootstrap sample by resampling the original observations with replacement. To avoid empty cells in the trials defined by $G$, we use a stratified bootstrap: resampling is performed separately within each trial $g$. 
                
                 \item From the bootstrap sample, compute $\hat{Y}_g^{(b)}$, $\hat{X}_{g}^{(b)}$, and $\hat{\theta}^{(b)}$ in exactly the same way as from the original data. 
               
                \item Form the bootstrap residual vector $\hat{\boldsymbol{\epsilon}}^{(b)}$ with components
                \[
         \hat{\epsilon}_{g}^{(b)} = \hat{Y}_g^{(b)} -  \hat  X_g^\intercal \hat \theta^{(b)}.
                \]
            \end{enumerate}
            \item Compute the sample covariance matrix of the $\sqrt{n}\,\hat{\boldsymbol{\epsilon}}^{(b)}$ vectors across the $B$ bootstrap replicates:
            \[
    \hat{\boldsymbol{\Sigma}}_*^{\text{boot}}= n \cdot \frac{1}{B-1} \sum_{b=1}^{B} \bigl(\hat{\boldsymbol{\epsilon}}^{(b)} - \bar{\hat{\boldsymbol{\epsilon}}}\bigr)\bigl(\hat{\boldsymbol{\epsilon}}^{(b)} - \bar{\hat{\boldsymbol{\epsilon}}}\bigr)^\intercal,
            \]
            where $\bar{\hat{\boldsymbol{\epsilon}}} = \frac{1}{B}\sum_{b=1}^{B} \hat{\boldsymbol{\epsilon}}^{(b)}$.
        \end{enumerate}
       
    
    \item Compute the pseudo-inverse $\hat{\boldsymbol{\Sigma}}_*^+$ and then compute $J = n \cdot \hat{\boldsymbol{\epsilon}}^\intercal \hat{\boldsymbol{\Sigma}}_*^+ \hat{\boldsymbol{\epsilon}}$. Finally, compare the $J$-statistic to the $\chi^2_{m-2}$ critical value. 
\end{enumerate}

\section{Estimation Method Based on Corollary \ref{coro3} and Theorem \ref{thm-add}}  \label{appendix-s3}

In this section, we present the estimation method for the joint distribution $\P(S^0, S^1, Y^0, Y^1 \mid G=g)$ and the principal stratification average causal effect $\textup{PSACE}_{ab|g}$    based on Corollary \ref{coro3} and Theorem \ref{thm-add}, respectively.

\subsection{Estimation Method Based on Corollary \ref{coro3}}
    We first give the estimation method for $\P(S^0, S^1, Y^0, Y^1 \mid G=g)$ and $\textup{PSACE}_{ab|g}$  for $g \in \mathcal{G}$ based on Corollary \ref{coro3}. 
Assumption \ref{assump6} implies that $\P(S^1 =c, Y^1 = d \mid S^0 = a, Y^0 =b)$  for $a, b, c, d = 0, 1$ are invariant parameters across trials.  The proposed estimation method consists of the following steps. 
   \begin{itemize}  
   \item {\bf Step 1.} estimate $\P(S^1 =c, Y^1 = d \mid G=g)$ and $\P(S^0 =a, Y^0 = b \mid G=g)$ for  $a, b, c, d = 0, 1$. We denote the estimators as $\hat \P(S^1 =c, Y^1 = d \mid G=g)$ and $\hat \P(S^0 =a, Y^0 = b \mid G=g)$.  
   
   \item {\bf Step 2.}  
   estimate $\P(S^1 =0, Y^1 = 0 \mid S^0 = a, Y^0 =b)$ for $a, b = 0, 1$ by conducting a linear regression of  $\hat \P(S^1 =0, Y^1 = 0 \mid G=g) $ on $( \hat \P(S^0 = 0, Y^0 = 0 | G=g), \hat \P(S^0 = 0, Y^0 = 1 | G=g), \hat \P(S^0 = 1, Y^0 = 0 | G=g),\hat \P(S^0 = 1, Y^0 = 1 | G=g))$. 
 Similarly, we can estimate $\P(S^1 =0, Y^1 = 1 \mid S^0 = a, Y^0 =b)$, $\P(S^1 =1, Y^1 = 0 \mid S^0 = a, Y^0 =b)$, and $\P(S^1 =1, Y^1 = 1 \mid S^0 = a, Y^0 =b)$ for $a, b = 0, 1$. We denote the estimator as 
  $\hat \P(S^1 =d, Y^1 = d \mid S^0 = a, Y^0 =b)$.  
  
     \item {\bf Step 3.} estimate   the joint distribution 
  $\P(S^0=a, Y^0=b, S^1=c, Y^1=d \mid G=g)$. Based on $\hat \P(S^1 =c, Y^1 = d \mid S^0 = a, Y^0 =b)$, the estimator of 
  $\P(S^0=a, Y^0=b, S^1=c, Y^1=d \mid G=g)$ is given as 
   \begin{align*} 
      & \hat \P(S^0=a, Y^0=b, S^1=c, Y^1=d \mid G=g) \\
     ={}& \hat \P(S^1 =c, Y^1 = d \mid S^0 = a, Y^0 =b) \cdot \hat \P(  S^0=a, Y^0=b \mid G=g)  
     \end{align*}
     
     \item {\bf Step 4.}  estimate  $\text{PSACE}_{ab|g}$, which is defiend by
       \[     \text{PSACE}_{ab|g} = \P( Y^1 = 1 \mid S^0 = a, S^1 = b, G=g) - \P( Y^0 = 1 \mid S^0 = a, S^1 = b, G=g),~ a = 0, 1; b = 0, 1.   \] 
  The estimator for  $ \P( Y^1 = 1 \mid S^0 = a, S^1 = b, G=g) $ is given as 
        \begin{align*}
        \hat \P( Y^1 = 1 \mid & S^0 = a, S^1 = b, G=g) ={} \frac{ \hat \P(Y^1 = 1, S^0 = a, S^1 = b \mid G=g)  }{\hat \P( S^0 = a, S^1 = b \mid G=g) }  \\
                                      ={}&  \frac{ \sum_{d=0,1} \hat \P(Y^1 = 1, Y^0 = d, S^0 = a, S^1 = b \mid G=g)  }{ \sum_{c,d=0,1} \hat \P( S^0 = a, S^1 = b, Y^0 = c, Y^1 = d \mid G=g) }             
        \end{align*}    
Likewise, we can estimate $\P( Y^0 = 1 \mid S^0 = a, S^1 = b, G=g)$ with 
                     \begin{align*}
        \hat \P( Y^0 = 1 \mid & S^0 = a, S^1 = b, G=g) ={} \frac{ \hat \P(Y^0 = 1, S^0 = a, S^1 = b \mid G=g)  }{\hat \P( S^0 = a, S^1 = b \mid G=g) }  \\
                                      ={}&  \frac{ \sum_{d=0,1} \hat \P(Y^0 = 1, Y^1 = d, S^0 = a, S^1 = b \mid G=g)  }{ \sum_{c,d=0,1} \hat \P( S^0 = a, S^1 = b, Y^0 = c, Y^1 = d \mid G=g) }             
        \end{align*}   
   \end{itemize}

\medskip 
Similar to the estimation method in Section \ref{appendix-s2-3}, another robust approach to estimate the invariance parameters is to solve a least squares optimization problem under several equality and inequality constraints. 

\subsection{Estimation Method Based on Theorem \ref{thm-add}} 
We then present the estimation method for $\P(S^0, S^1, Y^0, Y^1 \mid G=g)$ and  $\textup{PSACE}_{ab|g}$  for $g \in \mathcal{G}$ based on Theorem \ref{thm-add}.     
Under Assumption \ref{assump6},   
$$\pi_{cd|ab} = \P( S^1 = c, Y^1  = d  \mid S^0 = a, Y^0 =b), \quad a,b,c,d=0,1,$$
 are invariant across trials. When the monotonicity condition $S^1 \geq S^0$ and $Y^1 \geq Y^0$ holds, 
   $\pi_{00 | 01} = \pi_{10 | 01} = \pi_{ 00 |10 } =\pi_{ 01 |10 } =\pi_{ 00 |11 }  = \pi_{ 01 |11 } = \pi_{ 10 |11 }   \equiv 0$, and $\pi_{11|11}= 1$,  as shown in Table \ref{tab-s1}. 
We can estimate the other invariant parameters and $\textup{PSACE}_{ab|g}$  with the following steps. 
   \begin{itemize}
   \item {\bf Step 1.} estimate $\P(S^1 =c, Y^1 = d \mid G=g)$ and $\P(S^0 =a, Y^0 = b \mid G=g)$ for  $a, b, c, d = 0, 1$. We denote the estimators as $\hat \P(S^1 =c, Y^1 = d \mid G=g)$ and $\hat \P(S^0 =a, Y^0 = b \mid G=g)$. 
   
   \item {\bf Step 2.}  estimate the other invariant parameters and the joint distribution $\P(S^0=a, Y^0=b, S^1=c, Y^1=d \mid G=g)$.  
         \begin{itemize}
         	\item     estimate $\pi_{00|00}$ by linear regression of $\hat \P(S^1 =0, Y^1 = 0 \mid G=g) $ on $\hat \P(S^0 = 0, Y^0 = 0 | G=g)$; 
		\item     estimate $(\pi_{01 |00}, \pi_{01 | 01} )$ by  linear regression of $\hat \P(S^1 =0, Y^1 = 1 \mid G=g) $ on $( \hat \P(S^0 = 0, Y^0 = 0 | G=g), \hat \P(S^0 = 0, Y^0 = 1 | G=g))$; 
		\item   estimate $(\pi_{10 | 00}, \pi_{10 | 10} )$ by linear regression of $\hat \P(S^1 =1, Y^1 = 0 \mid G=g) $ on $( \hat \P(S^0 = 0, Y^0 = 0 | G=g), \hat \P(S^0 = 1, Y^0 = 0 | G=g))$;
		
		\item estimate $\pi_{11|00}$ with $\pi_{11|00} = 1 - \pi_{00|00} -\pi_{01 |00} - \pi_{10 | 00}$,  $\pi_{11|01}$ with $\pi_{11|01} = 1 -  \pi_{01 | 01}$, and  $\pi_{11|01}$ with $\pi_{11|10} = 1 -   \pi_{10 | 10}$. 
         \end{itemize}
Let $\hat \pi_{cd|ab}$ be the estimator of $\pi_{cd|ab}$, then the estimator of the joint distribution 
  $\P(S^0=a, Y^0=b, S^1=c, Y^1=d \mid G=g)$ is given as 
   \begin{align*} 
       \hat \P(S^0=a, Y^0=b, S^1=c, Y^1=d \mid G=g) = \hat \pi_{cd|ab} \cdot \hat \P(  S^0=a, Y^0=b \mid G=g)  
     \end{align*}
     
          \item {\bf Step 4.}  estimate  $\text{PSACE}_{ab|g}$. 
  The estimator for  $ \P( Y^1 = 1 \mid S^0 = a, S^1 = b, G=g) $ is  
        \begin{align*}
        \hat \P( Y^1 = 1 \mid & S^0 = a, S^1 = b, G=g) ={} \frac{ \hat \P(Y^1 = 1, S^0 = a, S^1 = b \mid G=g)  }{\hat \P( S^0 = a, S^1 = b \mid G=g) }  \\
                                      ={}&  \frac{ \sum_{d=0,1} \hat \P(Y^1 = 1, Y^0 = d, S^0 = a, S^1 = b \mid G=g)  }{ \sum_{c,d=0,1} \hat \P( S^0 = a, S^1 = b, Y^0 = c, Y^1 = d \mid G=g) }             
        \end{align*}    
Likewise, we can estimate $\P( Y^0 = 1 \mid S^0 = a, S^1 = b, G=g)$.   
   \end{itemize}

Similar to the estimation method in Section \ref{appendix-s2-3}, another robust approach to estimate the invariance parameters is to solve a least squares optimization problem under several equality and inequality constraints.

\subsection{\bcol{Variance Estimation via Bootstrap}}
\label{appendix-s3.3}

In both the simulation and the application, we use a nonparametric stratified bootstrap for variance estimation. The procedure is as follows:
	\begin{itemize}
		\item  {Step 1: Resample within each stratum.}  
    For each bootstrap replication $b = 1, \ldots, B$:
    \begin{itemize}
        \item For each trial $g$, draw a bootstrap sample by sampling $n_g$ observations with replacement from the original $n_g$ observations in that trial.
        \item Combine the resampled observations to form the complete bootstrap dataset.
    \end{itemize}
		
       \item {Step 2: Recompute the estimator.}  
    Using each bootstrap dataset $b$ ($b = 1, ..., B$), re-estimate the parameter $\theta$, denoted as $\hat{\theta}^{(b)}$. 

 \item {Step 3: Variance estimation.}  
    After obtaining $\{\hat{\theta}^{(b)}\}_{b=1}^B$, compute the estimated variance:
    \[
        \widehat{\mathrm{Var}}(\hat{\theta}) 
        = \frac{1}{B-1}\sum_{b=1}^B \left(\hat{\theta}^{(b)} - \bar{\theta} \right)^2,
        \qquad 
        \bar{\theta} = \frac{1}{B}\sum_{b=1}^B \hat{\theta}^{(b)}.
    \]
 	
	\end{itemize}

\section{Principal Causal Effects under Partial Monotonicity}   \label{appendix-s2}  
In this section, we extend the results of Corollary \ref{coro3}  under partial monotonicity condition ($S^1 \geq S^0$ or $Y^1 \geq Y^0$, but not both).  

\subsection{Under the Monotonicity Condition $S^1 \geq S^0$}  \label{appendix-s2-1}

We first extend the results of Corollary \ref{coro3} under $S^1 \geq S^0$. Let 
$$\pi_{cd|ab} = \P( S^1 = c, Y^1  = d  \mid S^0 = a, Y^0 =b), \quad a,b,c,d=0,1,$$
which are invariant parameters across trials by Assumption \ref{assump6}. When the monotonicity condition $S^1\geq S^0$ holds, $\pi_{00|10} = \pi_{01|10}  =  \pi_{00|11} = \pi_{01|11}  = 0$, and it has 12 possible values of $(a, b, c, d)$, as shown in Table \ref{tab-s2}.

\begin{table}[h!]\centering
\caption{Parameters under Assumption \ref{assump6} and the monotonicity condition $S^1\geq S^0$}
\begin{tabular}{cccc | c|c}
\hline
$S^0=a$ & $Y^0=b$ & $S^1=c$  & $Y^1=d$  &  Restriction &  Number of Free Parameters                                  \\\hline
0 & 0 & 0  & 0  &    \multirow{4}{*}{$\sum_{c=0}^1 \sum_{d=0}^1 \pi_{cd|00}= 1$} & \multirow{4}{*}{3}          \\
0 & 0 & 0  & 1  &        \\
0 & 0 & 1  & 0  &        \\
0 & 0 & 1  & 1  &      \\ \hline 
0 & 1 & 0  & 0  &     \multirow{4}{*}{$\sum_{c=0}^1 \sum_{d=0}^1 \pi_{cd|01}= 1$} & \multirow{4}{*}{3}       \\
0 & 1 & 0  & 1  &        \\
0 & 1 & 1  & 0  &        \\
0 & 1 & 1  & 1  &        \\  \hline 
1 & 0 & 1  & 0  &     \multirow{2}{*}{$\sum_{d=0}^1 \pi_{1d|10}= 1$} & \multirow{2}{*}{1}       \\
1 & 0 & 1  & 1  &        \\ \hline 
1 & 1 & 1  & 0  &     \multirow{2}{*}{$\sum_{d=0}^1 \pi_{1d|11}= 1$} & \multirow{2}{*}{1}       \\
1 & 1 & 1  & 1  &        \\
\hline 
\end{tabular}
\label{tab-s2}
\end{table}

On the other hand, under Assumption \ref{assump6}, we have the following system of equations 
    \begin{equation} \label{eq-s1}  
 	\begin{cases} 
	\P(S^1 = 0, Y^1 = 0  \mid G=g) ={}& \pi_{00 | 00}  \cdot  \P(S^0 =0, Y^0 = 0 \mid  G=g) \\
	  {}& + \pi_{00 | 01}  \cdot  \P(S^0 =0, Y^0 = 1 \mid  G=g),    \quad \text{for }  g = 1, ..., m.  \\
    \P(S^1 = 0, Y^1 = 1  \mid G=g) ={}& \pi_{01 | 00}  \cdot  \P(S^0 =0, Y^0 = 0 \mid  G=g)  \\
    {}& + \pi_{01 | 01}  \cdot  \P(S^0 =0, Y^0 = 1 \mid  G=g),   \quad \text{for }  g = 1, ..., m.    \\
    \P(S^1 = 1, Y^1 = 0  \mid G=g) ={}& \pi_{10 | 00}  \cdot  \P(S^0 =0, Y^0 = 0 \mid  G=g)   + \pi_{10 | 01}  \cdot  \P(S^0 =0, Y^0 = 1 \mid  G=g)   \\
    {}& +  \pi_{10 | 10}  \cdot  \P(S^0 =1, Y^0 = 0 \mid  G=g) \\
    {}& + \pi_{10 | 11}  \cdot  \P(S^0 =1, Y^0 = 1 \mid  G=g), \quad \text{for }  g = 1, ..., m.    \\
        \P(S^1 = 1, Y^1 = 1  \mid G=g) ={}& \pi_{11 | 00}  \cdot  \P(S^0 =0, Y^0 = 0 \mid  G=g) + \pi_{11 | 01}  \cdot  \P(S^0 =0, Y^0 = 1 \mid  G=g) \\
        {}& +  \pi_{11 | 10}  \cdot  \P(S^0 =1, Y^0 = 0 \mid  G=g) \\
        {}& + \pi_{11 | 11}  \cdot  \P(S^0 =1, Y^0 = 1 \mid  G=g),      \quad \text{for }  g = 1, ..., m.  \\
	\end{cases} 
\end{equation}	  
where $\P(S^1 = c, Y^1 = d  \mid G=g)$ and $ \P(S^0 =a, Y^0 = b  \mid  G=g)$ are identifiable quantities under Assumption \ref{assump5}.

The above system of equations include a total of $m\times 3$ linearly independent equations due to $\sum_{c, d} \P(S^1 = c, Y^1 = d  \mid G=g) = 1$.  However, it contains 
 $8$ unknown free parameters as shown in the last column of Table \ref{tab-s2}. 
It seems that these invariant parameters $\pi_{cd|ab}$ for $a,b,c,d=0,1$ are identifiable when $m \geq 3$ under regular conditions. However, it is not this case due to the fact that the  first  two rows of the system of equations including only two unknown parameters.  
Formally, we have the following conclusion, as shown in Corollary \ref{coro-s2}. 

\begin{corollary} \label{coro-s2}
If $S^1 \geq S^0$, then under Assumptions \ref{assump5}, \ref{assump6}, and Condition \ref{cond-add}(i), 
$\pi_{00|10} = \pi_{01|10}  =  \pi_{00|11} = \pi_{01|11}  = 0$, and 
$\{ \pi_{00|00},  \pi_{00|01}, \pi_{01|00}, \pi_{01|01} \}$ are identifiable. 
\end{corollary}

In addition, observing that the third row of the system of equations has four free parameters. To identify them, it still requires Condition \ref{cond3},   
Thus, we cannot relax Condition \ref{cond3} for identifying the additional parameters, including 
$\{ \pi_{10|00},  \pi_{10|01}, \pi_{10|10}, \pi_{10|11}, \pi_{11|00},  \pi_{11|01}, \pi_{10110}, \pi_{11|11} \}$.  
As a result, we cannot relax the Condition \ref{cond3} for identifying the joint distribution $\P(S^0 = a, Y^0 =b, S^1 = c, Y^1  = d \mid G=g)$ under the additional monotonicity condition $S^1 \geq S^0$. 
Nevertheless, the additional monotonicity condition $S^1 \geq S^0$ allows us to slightly simplify the estimation method, see Section \ref{appendix-s2-3} for details. 



\subsection{Under the Monotonicity Condition $Y^1 \geq Y^0$}  \label{appendix-s2-2} 

We then extend the results of Corollary \ref{coro3} under the monotonicity condition $Y^1 \geq Y^0$.  
Under Assumption \ref{assump6},  
$$\pi_{cd|ab} = \P( S^1 = c, Y^1  = d  \mid S^0 = a, Y^0 =b), \quad a,b,c,d=0,1,$$
 are invariant across trials. 
 If $Y^1 \geq Y^0$ holds, then  
   $\pi_{00 | 01} = \pi_{10 | 01}  \equiv 0$ and $\pi_{ 00 |11 }  = \pi_{ 10 |11 } \equiv 0$.
     It has 12 possible values of $(a, b, c, d)$, as shown in Table \ref{tab-s3}.

   \begin{table}[h!]\centering
\caption{Parameters under Assumption \ref{assump6} and the monotonicity condition $Y^1\geq Y^0$}
\begin{tabular}{cccc | c|c}
\hline
$S^0=a$ & $Y^0=b$ & $S^1=c$  & $Y^1=d$  &  Restriction &  Number of Free Parameters                                  \\\hline
0 & 0 & 0  & 0  &    \multirow{4}{*}{$\sum_{c=0}^1 \sum_{d=0}^1 \pi_{cd|00}= 1$} & \multirow{4}{*}{3}          \\
0 & 0 & 0  & 1  &        \\
0 & 0 & 1  & 0  &        \\
0 & 0 & 1  & 1  &      \\ \hline 
0 & 1 & 0  & 1  &      \multirow{2}{*}{$\sum_{c=0}^1  \pi_{c1|01}= 1$} & \multirow{2}{*}{1}       \\
0 & 1 & 1  & 1  &        \\  \hline 
 1 & 0 & 0  & 0  &        \multirow{4}{*}{ $ \sum_{d=0}^1 \sum_{d=0}^1 \pi_{cd|10}= 1$} & \multirow{3}{*}{3}              \\
 1 & 0 & 0  & 1  &        \\
1 & 0 & 1  & 0  &       \\
1 & 0 & 1  & 1  &        \\ \hline 
 1 & 1 & 0  & 1  &      \multirow{2}{*}{$\sum_{d=0}^1 \pi_{c1|11}= 1$} & \multirow{2}{*}{1}        \\
1 & 1 & 1  & 1  &        \\
\hline 
\end{tabular}
\label{tab-s3}
\end{table}
On the other hand, we have the following system of equations 
    \begin{equation}    \label{eq-s2}
 	\begin{cases} 
	\P(S^1 = 0, Y^1 = 0  | G=g) ={}& \pi_{00 | 00}  \cdot  \P(S^0 =0, Y^0 = 0 |  G=g) + \pi_{00 | 10}  \cdot  \P(S^0 =1, Y^0 = 0 |  G=g),  \\
    \P(S^1 = 0, Y^1 = 1  | G=g) ={}& \pi_{01 | 00}  \cdot  \P(S^0 =0, Y^0 = 0 |  G=g) + \pi_{01 | 01}  \cdot  \P(S^0 =0, Y^0 = 1 |  G=g)    \\
         \qquad \qquad \qquad \qquad \qquad \quad  +{}&   \pi_{01 | 10}  \cdot  \P(S^0 =1, Y^0 = 0 |  G=g) + \pi_{01 | 11}  \cdot  \P(S^0 =1, Y^0 = 1 |  G=g)  \\
    \P(S^1 = 1, Y^1 = 0  | G=g) ={}& \pi_{10 | 00}  \cdot  \P(S^0 =0, Y^0 = 0 |  G=g) + \pi_{10 | 10}  \cdot  \P(S^0 =1, Y^0 = 0 |  G=g)   \\
        \P(S^1 = 1, Y^1 = 1  | G=g) ={}& \pi_{11 | 00}  \cdot  \P(S^0 =0, Y^0 = 0 |  G=g) + \pi_{11 | 01}  \cdot  \P(S^0 =0, Y^0 = 1 |  G=g) \\
    \qquad \qquad \qquad \qquad \qquad \quad  +{}& \pi_{11 | 10}  \cdot  \P(S^0 =1, Y^0 = 0 |  G=g) + \pi_{11 | 11}  \cdot  \P(S^0 =1, Y^0 = 1 |  G=g), \\
      \quad \text{for }  g = 1, ..., m. 
	\end{cases} 
\end{equation}	  
where $\P(S^1 = c, Y^1 = d  \mid G=g)$ and $ \P(S^0 =a, Y^0 = b  \mid  G=g)$ are identifiable quantities under Assumption \ref{assump5}.

Similar to the conclusion in Section \ref{appendix-s2-1}, we have the following conclusion, as shown in Corollary \ref{coro-suppl-add}.

\begin{corollary} \label{coro-suppl-add}
If $Y^1 \geq Y^0$, then under Assumptions \ref{assump5}, \ref{assump6}, and Condition \ref{cond-add}(ii), 
$\pi_{00 | 01} = \pi_{10 | 01}  = \pi_{ 00 |11 }  = \pi_{ 10 |11 } \equiv 0$, and  
$\{ \pi_{00|00},  \pi_{00|10}, \pi_{10|00}, \pi_{10|10} \}$ are identifiable. 
\end{corollary}

Under the additional monotonicity condition $Y^1 \geq Y^0$,  
Corollary \ref{coro3} also requires Condition \ref{cond3} for identifying the full joint distribution $\P(S^0 = a, Y^0 = b, S^1 = c, Y^1 = d \mid G = g)$. 
 Nevertheless, it can be used to simplify the estimation method for $\P(S^0 = a, Y^0 =b, S^1 = c, Y^1  = d \mid G=g)$ and $\text{PSACE}_{ab|g}$. See Section \ref{appendix-s2-3} for details.

\subsection{Estimation Method under Partial Monotonicity}  \label{appendix-s2-3}

We give a detailed description on the corresponding estimation method for $\P(S^0, S^1, Y^0, Y^1 \mid G=g)$ and $\textup{PSACE}_{ab|g}$  for $g \in \mathcal{G}$ based on the identification in Corollary  \ref{coro3}, together with additional partial monotonicity $Y^1 \ge Y^0$ or $S^1 \ge S^0$.      These additional monotonicity conditions simplify the identification following the analyses in Sections \ref{appendix-s2-1} and \ref{appendix-s2-2}, so the estimation method can be also simplified accordingly. 
 We mainly present the estimation method under $Y^1 \geq Y^0$, as the method under $S^1 \geq S^0$ is similar.

\bigskip 
{\bf 1. Under the monotonicity condition $Y^1 \geq Y^0$}.  

\smallskip  \noindent 
Denote  
$\pi_{cd|ab} = \P( S^1 = c, Y^1  = d  \mid S^0 = a, Y^0 =b), \quad a,b,c,d=0,1,$
 as invariant parameters across trials. 
We set $\pi_{00 | 01} = \pi_{10 | 01}  \equiv 0$ and $\pi_{ 00 |11 }  = \pi_{ 10 |11 } \equiv 0$ due to $Y^1 \geq Y^0$. Then, the estimation procedures are given as follows:  
   \begin{itemize}
   \item {\bf Step 1.} estimate $\P(S^1 =c, Y^1 = d \mid G=g)$ and $\P(S^0 =a, Y^0 = b \mid G=g)$ for  $a, b, c, d = 0, 1$. We denote the estimators as $\hat \P(S^1 =c, Y^1 = d \mid G=g)$ and $\hat \P(S^0 =a, Y^0 = b \mid G=g)$. 
   
   \item {\bf Step 2.}  estimate the other invariant parameters and the joint distribution $\P(S^0=a, Y^0=b, S^1=c, Y^1=d \mid G=g)$.  
         \begin{itemize}
         	\item  from equation \eqref{eq-s2},  we estimate $(\pi_{00|00}, \pi_{00|10})$ by linear regression of $\hat \P(S^1 =0, Y^1 = 0 \mid G=g) $ on $( \hat \P(S^0 = 0, Y^0 = 0 | G=g),  \hat \P(S^0 = 1, Y^0 = 0 | G=g) )$; 
		\item    from equation \eqref{eq-s2},  we  estimate $(\pi_{01 |00}, \pi_{01 | 01}, \pi_{01 | 10}, \pi_{01 | 11} )$ by  linear regression of $\hat \P(S^1 =0, Y^1 = 1 \mid G=g) $ on $( \hat \P(S^0 = 0, Y^0 = 0 | G=g), \hat \P(S^0 = 0, Y^0 = 1 | G=g),  \hat \P(S^1 = 0, Y^0 = 0 | G=g), \hat \P(S^0 = 1, Y^0 = 1 | G=g) )$; 
		\item   from equation \eqref{eq-s2}, we estimate $(\pi_{10 | 00}, \pi_{10 | 10} )$ by linear regression of $\hat \P(S^1 =1, Y^1 = 0 \mid G=g) $ on $( \hat \P(S^0 = 0, Y^0 = 0 | G=g), \hat \P(S^0 = 1, Y^0 = 0 | G=g))$;
		
		\item estimate $\pi_{11|ab}$ with $\pi_{11|ab} = 1 - \pi_{00|ab} -\pi_{01 |ab} - \pi_{10 | ab}$ for $a, b= 0 ,1$.  
         \end{itemize}
Based on $\hat \pi_{cd|ab}$, the estimator of 
  $\P(S^0=a, Y^0=b, S^1=c, Y^1=d \mid G=g)$ is given as 
   \begin{align*} 
       \hat \P(S^0=a, Y^0=b, S^1=c, Y^1=d \mid G=g) = \pi_{cd|ab} \cdot \hat \P(  S^0=a, Y^0=b \mid G=g)  
     \end{align*}
     
          \item {\bf Step 4.}  estimate  $\text{PSACE}_{ab|g} =\P( Y^1 = 1 \mid S^0 = a, S^1 = b, G=g) - \P( Y^0 = 1 \mid S^0 = a, S^1 = b, G=g)$.   
  The estimator for  $ \P( Y^1 = 1 \mid S^0 = a, S^1 = b, G=g) $ is given as 
        \begin{align*}
        \hat \P( Y^1 = 1 \mid & S^0 = a, S^1 = b, G=g) ={} \frac{ \hat \P(Y^1 = 1, S^0 = a, S^1 = b \mid G=g)  }{\hat \P( S^0 = a, S^1 = b \mid G=g) }  \\
                                      ={}&  \frac{ \sum_{d=0,1} \hat \P(Y^1 = 1, Y^0 = d, S^0 = a, S^1 = b \mid G=g)  }{ \sum_{c,d=0,1} \hat \P( S^0 = a, S^1 = b, Y^0 = c, Y^1 = d \mid G=g) }             
        \end{align*}    
Likewise, we can estimate $\P( Y^0 = 1 \mid S^0 = a, S^1 = b, G=g)$ with 
                     \begin{align*}
        \hat \P( Y^0 = 1 \mid & S^0 = a, S^1 = b, G=g) ={} \frac{ \hat \P(Y^0 = 1, S^0 = a, S^1 = b \mid G=g)  }{\hat \P( S^0 = a, S^1 = b \mid G=g) }  \\
                                      ={}&  \frac{ \sum_{d=0,1} \hat \P(Y^0 = 1, Y^1 = d, S^0 = a, S^1 = b \mid G=g)  }{ \sum_{c,d=0,1} \hat \P( S^0 = a, S^1 = b, Y^0 = c, Y^1 = d \mid G=g) }             
        \end{align*}   

   \end{itemize}


\bigskip 
{\bf Further Consideration}. In the estimation procedures above, we do not impose the restriction that $\pi_{cd|ab}$ must belong to the interval [0, 1]. To address this issue, a straightforward approach is to use the restricted least-squares method in Step 2 instead of directly applying the standard least-squares method. However, this approach may not guarantee that $\sum_{c,d=0,1} \hat \pi_{cd|ab} = 1$.

To avoid this problem, we can estimate the invariant parameters $\pi_{cd|ab}$ by solving a least squares programming under equality and inequality constraints. Specifically, we denote $\theta = (\pi_{00|00}, \pi_{00|01}, \pi_{00|10}, \pi_{00|11}, \pi_{01|00}, \pi_{01|01}, \pi_{01|10}, \pi_{01|11},
  \pi_{10|00}, \pi_{10|01}, \pi_{10|10}, \pi_{10|11},
  \pi_{11|00}, \pi_{11|01}, \pi_{11|10}, \pi_{11|11})^\intercal$ as the vector of all invariant parameters, let 
\[ \begin{cases}
    \tilde Y_{00} ={}& (\P(S^1=0, Y^1=0\mid G=1), \P(S^1=0, Y^1=0\mid G=2), \cdots, \P(S^1=0, Y^1=0\mid G=m))^\intercal \\
    \tilde Y_{01} ={}& (\P(S^1=0, Y^1=1\mid G=1), \P(S^1=0, Y^1=1\mid G=2), \cdots, \P(S^1=0, Y^1=1\mid G=m))^\intercal \\
    \tilde Y_{10} ={}& (\P(S^1=1, Y^1=0\mid G=1), \P(S^1=1, Y^1=0\mid G=2), \cdots, \P(S^1=1, Y^1=0\mid G=m))^\intercal \\
    \tilde Y_{11} ={}& (\P(S^1=1, Y^1=1\mid G=1), \P(S^1=1, Y^1=0\mid G=2), \cdots, \P(S^1=1, Y^1=1\mid G=m))^\intercal
\end{cases}
\]
 
{\small\begin{align*}
   & \tilde X =   \\
 & \begin{pmatrix}
    \P(S^0 = 0, Y^0 =0\mid G=1) &  \P(S^0 = 0, Y^0 =1\mid G=1) & \P(S^0 = 1, Y^0 =0\mid G=1) & \P(S^0 = 1, Y^0 =1\mid G=1) \\
    \P(S^0 = 0, Y^0 =0\mid G=2) &  \P(S^0 = 0, Y^0 =1\mid G=2) & \P(S^0 = 1, Y^0 =0\mid G=2) & \P(S^0 = 1, Y^0 =1\mid G=2) \\
    \vdots \\
    \P(S^0 = 0, Y^0 =0\mid G=m) &  \P(S^0 = 0, Y^0 =1\mid G=m) & \P(S^0 = 1, Y^0 =0\mid G=m) & \P(S^0 = 1, Y^0 =1\mid G=m) 
\end{pmatrix}, 
\end{align*} } 

\[
{\bf \tilde X} = \begin{pmatrix}
    \tilde X & \bm{0} & \bm{0}  & \bm{0}   \\
    \bm{0}    & \tilde X & \bm{0} & \bm{0}  \\
    \bm{0}    & \bm{0}   & \tilde X & \bm{0}  \\
    \bm{0}   & \bm{0}   & \bm{0}  & \tilde X
\end{pmatrix}_{4m \times 16}, \quad  {\bf \tilde Y} = \begin{pmatrix}
    \tilde Y_{00} \\
    \tilde Y_{01} \\
    \tilde Y_{10} \\
    \tilde Y_{11}
\end{pmatrix}. 
\] 
Under the condition $Y^1 \geq Y^0$, solving $\theta$ can be formulated as follows: 
 \begin{equation} \label{eq-temp}
 \begin{cases}
     \text{minimize}_{\theta} ~~ || {\bf \tilde Y} - {\bf \tilde X} \theta  ||_2^2 \\
     \text{subject to } {\bf C} \theta = {\bf d}, \quad {\bf E} \theta \geq {\bf f},
 \end{cases}
 \end{equation}
where 
\[
{\bf C} =  
\left(
\begin{array}{*{16}{c}}
 0 & 1 & 0 & 0 & 0 & 0 & 0 & 0 & 0 & 0 & 0 & 0 &0 & 0 & 0  &   0  \\ 
  0   & 0 &    0  &  1  &  0  &  0  &   0   & 0  &  0   &  0  &   0  &   0   &  0  &   0  &   0  &   0  \\
 0   & 0 &    0  &  0  &  0  &  0  &   0   & 0  &  0   &  1  &   0  &   0   &  0  &   0  &   0  &   0  \\
    0   & 0 &    0  &  0  &  0  &  0  &   0   & 0  &  0   &  0  &   0  &   1   &  0  &   0  &   0  &   0  \\
    1   & 0 &    0  &  0  &  1  &  0  &   0   & 0  &  1   &  0  &   0  &   0   &  1 &   0  &   0  &   0  \\
   0 &  1   & 0 &    0  &  0  &  1  &  0  &   0   & 0  &  1   &  0  &   0  &   0   &  1 &   0  &   0    \\
  0 &  0 &  1   & 0 &    0  &  0  &  1  &  0  &   0   & 0  &  1   &  0  &   0  &   0   &  1 &   0     \\
  0  &   0 &  0 &  1   & 0 &    0  &  0  &  1  &  0  &   0   & 0  &  1   &  0  &   0  &   0   &  1      
\end{array}
\right)
 \] 

\[ {\bf d} =  \begin{pmatrix}
    0 \\
    0 \\
    0 \\
    0 \\
        1 \\
    1 \\
    1 \\
    1 
\end{pmatrix},
\]
\[
{\bf E} = \begin{pmatrix}
    {\bf I}_{16\times 16}  \\
    - {\bf I}_{16\times 16}
\end{pmatrix}, \quad {\bf f} =  \begin{pmatrix}
    {\bf 0}_{16\times 1} \\
     -{\bf 1}_{16\times 1}
\end{pmatrix},
\]
where $  {\bf I}_{16\times 16} $ is is the identity matrix of dimension $16 \times 16$, ${\bf 0}_{16\times 1}$ is a $16$-dimensional vector with each element being 0, and  ${\bf I}_{16\times 1}$ is a $16$-dimensional vector with each element being -1.  

In equation \eqref{eq-temp}, ${\bf C} \theta = {\bf d}$ corresponds to 8 equality constraints:  $\pi_{00 | 01} = 0, \pi_{ 00 |11 } = 0, \pi_{10 | 01} = 0, \pi_{ 10 |11 }  = 0$, $\sum_{c,d=0,1} \hat \pi_{cd|00} = 1$, $\sum_{c,d=0,1} \hat \pi_{cd|01} = 1$, $\sum_{c,d=0,1} \hat \pi_{cd|10} = 1$, $\sum_{c,d=0,1} \hat \pi_{cd|11} = 1$. In addition, ${\bf E} \theta \geq {\bf f}$ corresponds to 32 inequality constraints: $\pi_{cd|ab} \geq 0$, $\pi_{cd|ab} \leq 1$ for $a, b, c, d = 0, 1$.

We can solve the optimization problem \eqref{eq-temp} using the 
 \texttt{lsei} function in \texttt{R} package \texttt{lsei}~\citep{Wang-etal2020}.

\bigskip 
{\bf 2. Under the monotonicity condition $S^1 \geq S^0$}.  
The estimation method under $S^1 \geq S^0$ is similar to that under $Y^1 \geq Y^0$, and is therefore omitted to avoid redundancy.


\section{Further Discussion under a Weaker Version of Assumption \ref{assump6}}  \label{appendix-s4}

If we relax Assumption \ref{assump6} to Assumption \ref{a1} below, we can obtain the identifiability for $\P(S^0, S^1, Y^0 \mid G=g)$ and  $\P(S^0, Y^0, Y^1 \mid G=g)$, as shown in Corollary \ref{coro2}.

\begin{assumption}  \label{a1}
   (i) $G \indep S^1 \mid S^0, Y^0$; (ii) $G \indep  Y^1 \mid S^0, Y^0$. 
\end{assumption}

\medskip 
\begin{corollary} \label{coro2} Under Assumption \ref{assump5}, Assumption \ref{a1}, and Condition \ref{cond3}, 
 the joint distributions $\P(S^0, S^1, Y^0 \mid G=g)$ and $\P(S^0, Y^0, Y^1 \mid G=g)$ for $g \in \mathcal{G}$ are identifiable. 
\end{corollary}


It is noteworthy that $\P(S^0, S^1, Y^1 \mid G=g)$ is not identifiable under the conditions outlined in Corollary \ref{coro2}, and therefore, we cannot identify $\text{PSACE}_{ab|g}$.    
Also, similar to Theorem \ref{thm-add},  if we impose the monotonicity assumption ($Y^1 \geq Y^0$ and $S^1 \geq S^0$), we can relax Condition \ref{cond3}. 

\medskip 
\begin{corollary}  \label{coro-s3} If $Y^1 \geq Y^0$ and $S^1 \geq S^0$, then under Assumptions \ref{assump5},  \ref{a1}, and Condition \ref{cond-add}, 
the joint distributions $\P(S^0, S^1, Y^0 \mid G=g)$ and $\P(S^0, Y^0, Y^1 \mid G=g)$ for $g \in \mathcal{G}$ are identifiable. 

\end{corollary}

\medskip 

 \begin{proof}[Proof of Corollary \ref{coro2}]  
Let
  \[    \pi_{c | ab} := \P(S^1 = c \mid  S^0 = a, Y^0 =b) = \P(S^1 = c \mid  S^0 = a, Y^0 =b, G=g), \quad a,b,c=0,1, \]
which are invariant parameters across trials.
    Under Assumption \ref{a1}(i), we have that 
    \begin{equation*}  
	\P(S^1 = 1 | G=g) = \sum_{a, b}  \pi_{1 | ab}  \cdot  \P(S^0 =a, Y^0 = b |  G=g),  \quad \text{for } g = 1, ..., m,
\end{equation*}	  
where $\P(S^1 = 1  | G=g)$ and $\P(S^0 =a, Y^0 = b |  G=g)$ are identifiable quantities under Assumption \ref{assump5}.
The above system of equations contain a total of $m$ linearly independent equations, 
and $4$ unknown free parameters (i.e., $\{ (\pi_{1|ab}, a, b = 0, 1\}$). Then under Assumption \ref{assump5}, \ref{a1}(i), and Condition \ref{cond3}, $ \pi_{1 | ab} $ for $a=0,1; b = 0, 1$ are  identifiable by solving the system of equations.   
This also implies the identifiability of $\P(S^0, S^1, Y^0 | G=g)$ by noting that 
         \[   \P(S^0 =a, S^1=1, Y^0 = b | G=g) =   \pi_{ 1 | ab} \cdot \P(S^0 =a, Y^0 = b|  G=g).     \]
similarly, we can show the identifiability of $\P(S^0, Y^0, Y^1\mid G=g)$. 

\end{proof}

 \begin{proof}[Proof of Corollary \ref{coro-s3}]  
Under Assumption \ref{a1}(i), $ \pi_{c | ab} := \P(S^1 = c \mid  S^0 = a, Y^0 =b), \quad a,b,c = 0, 1$ 
are invariant parameters across trials. Due to $S^1 \geq S^0$, we have
        $\pi_{0 | 10} = \pi_{0 | 11} = 0$, which implies  $\pi_{1 | 10} = \pi_{1 | 11} = 1$. 
By Assumption \ref{a1}(i), 
    \begin{align*}  
	& \P(S^1 = 1 \mid G=g) = \sum_{a, b}  \pi_{1 | ab}  \cdot  \P(S^0 =a, Y^0 = b \mid  G=g),  \\
	 ={}&   \pi_{1 | 00}  \P(S^0 =0, Y^0 = 0 \mid  G=g) + \pi_{1 | 01}  \P(S^0 =0, Y^0 = 1 \mid G=g)  \\
	 {}& +    \P(S^0 =1, Y^0 = 0 \mid  G=g) +  \P(S^0 =1, Y^0 = 1 \mid  G=g)
	 \quad \text{for } g = 1, ..., m,
\end{align*}	  
where $\P(S^1 = 1  | G=g)$ and $\P(S^0 =a, Y^0 = b |  G=g)$ are identifiable under Assumption \ref{assump5}.  
The above system of equations contains a total of $m$ linearly independent equations, 
and $2$ unknown free parameters (i.e., $\{\pi_{1|00}, \pi_{1|01} \}$). Thus, when Condition \ref{cond-add}(i)  holds,  $ \pi_{1 | ab} $ for $a=0,1; b = 0, 1$ are  identifiable by solving the system of equations.  
This also implies the identifiability of $\P(S^0, S^1, Y^0 \mid G=g)$ by noting that 
         $\P(S^0 =a, S^1=1, Y^0 = b \mid G=g) =   \pi_{ 1 | ab} \cdot \P(S^0 =a, Y^0 = b \mid  G=g).$ 

Likewise,  under 
      Assumption \ref{a1}(ii), 
             \[   \bar \pi_{d | ab} := \P(Y^1 = d \mid  S^0 = a, Y^0 =b), \quad a,b, d = 0, 1, \]
are invariant parameters across trials. Due to $Y^1 \geq Y^0$, we have
        $\bar \pi_{0 | 01} = \bar \pi_{0 | 11} = 0$, which implies  $\bar \pi_{1 | 01} = \bar \pi_{1 | 11} = 1$. 
By Assumption \ref{a1}(ii),  
    \begin{align*}  
	& \P(Y^1 = 1 \mid G=g) = \sum_{a, b}  \bar \pi_{1 | ab}  \cdot  \P(S^0 =a, Y^0 = b \mid  G=g),  \\
	 ={}&  \bar \pi_{1 | 00}  \P(S^0 =0, Y^0 = 0 \mid  G=g) +   \P(S^0 =0, Y^0 = 1 \mid G=g)  \\
	 {}& +   \bar \pi_{1 | 10}    \P(S^0 =1, Y^0 = 0 \mid  G=g) +  \P(S^0 =1, Y^0 = 1 \mid  G=g)
	 \quad \text{for } g = 1, ..., m. 
\end{align*}	   
The above system of equations contains a total of $m$ linearly independent equations, 
and $2$ unknown free parameters (i.e., $\{\bar \pi_{1|00}, \bar \pi_{1|10} \}$). Thus, when Condition \ref{cond-add}(ii)  holds,  $\bar \pi_{1 | ab} $ for $a=0,1; b = 0, 1$ are  identifiable by solving the system of equations.  
This also implies the identifiability of $\P(S^0, Y^0, Y^1 \mid G=g)$ by noting that 
         $\P(S^0 =a, Y^0 = b, Y^1=1  \mid G=g) =   \bar \pi_{ 1 | ab} \cdot \P(S^0 =a, Y^0 = b \mid  G=g).$

\end{proof}

\section{\bcol{Monotonicity-Based Route of \citet{Jiang-etal2016}}}  \label{app-s7}

In this section, we formally restate the monotonicity-based approach of \citet{Jiang-etal2016} and provide an alternative proof of their identification theorem, which in turn motivates a least-squares estimator.

\subsection{\bcol{Restatement of the Route of \citet{Jiang-etal2016}}}
\citet{Jiang-etal2016} adopt Assumption \ref{assump5}, together with the key Assumptions \ref{assump7}--\ref{assump8} and Condition \ref{cond4}, presented below.


\begin{assumption}  \label{assump7}
   $G \indep Y^a \mid S^0, S^1$ for $a = 0, 1$. 
\end{assumption}   

\begin{assumption} \label{assump8}  
 $S^1 \geq S^0$. 
\end{assumption}

\begin{condition}  \label{cond4}  
(i) The matrix $ (  \P(S^0 = 0, S^1 = 1 \mid  G=g), \P(S^0 = 1, S^1 = 1 \mid  G=g) )_{m\times 2}$
is full-column rank; 
(ii)  The matrix $( \P(S^0 = 0, S^1 = 1 \mid  G=g), \P(S^0 = 0, S^1 = 0 \mid  G=g))_{m\times 2}$ 
is full-column rank.  
\end{condition}

Assumption \ref{assump7} means that there is no dependence between $G$ and $Y^0$ or $Y^1$ within the principal stratum defined by $(S^0, S^1)$.  This implies that $\text{PSACE}_{ab|g}$ for $g \in \cG$ are invariant across trials. In contrast, the assumptions in Section \ref{sec6-2} allow different principal effects across trials. 

The monotonicity condition in Assumption \ref{assump8}, together with the unconfoundedness Assumption \ref{assump5}, imply the identifiability of principal scores $\delta_{ab|g} = \P(S^0 = a, S^1 = b \mid  G=g)$ for $a , b \in \{0, 1\}$. Specifically, Assumption \ref{assump8} implies $\delta_{10 \mid g} = 0$,  thus the joint distribution $\P(S^0, S^1 \mid  G=g)$ involves only three free parameters ($\delta_{00 \mid g}, \delta_{01 \mid g }, \delta_{11 \mid g}$). These parameters satisfy three equations 
	$ 
	   \delta_{10 \mid g} + \delta_{11 \mid g}  = \P(S^0 = 1 \mid  G=g), ~  \delta_{01 \mid g} + \delta_{11 \mid g}  = \P(S^1 = 1 \mid  G=g),~
	\sum_{a=0}^1 \sum_{b=0}^1  \delta_{ab \mid g} = 1.
	 $
Solving these equations easily gives the identification of the principal scores.

Condition \ref{cond4}  is also a full rank condition, which, like Condition \ref{cond-add}, also requires $m \ge 2$. Under Assumption \ref{assump5} and the three above assumptions, \citet{Jiang-etal2016} show the identifiability of $ \text{PSACE}_{ab|g}$ in their Theorem 1.  We reproduce their results in Theorem \ref{thm3} below and provide an alternative proof (see Supplementary Material \ref{app-s1.6}). This alternative proof will motivate our new estimator in Section \ref{sec6-4}. 


\begin{theorem} \label{thm3}  Under Assumptions \ref{assump5}, \ref{assump7}, and \ref{assump8}, for  $g = 1, ..., m$, we have that 

(a) $\P(Y^1 \mid S^0, S^1, G=g)$ is identifiable if Condition \ref{cond4}(i) holds. 

(b) $\P(Y^0 \mid S^0, S^1,  G=g)$ is identifiable if Condition \ref{cond4}(ii) holds. 
   
 (c) $\text{PSACE}_{ab|g}$ for $a,b \in \{0,1\}$ are identifiable if  Condition \ref{cond4} holds. 
\end{theorem}

In Theorem \ref{thm3}, the monotonicity condition $S^1 \ge S^0$ in Assumption \ref{assump8} is used to identify the principal scores as discussed above. One may wonder whether this condition can be replaced by an alternative  condition,  $G \indep S^1 \mid S^0$, that is also relevant for identifying 
the principal scores\footnote{According to our theory in Section \ref{sec3}, the condition $G \indep S^1 \mid S_0$ is relevant for the identification of the joint distribution of $(S^1, S^0)$ and thus the principal scores.}. Section  \ref{appendix-s5} gives a negative answer to this question, showing the importance of the monotonicity condition.  Moreover, 
Theorem \ref{thm-add} also considers another monotonicity condition $Y^1 \ge Y^0$. This alternative monotonicity condition is not relevant here because the distributions $\P(Y^1 \mid S^0, S^1, G=g)$ and $\P(Y^0 \mid S^0, S^1,  G=g)$ and the principal effects $\text{PSACE}_{ab|g}$ do not involve the joint distribution of $Y^1$ and $Y^0$.


Without the monotonicity condition $S^1 \ge S^0$, Proposition 2 of \citet{Jiang-etal2016} further shows that a necessary condition for local identifiability of the principal effects is $m \ge 3$. In contrast, our previous Corollary \ref{coro3}, which also considers identification without any monotonicity condition, establishes sufficient conditions for global identifiability when $m \ge 4$. 
In this sense, our new identification results provide some alternative conditions that lead to a stronger identification, which complements the findings of \citet{Jiang-etal2016}.

\subsection{\bcol{Least-squares Estimation Based on Theorem \ref{thm3}}}  \label{sec6-4}

Under the identification in Theorem \ref{thm3}, \citet{Jiang-etal2016} propose a Bayesian estimation method. In this part, we follow our new proof for  Theorem \ref{thm3} and extend the least-square estimation method in Section \ref{sec4} to the setting of Theorem \ref{thm3}. 

We first define the following parameters for $a,b \in \{0, 1\}$:
      \[        \pi_{1|ab} = \P(Y^1 = 1 \mid S^0 =a , S^1 = b),~    \tilde \pi_{1|ab} = \P(Y^0 = 1 \mid S^0 =a , S^1 = b).      \]
These are key invariant parameters across trials.  
Clearly, under the monotonicity condition $S^1 \geq S^0$, $\pi_{1|10}$ and $\tilde \pi_{1|10}$ are undefined and do not need estimation.  
Now we describe how to estimate $\beta :=  (\pi_{1|00}, \pi_{1|01}, \pi_{1|11})$ and $\gamma := (\tilde \pi_{1|00}, \tilde  \pi_{1|01}, \tilde  \pi_{1|11})$.

{\bf Step 1:} estimate the principal scores $\delta_{ab|g}:= \P(S^0=a, S^1 = b \mid G=g)$, and the  probabilities $\P(Y^1=1 \mid G=g)$, and $\P(Y^0 = 1 \mid G=g)$. Specifically, for any given $g$, $\delta_{10 |g} = 0$ by Assumption \ref{assump8}, and 
    \[
\begin{cases}
      \delta_{11|g} ={}& \P(S^0=1, S^1 = 1 | G=g) = \P(S^0 = 1| G=g)  = \P(S =1 | A=0, G=g), \\
      \delta_{01|g} ={}&  \P(S^0=0, S^1 = 1 | G=g) = \P(S^1=1|G=g) - \delta_{11|g} \\
                       \quad  \quad   ={}&  \P(S=1| A=1, G=g) - \delta_{11|g}, \\
       \delta_{00|g}  ={}& 1-  \delta_{11|g} - \delta_{01|g}.  
\end{cases}
    \]
In addition, $\P(Y^1=1 \mid G=g) = \P(Y=1 \mid G=g, A=1)$ and $\P(Y^0 = 1 \mid G=g)= \P(Y=1 \mid  G=g, A=0)$. We can construct their estimators $\hat \delta_{11|g}, \hat \delta_{01|g}, \hat \delta_{00|g}$, $\hat \P(Y^1=1 \mid G=g)$, and $\hat \P(Y^0 = 1 \mid G=g)$ via replacing the population probabilities by the empirical frequencies and solving the resulting linear equations.

{\bf Step 2:} estimate $\pi_{1| 0 0} $ and $\tilde \pi_{1| 11}$.  
Clearly, due to $S^1 \geq S^0$, we have 
 $\pi_{1| 0 0} =  \P(Y^1 = 1 \mid S^0 =0 , S^1 = 0) =  \P(Y^1 = 1 \mid S^1 = 0)  =  \P(Y = 1 \mid S = 0, A  =1 ),$
and 
$\tilde \pi_{1| 11} =  \P(Y^0 = 1  \mid S^0 =1 , S^1 = 1) =  \P(Y^0 = 1 \mid S^0 = 1)  =  \P(Y = 1 \mid S = 1, A  =0 ).$
These two can again be estimated by the empirical frequencies. We denote the estimators by $\hat \pi_{1| 0 0} $ and $\hat{\tilde \pi}_{1| 11}$ respectively.

{\bf Step 3:} estimate $(\pi_{1|01}, \pi_{1|11})$ and $(\tilde \pi_{1|00}, \tilde \pi_{1|01})$.  
Note that 
    \begin{align*} 
       \P(Y^1 =1 | G = g) - \pi_{1| 0 0}  \delta_{00|g}
      ={}&    \pi_{1| 0 1}  \delta_{01|g} + \pi_{1| 1 1}  \delta_{11|g},  \quad g = 1, ..., m,      \\
             \P(Y^0 =1 | G = g) - \tilde  \pi_{1| 1 1}  \delta_{11|g}
      ={}&     \tilde \pi_{1| 0 0}  \delta_{00|g} + \tilde \pi_{1| 0 1}  \delta_{01|g},  \quad g = 1, ..., m.     
      \end{align*}  
Then we can estimate $(\pi_{1|01}, \pi_{1|11})$ by running linear regression of $ \hat \P(Y^1 =1  \mid G = g) -  \hat \pi_{1| 0 0}  \hat \delta_{00|g} $ on $(\hat \delta_{01|g},  \hat \delta_{11|g} )$, and estimate  $(\tilde \pi_{1|00}, \tilde \pi_{1|01})$ by running linear regression of $ \hat \P(Y^0 =1  \mid G = g) - \hat{\tilde  \pi}_{1| 1 1} \hat \delta_{11|g}$ on $(\hat \delta_{00|g},  \hat \delta_{01|g} )$.  
Let $(\hat \pi_{1|01}, \hat \pi_{1|11})$ and $(\hat{\tilde \pi}_{1|00},  \hat{\tilde \pi}_{1|01})$ be the corresponding estimators, and denote  $\hat \beta =  (\hat \pi_{1|00}, \hat \pi_{1|01}, \hat \pi_{1|11})$ and $\hat \gamma = (\hat{\tilde \pi}_{1|00}, \hat{\tilde  \pi}_{1|01}, \hat{\tilde  \pi}_{1|11})$. 

{\bf Step 4:} {estimate principal effects $\text{PSACE}_{ab|g}$. The estimators are given by $\hat \pi_{1|ab} - \hat {\tilde  \pi}_{1|ab}$ for $ab = 00, 01$, and 11. Due to the monotonicity condition $S^1 \ge S^0$, $\text{PSACE}_{ab|g}$ for  $ab = 10$ is undefined and needs no  estimation.}

Similar to Theorem \ref{thm2}, we can show the large sample  properties of the proposed estimators  
$\hat \beta :=  (\hat \pi_{1|00}, \hat \pi_{1|01}, \hat \pi_{1|11})$ and $\hat \gamma := ( \hat {\tilde \pi}_{1|00},  \hat {\tilde \pi}_{1|01}, \hat {\tilde \pi}_{1|11})$, respectively.
The corresponding results are presented in Theorem \ref{thm-s2}. 
 For notational convenience, we slightly abuse notation for $X_g$, $Y_g$, and $C$ below.

\begin{theorem}  \label{thm-s2}
It follows that 

(a) $\sqrt{n}(\hat \beta - \beta) \xrightarrow{d} N(0, \sigma_{\beta}^2),$ 
where $\sigma_{\beta}^2$ is the covariance matrix of
   \[ \begin{pmatrix}
     \dfrac{\mathbb{I}(Y_i = 1, S_i = 0, A_i = 1) - 
        \P(Y=1, S=0, A=1)}{\P(S=0, A=1)} \\
         C^{-1}  
 \left\{  \dfrac{1}{m}  \sum_{g=1}^m  \Big( \phi_g(X_i, A_i, S_i, Y_i) Y_g +  
 \varphi_g(X_i, A_i, S_i, Y_i) X_g \Big) \right\} 
   \end{pmatrix}, 
   \]
with $Y_g =  \P(Y =1 \mid G = g, A=1) - \pi_{1| 0 0}  \delta_{00|g}$, $X_g =  (\delta_{01|g}, \delta_{11|g})^\intercal$, $C = m^{-1} \sum_{g=1}^m  X_{g} X_{g}^\intercal$, 
 \begin{align*}
     \varphi_g(X_i, A_i, S_i, Y_i) ={}&  \frac{ \mathbb{I}(Y_i =1, A_i = 1, G_i = g) - \P(Y=1,  A=1, G=g) }{ \P(A=1, G=g) } \\
	-{}&   \pi_{1|00} \frac{  \mathbb{I}(S_i = 0, A_i = 1, G_i = g)-    \P(S=0, A=1, G=g)}{\P(A=1, G=g)} \\
-{}&   \delta_{00|g} \frac{ \mathbb{I}(Y_i = 1, S_i = 0, A_i = 1) - 
        \P(Y=1, S=0, A=1)}{\P(S=0, A=1)},
 \end{align*}
 and 
 \begin{align*}
    & \phi_g(X_i, A_i, S_i, Y_i) = \\
    & \begin{pmatrix}    
    \dfrac{\mathbb{I}(S_i = 1, A_i = 1, G_i = g) - 
        \P(S=1, A=1, G=g)}{\P(A=1, G=g)}  - \dfrac{\mathbb{I}(S_i = 1, A_i = 0, G_i = g) - 
        \P(S=1, A=0, G=g)}{\P(A=0, G=g)}   \\
        \dfrac{\mathbb{I}(S_i = 1, A_i = 0, G_i = g) - 
        \P(S=1, A=0, G=g)}{\P(A=0, G=g)} 
        \end{pmatrix}. 
 \end{align*}

(b) $\sqrt{n}(\hat \gamma - \gamma) \xrightarrow{d} N(0, \sigma_{\gamma}^2),$ 
where $\sigma_{\gamma}^2$ is the covariance matrix of
   \[ \begin{pmatrix}
         C^{-1}  
 \left\{  \dfrac{1}{m}  \sum_{g=1}^m  \Big( \tilde \phi_g(X_i, A_i, S_i, Y_i) Y_g +  
 \tilde \varphi_g(X_i, A_i, S_i, Y_i) X_g \Big) \right\} \\
        \dfrac{\mathbb{I}(Y_i = 1, S_i = 1, A_i = 0) - 
        \P(Y=1, S=1, A=0)}{\P(S=1, A=0)}
   \end{pmatrix}, 
   \]   
with $Y_g =  \P(Y=1 \mid A=0, G = g) - \tilde  \pi_{1| 1 1}  \delta_{11|g}$, $X_g = (\delta_{00|g}, \delta_{01|g})^\intercal$, $C = m^{-1} \sum_{g=1}^m  X_{g} X_{g}^\intercal$, 
 \begin{align*}
     \tilde \varphi_g(X_i, A_i, S_i, Y_i) ={}&   \frac{ \mathbb{I}(Y_i =1, A_i = 0, G_i = g) - \P(Y=1, A=0,  G=g) }{ \P(A=0, G=g) } \\
	-{}&  \tilde \pi_{1|11}\frac{ \mathbb{I}(S_i = 1, A_i = 0, G_i = g) - 
        \P(S=1, A=0, G=g)}{\P(A=0, G=g)} \\
-{}&   \delta_{11|g} \frac{ \mathbb{I}(Y_i = 1, S_i = 1, A_i = 0) - 
        \P(Y=1, S=1, A=0)}{\P(S=1, A=0)}
 \end{align*}
 and 
 \begin{align*}
    & \tilde \phi_g(X_i, A_i, S_i, Y_i) = \\
    & \begin{pmatrix}    
        \dfrac{ \mathbb{I}(S_i = 0, A_i = 1, G_i = g) - 
        \P(S=0, A=1, G=g)}{\P(A=1, G=g)} \\
    \dfrac{\mathbb{I}(S_i = 1, A_i = 1, G_i = g) - 
        \P(S=1, A=1, G=g)}{\P(A=1, G=g)}  - \dfrac{\mathbb{I}(S_i = 1, A_i = 0, G_i = g) - 
        \P(S=1, A=0, G=g)}{\P(A=0, G=g)} 
        \end{pmatrix}. 
 \end{align*}

\end{theorem}

The proof of Theorem \ref{thm-s2} is provided in Section \ref{app-s1.7}

\subsection{Further Discussion on Theorem \ref{thm3} without Monotonicity}  \label{appendix-s5}

In this section, we examine the identifiability of  $\P(Y^1, S^0, S^1| G=g)$ and  $\P(Y^0, S^0, S^1| G=g)$,  replacing the monotonicity assumption  $S^1 \geq S^0$ with the following Assumption  \ref{assump-s1}. 
We primarily show that $\P(Y^1, S^0, S^1| G=g)$ and  $\P(Y^0, S^0, S^1| G=g)$ are not identifiable in this case. 

 \begin{assumption} 
  \label{assump-s1}
  $S^1\indep G\mid S^0$.
\end{assumption}

According the proposed method in Section \ref{sec3-2} of the manuscript, under Assumptions \ref{assump5} and  \ref{assump-s1}, 
  \[     \P(S^1 = b \mid S^0 = a )      \]
are invariant parameters across trials, and they are identifiable if $( \P(S^0=0 \mid G=g), \P(S^0=1 \mid G=g) )_{m \times 2}$ is full column rank. 
In addition, $\P(S^0 = a \mid G=g )$ is identifiable under Assumption  \ref{assump5}. 
Thus, 
  the principal score $\P( S^0, S^1 \mid G=g)$ is identifiable with 
   	\[   \delta_{ab|g} := \P( S^0 = a, S^1 = b \mid G=g) =    \P(S^1 = b \mid S^0 = a )  \cdot \P(S^0 = a \mid G=g ),           \]

On the other hand, under Assumption \ref{assump7}, 
   \[      \pi_{1 |ab} :=   \P(Y^1 =1 | S^0 = a, S^1 = b, G = g)   \]
are also invariant parameters  across trials. In addition, 
 we have the following decomposion, 
       \begin{align*} 
       \P(Y^a =1 \mid G = g)  
      ={}& \pi_{1| 0 0}  \delta_{00|g} +  \pi_{1| 0 1}  \delta_{01|g}  +  \pi_{1|  10}  \delta_{ 10 |g} +  \pi_{1| 1 1}  \delta_{11|g},  \quad g = 1, ..., m,     
      \end{align*} 

It seems that the parameters $\{ \pi_{1|ab}: a, b = 0, 1 \}$ are identifiable by solving the above system of equations, under Condition \ref{cond-s1} below. 

\begin{condition} \label{cond-s1} $m\geq 4$, the matrix $( \P(S^0=0, S^1 =0 | G=g),   \P(S^0=0, S^1 =1 | G=g),  \P(S^0=1, S^1 =0 | G=g),  \P(S^0=1, S^1 =1 | G=g) )_{m\times 4}$ is full column rank.  
\end{condition} 
 
 However, Condition \ref{cond-s1} can never hold under Assumption \ref{assump-s1} due to the collinearity among $\delta_{ab|g}$. Specifically, 
     \[     \frac{  \delta_{01 |g}  }{  \delta_{ 00 |g}  } = \frac{  \P(S^1 = 1 \mid S^0 = 0 )  \cdot \P(S^0 = 0 \mid G=g )  }{    \P(S^1 = 0 \mid S^0 = 0 )  \cdot \P(S^0 = 0 \mid G=g )} =  \frac{  \P(S^1 = 1 \mid S^0 = 0 )  }{    \P(S^1 = 0 \mid S^0 = 0 ) },    \]
 which is a constant independent of $g$. Similarly,  $ \delta_{11 |g} / \delta_{10 |g}$ is a constant. 

 \section{\bcol{Discussion: Extension to General Treatments}}  \label{app-multivalued-treatment}

In this section, we discuss extensions to more general treatment settings. 
For a multivalued treatment, one would need a compatible collection of transportability assumptions linking several potential outcomes, together with corresponding rank conditions, and the compatibility of these assumptions is nontrivial.
As an example, suppose the treatment takes values in $\{0, 1, 2\}$, binary outcome $Y\in \{0, 1\}$, and let $\{Y^0, Y^1, Y^2\}$ denote the corresponding potential outcomes.
Following the proposed method in the manuscript, a plausible set of identifiability assumptions can be formulated as follows.

\begin{assumption} \label{assump-R3}  \quad \\
 (i)  $(Y^0, Y^1, Y^2)\indep A \mid G=g$,  $0<\P(A=1\mid G=g) <1$; \\
 (ii) $Y^1 \indep G\mid Y^0$; \\
 (iii) The matrix  $(\P(Y^0 = 0 \mid G= g),  \P(Y^0 = 1  \mid G= g ))_{m \times 2}$
     has a full-column rank; \\
 (iv)  $Y^2 \indep G\mid (Y^0, Y^1)$; \\
 (v)   The matrix  $(\P(Y^0 = 0, Y^1 = 0 \mid G= g), \P(Y^0 = 0, Y^1 = 1 \mid G= g), \P(Y^0 = 1, Y^1 = 0 \mid G= g), \P(Y^0 = 1, Y^1 = 1 \mid G= g))_{m \times 4}$ has a full-column rank.
\end{assumption}
Intuitively, Assumption \ref{assump-R3}(i) identifies the marginal distributions of the potential outcomes. Assumptions \ref{assump-R3}(i)--(iii), which are exactly the same as the conditions in Theorem 1 of the manuscript, identify the joint distribution $\P(Y^0, Y^1 \mid G = g)$. Following the same rationale, one might consider further using Assumptions \ref{assump-R3}(iv)--(v) to identify the conditional distribution $\P(Y^2 \mid Y^0, Y^1, G = g)$ and hence
the joint distribution $\P(Y^0, Y^1, Y^2 \mid G = g)$. However, these assumptions are not compatible: Assumption \ref{assump-R3}(ii) in fact  contradicts Assumption \ref{assump-R3}(v).  Specifically, if Assumption \ref{assump-R3}(ii) holds, then
     \[     \frac{  \P(Y^0 = 0, Y^1 = 1\mid G=g)  }{   \P(Y^0 = 0, Y^1 = 0\mid G=g)  } = \frac{  \P(Y^1 = 1 \mid Y^0 = 0 )  \cdot \P(Y^0 = 0 \mid G=g )  }{    \P(Y^1 = 0 \mid Y^0 = 0 )  \cdot \P(Y^0 = 0 \mid G=g )} =  \frac{  \P(Y^1 = 1 \mid Y^0 = 0 )  }{    \P(Y^1 = 0 \mid Y^0 = 0 ) },    \]
     which is a constant independent of $g$. This violates the rank condition in Assumption \ref{assump-R3}(v).


One possible way to identify  $\P(Y^0, Y^1, Y^2 \mid G = g)$ is to  replace Assumptions \ref{assump-R3}(ii)–(iii) with a monotonicity condition $Y^1 \ge Y^0$. The associated results are shown in Lemma \ref{lemma-R1}.
\begin{lemma}  \label{lemma-R1}
Under Assumptions \ref{assump-R3}(i), (iv), and (v), together with the condition $Y^1 \ge Y^0$,  $\P(Y^0, Y^1, Y^2 \mid G = g)$ is  identifiable.
\end{lemma}

Considering the role of the monotonicity condition, an alternative set of identifiability assumptions can be formulated as follows.
\begin{assumption}[Second Set of Possible Assumptions] \label{assump-R4}  \quad \\
 (i)  $(Y^0, Y^1, Y^2)\indep A \mid G=g$,  $0<\P(A=1\mid G=g) <1$; \\
 (ii) $Y^2 \geq Y^1 \geq Y^0$.
\end{assumption}

\begin{proposition} \label{thm-R2}
Under Assumption \ref{assump-R4},  $\P(Y^0, Y^1, Y^2 \mid G = g)$ is identifiable.
\end{proposition}

Although sequential monotonicity ($Y^2 \ge Y^1 \ge Y^0$) can be used to identify the joint distribution of the potential outcomes, it does not follow the same rationale as the proposed method in the manuscript. 

For continuous treatment, the potential outcome process $\{Y^a: a \in \mathcal{A}\}$ is infinite-dimensional. Our finite linear-system identification argument does not extend directly to identify the joint distribution of the whole potential outcome process. 

\section{Additional Numerical Results} \label{app-s9}

\subsection{\bcol{Additional Results for Simulation}}  \label{app-additional-simulation}

{\bf Additional simulations for the just-identified case.} We consider the just-identified case helps better evaluate the finite-sample performance of the proposed estimators. We include two additional simulation cases (C5)--(C6). These two cases follow the same data-generating processes as cases (C1)--(C2) in the manuscript, except that the number of trials is set to 2  to ensure just-identified settings. The results are presented in Table \ref{tab-R1}. As shown, the Bias is small and the CP95 values are close to 0.95, indicating that the proposed method performs well \emph{overall}. 

However, by comparing Table \ref{tab-R1} with Table 1 of the manuscript, we observe that:
(1) the SD and ESE are substantially larger than those in Table 1 (cases (C1)--(C2)) of the manuscript; and
(2) The SE and ESE are very close in Table 1 even when $n_g = 100$, whereas the SE and ESE in Table \ref{tab-R1} are unstable when $n_g = 100$, indicating that a larger sample size is required to obtain stable results in just-identified cases.

\begin{table*}[h!]
 \centering  
\caption{\centering Simulation results for cases (C5)--(C6).  \label{tab-R1}}
\resizebox{1\textwidth}{!}{\begin{tabular}{cc | rrrr | rrrr | rrrr}
  \hline
      & &     \multicolumn{4}{c|}{$n_g =100$}    & \multicolumn{4}{c|}{$n_g =200$} &  \multicolumn{4}{c}{$n_g =500$}  \\
Case & $\theta$ &  Bias & SD & ESE & CP95 & Bias & SD & ESE & CP95 & Bias & SD & ESE & CP95 \\ 
  \hline
\multirow{2}{*}{(C5)}  & $\pi_{1|0}$ & -0.014 & 0.289 & 0.312 & 0.964 & -0.006 & 0.181 & 0.184 & 0.952 & -0.001 & 0.104 & 0.107 & 0.945 \\ 
      &  $\pi_{1|1}$ & 0.009 & 0.154 & 0.164 & 0.964 & 0.003 & 0.098 & 0.102 & 0.951 & -0.001 & 0.059 & 0.060 & 0.948 \\  
   \hline  
\multirow{2}{*}{(C6)}  & $\pi_{1|0}$ & -0.016 & 0.345 & 0.296 & 0.960 & -0.008 & 0.166 & 0.166 & 0.958 & -0.004 & 0.090 & 0.093 & 0.945\\ 
 & $\pi_{1|1}$ & 0.006 & 0.161 & 0.146 & 0.965 & 0.004 & 0.089 & 0.088 & 0.950 & 0.001 & 0.049 & 0.050 & 0.952  \\  
   \hline 
\end{tabular}}
\begin{flushleft} \footnotesize 
Note: Bias and SD are the Monte Carlo bias and standard deviation over the 1000 simulations of the point estimates,  ESE and CP95 are the estimated asymptotic variances and coverage proportions of the 95\% confidence intervals based on 100 bootstraps, respectively. 
\end{flushleft}  
\end{table*}

\bigskip \noindent 
{\bf Additional simulations on the violation of Assumption 2.} 
 We also perform an additional simulation design in which Assumption~2 is violated by allowing the transition probabilities $\P(Y^1=1 \mid Y^0,G=g)$ to vary between trials to different degrees. 
  We consider the following data-generating process:  
\begin{itemize}
	\item  (C7):   $\P(Y^1=1\mid Y^0, G=g) = \text{expit}(Y^0 - 0.5 + \gamma_g)$ for $g = 1, ..., 10$,  $-\Gamma \leq \gamma_g \leq \Gamma$,
    where $\text{expit}(x) = \exp(x)/\{1 + \exp(x)\}$ is the standard logistic function, $\Gamma \geq 0$ is a constant, set to 0.1, 0.2, and 0.3 in our analysis. 
    For each simulation, 
$\gamma_g$  is drawn from the uniform distribution
 $\text{Uniform}(-\Gamma, \Gamma)$. 
    For each trial $g = 1, 2, ..., 10$, the potential outcome $Y^0$ follows from a Bernoulli distribution with $\P(Y^0 = 1  \mid G=g) = 0.5 + (g-1)/30$, i.e., 
    taking evenly spaced values at equal intervals from 0.5 to 0.8,   the binary treatment $A$ is randomly assigned with probability $\P(A=1 \mid G = g) = 0.5$, the sample size is set to 200.  
\end{itemize}

\begin{table*}[h!]
 \centering
\caption{\centering Simulation results for case (C7).  \label{tab-R1}}
\resizebox{1\textwidth}{!}{\begin{tabular}{cc | rrrr | rrrr | rrrr}
  \hline
      & &     \multicolumn{4}{c|}{$\Gamma =0.1$}    & \multicolumn{4}{c|}{$\Gamma =0.2$} &  \multicolumn{4}{c}{$\Gamma =0.3$}  \\
Trial & $\theta_g$ &  Bias & SD & ESE & CP95 & Bias & SD & ESE & CP95 & Bias & SD & ESE & CP95 \\
  \hline
\multirow{2}{*}{1}  & $\pi_{1|0,1}$ & 0.038 & 0.106 & 0.108 & 0.937 & 0.043 & 0.124 & 0.109 & 0.924 & -0.050 & 0.129 & 0.110 & 0.896   \\
      &  $\pi_{1|1,1}$ & 0.006 & 0.058 & 0.059 & 0.929 & 0.015 & 0.068 & 0.059 & 0.856 & -0.076 & 0.072 & 0.060 & 0.810     \\
   \hline
\multirow{2}{*}{2}  & $\pi_{1|0,2}$ & 0.020 & 0.106 & 0.108 & 0.942 & 0.055 & 0.124 & 0.109 & 0.914 & 0.077 & 0.129 & 0.110 & 0.906   \\
 & $\pi_{1|1,2}$ &  -0.013 & 0.058 & 0.059 & 0.933 & 0.028 & 0.068 & 0.059 & 0.856 & 0.048 & 0.072 & 0.060 & 0.802 \\
   \hline
\multirow{2}{*}{3}  & $\pi_{1|0,3}$ & 0.039 & 0.106 & 0.108 & 0.936 & 0.018 & 0.124 & 0.109 & 0.905 & -0.009 & 0.129 & 0.110 & 0.894    \\
      &  $\pi_{1|1,3}$ & 0.007 & 0.058 & 0.059 & 0.930 & -0.010 & 0.068 & 0.059 & 0.879 & -0.040 & 0.072 & 0.060 & 0.818      \\
   \hline
\multirow{2}{*}{4}  & $\pi_{1|0,4}$ &-0.003 & 0.106 & 0.108 & 0.936 & 0.049 & 0.124 & 0.109 & 0.900 & -0.018 & 0.129 & 0.110 & 0.885   \\
 & $\pi_{1|1,4}$ & -0.035 & 0.058 & 0.059 & 0.926 & 0.021 & 0.068 & 0.059 & 0.873 & -0.048 & 0.072 & 0.060 & 0.807  \\
   \hline 
\multirow{2}{*}{5}  & $\pi_{1|0,5}$ &  0.010 & 0.106 & 0.108 & 0.938 & 0.003 & 0.124 & 0.109 & 0.905 & -0.040 & 0.129 & 0.110 & 0.878   \\
      &  $\pi_{1|1,5}$ & -0.023 & 0.058 & 0.059 & 0.942 & -0.026 & 0.068 & 0.059 & 0.876 & -0.068 & 0.072 & 0.060 & 0.851   \\
   \hline
\multirow{2}{*}{6}  & $\pi_{1|0,6}$ &  0.044 & 0.106 & 0.108 & 0.939 & 0.008 & 0.124 & 0.109 & 0.899 & 0.074 & 0.129 & 0.110 & 0.868   \\
 & $\pi_{1|1,6}$ &  0.012 & 0.058 & 0.059 & 0.932 & -0.021 & 0.068 & 0.059 & 0.881 & 0.045 & 0.072 & 0.060 & 0.855  \\
   \hline
\multirow{2}{*}{7}  & $\pi_{1|0,7}$ &  0.016 & 0.106 & 0.108 & 0.939 & 0.040 & 0.124 & 0.109 & 0.896 & 0.062 & 0.129 & 0.110 & 0.863     \\
      &  $\pi_{1|1,7}$ &  -0.017 & 0.058 & 0.059 & 0.936 & 0.012 & 0.068 & 0.059 & 0.893 & 0.033 & 0.072 & 0.060 & 0.877   \\
   \hline
\multirow{2}{*}{8}  & $\pi_{1|0,8}$ &  0.025 & 0.106 & 0.108 & 0.938 & -0.025 & 0.124 & 0.109 & 0.895 & 0.049 & 0.129 & 0.110 & 0.863  \\
 & $\pi_{1|1,8}$ &   -0.008 & 0.058 & 0.059 & 0.933 & -0.052 & 0.068 & 0.059 & 0.902 & 0.018 & 0.072 & 0.060 & 0.886  \\
    \hline
\multirow{2}{*}{9}  & $\pi_{1|0,9}$ &  0.001 & 0.106 & 0.108 & 0.936 & -0.017 & 0.124 & 0.109 & 0.889 & 0.011 & 0.129 & 0.110 & 0.858    \\
      &  $\pi_{1|1,9}$ &   -0.031 & 0.058 & 0.059 & 0.939 & -0.044 & 0.068 & 0.059 & 0.912 & -0.020 & 0.072 & 0.060 & 0.883    \\
   \hline
\multirow{2}{*}{10}  & $\pi_{1|0,10}$ &  0.021 & 0.106 & 0.108 & 0.931 & -0.010 & 0.124 & 0.109 & 0.890 & 0.036 & 0.129 & 0.110 & 0.846  \\
 & $\pi_{1|1,10}$ &  -0.012 & 0.058 & 0.059 & 0.939 & -0.038 & 0.068 & 0.059 & 0.919 & 0.004 & 0.072 & 0.060 & 0.892   \\  \hline
 \multicolumn{2}{c|}{\bf Average} & {\bf 0.005} & 0.082 & 0.083 & {\bf 0.936} & {\bf 0.002} & 0.096 & 0.084 & {\bf 0.893} & {\bf 0.004} & 0.101 & 0.085 & {\bf 0.862}\\
   \hline
\end{tabular}}
\begin{flushleft} \footnotesize
Note: Bias and SD are the Monte Carlo bias and standard deviation over the 1000 simulations of the point estimates,  ESE and CP95 are the estimated asymptotic variances and coverage proportions of the 95\% confidence intervals based on 100 bootstraps, respectively.
\end{flushleft}
\end{table*}

For case (C7), $\Gamma \geq 0$ is the sensitivity parameter, with larger values indicating greater violations of Assumption 2.
When $\Gamma > 0$, there are no invariant parameters across trials. We let $\theta_g = (\pi_{1 \mid 0,g}, \pi_{1 \mid 1,g})$ denote the state transition probabilities for each trial $g$, where $\pi_{1 \mid 0,g} = \P(Y^1 = 1\mid Y^0 = 0, G=g)$ and $\pi_{1 \mid 1,g} = \P(Y^1 = 1\mid Y^0 = 1, G=g)$.
If we nonetheless treat them as invariant parameters and apply the proposed estimation method, the corresponding numerical results are shown in Table \ref{tab-R1}. 
From Table \ref{tab-R1}, we observe that as the degree of violation ($\Gamma$) increases, the point estimates of $\pi_{1 \mid 0,g}$ and $\pi_{1 \mid 1,g}$ become increasingly distorted, and the coverage of the nominal $95\%$ confidence intervals deteriorates. On average, however, the bias remains small, while CP95 declines from 0.925 to 0.849. 

\subsection{Additional Results for Application}

\begin{figure}[H]
\centering
\vspace{-5pt}  
\subfloat[]{
\begin{minipage}[t]{0.85\linewidth}
\centering
\includegraphics[width=1\textwidth]{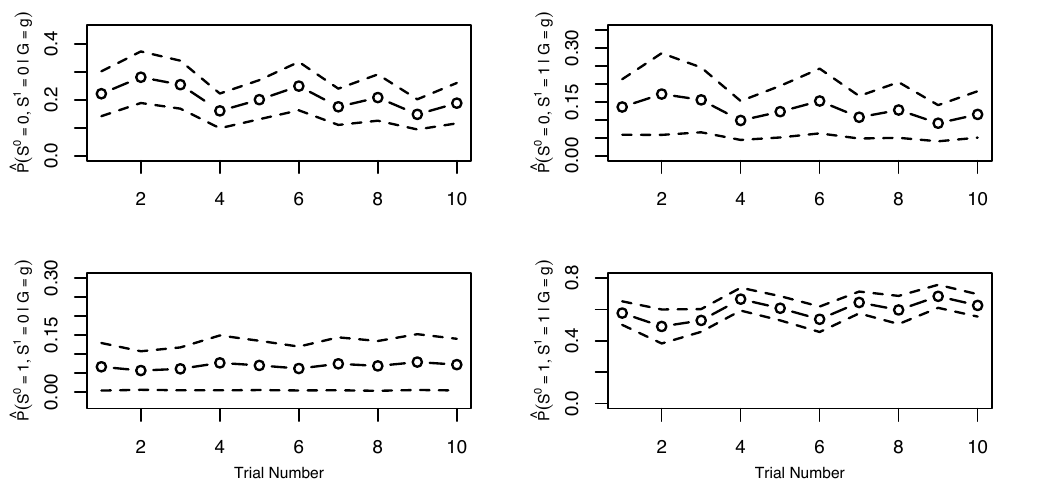}
\end{minipage}%
}%

\subfloat[]{
\begin{minipage}[t]{0.85\linewidth}
\centering
\includegraphics[width=1\textwidth]{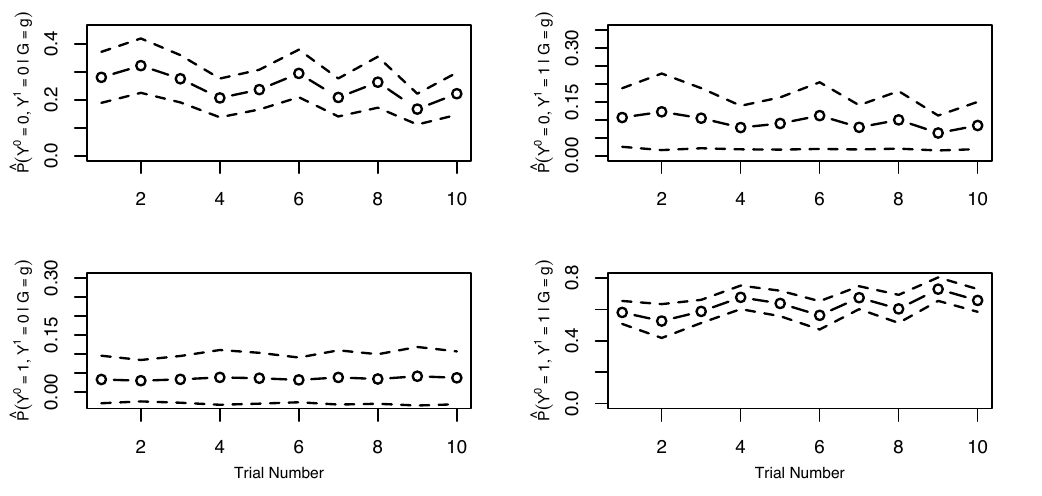}
\end{minipage}%
}%
\vspace{-2pt}  
\caption{(a) Estimated joint distribution $\P(S^0, S^1\mid G=g)$ for all $g \in \cG$; (b) Estimated joint distribution $\P(Y^0, Y^1\mid G=g)$ for all $g \in \cG$;}
\label{fig-s1} 
\end{figure}

We estimate the joint distributions $\P(S^0, S^1\mid G=g)$ and $\P(Y^0, Y^1\mid G=g)$  for $g \in \cG$. Figure \ref{fig-s1} displays the point estimates along with the corresponding pointwise 95\% confidence intervals.  From Figure \ref{fig-s1}, we observe that the point estimates of the joint distributions show only slight volatility across all trials, indicating low heterogeneity of these distributions among trials. Additionally, the estimated standard error of the point estimates is relatively small, ranging from 0.024 to 0.058. Moreover, it is noteworthy that the lower bounds of the 95\% confidence interval for $\hat \P(S^0 = 1, S^1 = 0 \mid G=g)$ for all trials are slightly greater than 0, suggesting that the monotonicity assumption (Assumption \ref{assump8}, $S^1 \geq S^0$) may not hold.  In contrast, the 95\% confidence intervals for $\hat \P(Y^0 = 1, Y^1 = 0 \mid G=g)$ cover 0 in all trials.  

\end{document}